%% file: main.tex
\documentclass{article}

\usepackage{graphicx}
\usepackage{amsmath}
\usepackage[margin=0.85in]{geometry}
\usepackage{xcolor}
\usepackage{bm}
\usepackage{hyperref}
\usepackage{subcaption}
\usepackage[ruled]{algorithm2e}
\usepackage{booktabs}
\usepackage{siunitx}
\usepackage{adjustbox}
\usepackage{arydshln}
\usepackage[numbers,sort&compress]{natbib}
\usepackage{enumitem}

\renewcommand{\textcolor}[2]{#2}

\input{title}

\begin{document}

\maketitle

\input{abstract}

\input{introduction}
\input{heat_conduction_problem}
\input{probabilistic_preliminaries}
\input{algorithm_methodology}
\input{creating_an_algorithm_with_synthetic_data}
\input{application_of_the_algorithm_to_real_data}
\input{conclusions_and_recommendations}

\input{acknowledgments}
\input{data_availability}

\input{apendices}

\bibliographystyle{unsrtnat}
\bibliography{references}

\end{document}

%% file: title.tex
\title{Discretization-optimized Bayesian model calibration for nonlinear constitutive modeling in heat conduction}
\author{Rodrigo L. S. Silva$^{1,2}$, Clemens Verhoosel$^{1}$, Erik Quaeghebeur$^{2}$
\\
\\
$^{1}$Department of Mechanical Engineering
\\
$^{2}$Department of Mathematics and Computer Science
\\
\\
Eindhoven University of Technology, The Netherlands}
\date{}

%% file: abstract.tex
\begin{abstract}
    We present a Bayesian model calibration framework for inferring nonlinear constitutive relationships in heat conduction problems, with a focus on temperature-dependent thermal conductivity.
    The proposed framework integrates gradient-based optimization and uncertainty quantification (UQ) to address the inverse problem of estimating the conductivity function from transient temperature measurements.
    A key contribution is an adaptive algorithm that sequentially refines both the numerical discretization for model simulation, and the model complexity used to represent the conductivity curve.
    The discretization is optimized through the minimization of a loss function, and Morozov’s discrepancy principle is used as an uncertainty-motivated stopping criterion.
    \textcolor{blue}{The model complexity is selected using an approach that balances maximizing the likelihood of the data with penalizing excessive model complexity.}
    As a result, the numerical and modeling biases remain of the same order as the uncertainty imposed by the measurement noise, leading to robust and computationally efficient inference.
    The methodology is demonstrated on both synthetic and experimental data, showing that it enables accurate calibration of nonlinear constitutive models with minimal overfitting and limited computational cost.
\end{abstract}

%% file: introduction.tex
\section{Introduction}
\label{sec-introduction}

In many engineering applications, physical systems are modeled by partial differential equations (PDEs), where some of the governing parameters or constitutive relations are not directly observable.
    These are estimated from measurement data, which inevitably contain measurement errors.
    Bayesian model calibration offers a powerful framework for estimating such parameters while also quantifying their uncertainties \citep{kaipio2006}.
    In these problems, numerical methods are typically required to solve the system of PDEs.
    These methods frequently require specifying discretization parameters, such as spatial and temporal resolution, which introduce approximations in the final solution due to discretization.
    Additionally, the complexity of the model used to represent the unknown quantities -- e.g., the number of segments of piecewise linear functions -- can significantly affect accuracy.
    Understanding how to select appropriate discretization and model complexity parameters is then crucial, so that the resulting model error remains small when compared to the measurement noise, while avoiding unnecessary computational effort from excessive refinement.

    A way to understand the impact of discretization and model fidelities in uncertainty quantification is through a decomposition of the observed data.
    Let $\bm{f}$  be the predictions obtained from the numerical model, and $\bm{t}$ the unknown true values of the physical quantity being measured.
    We can then write $\bm{t} = \bm{f} +  \bm{\epsilon}_{\rm pred}$, where $ \bm{\epsilon}_{\rm pred}$ is the discrepancy between the true values and the model prediction.
    This discrepancy captures the model bias, which occurs due to discretization and model approximations.
    Additionally, the observed data, denoted by $\bm{d}$, is represented as a noisy version of the truth, i.e., $\bm{d} = \bm{t} + \bm{\epsilon}_{\rm meas}$, where $\bm{\epsilon}_{\rm meas}$ represents the measurement noise.
    Combining these relations yields $\bm{d} = \bm{f} +  \bm{\epsilon}_{\rm pred} + \bm{\epsilon}_{\rm meas}$, highlighting that the total discrepancy between data and model predictions includes both measurement noise and numerical bias.
    For reliable uncertainty quantification, we aim at obtaining $ \bm{\epsilon}_{\rm pred}$ small relative to the measurement noise $ \bm{\epsilon}_{\rm meas}$, while ensuring it is not excessively small to avoid unnecessary computational effort.

    An example of model calibration involves estimating material properties that appear in constitutive models.
    To investigate the challenges of discretization, model complexity and uncertainty quantification in a well-defined yet practically relevant setting, we focus on the estimation of a temperature-dependent thermal conductivity.
    This property plays a central role in Fourier’s law, which relates heat flux to the temperature gradient \citep{ozisik1993}.
    Although often a constant conductivity is assumed, this assumption often fails in systems experiencing large thermal gradients, where conductivity can vary significantly with temperature \citep{stelzer1987, ozisik1993, mota2010, ramos2022}.
    This problem serves as a representative example that reflects conditions commonly encountered in real-world engineering systems, where data are noisy and direct measurement of material properties is infeasible.

    Many inverse problem approaches have been developed to estimate temperature-dependent constitutive models over the past decades \citep{yang1999, alifanov1978, huang1991, huang1995, sawaf1995, chen1996, martin2000, lin2001, kim2002, aquino2006, zueco2006, mierzwiczak2011, czel2012, mohebbi2017}.
    These methods calibrate the parameters of the constitutive model to match available data, using techniques such as gradient-based optimization \citep{yang1999, alifanov1978, huang1991, huang1995, sawaf1995, chen1996, kim2002, mohebbi2017}, neural networks \citep{aquino2006}, and genetic algorithms \citep{czel2012}.
    Most studies focus on point estimates and do not formally account for uncertainty, although inverse problems can also be framed probabilistically to allow uncertainty quantification \citep{ozisik2018}.

    Bayesian inference offers a natural approach for this purpose by incorporating prior knowledge and expressing the solution in probabilistic terms \citep{kaipio2006, ozisik2018, silva2024, rinkens2025squeeze}.
    \textcolor{blue}{
    Several Bayesian frameworks have been proposed for estimating constitutive models \citep{madireddy2015, madireddy2016, tao2021, aggarwal2023, yue2021, do2022, akintunde2019, rappel2018bayesian, rappel2019identifying}, including practical tutorials \cite{rappel2020tutorial}.
    }
    Bayesian applications to transient heat problems under linear assumptions can be found in references \citep{orlande2008, naveira2010, naveira2011, gnanasekaran2011, lanzarone2014bayesian, berger2016bayesian, woo2022estimation}.
    However, the nonlinear case, where the conductivity depends on temperature, has received less attention.
    Mota et al. \citep{mota2010} applied Bayesian inference to estimate the coefficients of an exponential model for temperature-dependent conductivity, while Ramos et al. \citep{ramos2022} considered a piecewise constant conductivity defined over temperature intervals.
    Both papers highlight that extending to the nonlinear case significantly changes the structure of the inverse problem, requiring a model to describe the full functional dependence of the conductivity on temperature.

    \textcolor{blue}{
    Alongside parameter estimation, there is growing interest in integrating mesh refinement into the inference process.
    For example, Calvetti et al. \citep{calvetti2020bayesian} proposed a hierarchical Bayesian method that treats mesh parameters as random variables, and the mesh is refined iteratively during inference.
    The refinement is guided by the behavior of the posterior distribution, concentrating resolution in regions where parameters that are spatially distributed exhibit sharp gradients.
    Such methods do not address the adaptation of model complexity -- for instance, determining the number of segments in piecewise linear functions used to model a parameter.
    This leaves open the need for methods that adapt both the discretization and the model complexity in a coordinated way.
    }

    \textcolor{blue}{In this work, we address the challenge of achieving accurate parameter estimation, while controlling computational cost through a coordinated adjustment of discretization and model complexity, in inverse problems governed by partial differential equations.
    }
    Our coordinated approach offers a more computationally feasible \textcolor{blue}{practical} alternative to fully Bayesian methods, such as evaluating the model evidence \citep{kaipio2006}, which are often intractable in practical situations.
    We propose a Bayesian framework that integrates uncertainty quantification with sequential refinement of both the numerical discretization and the model complexity.
    Rather than treating these elements as random variables, we adapt them iteratively, using the measurement noise as a reference to determine when additional refinement no longer improves the reliability of the inference.
    This approach ensures that the bias introduced by discretization and model approximation remains controlled relative to the uncertainty inherent in the data.
    We demonstrate this strategy through the estimation of a temperature-dependent thermal conductivity.
    This is a representative example, where model complexity and numerical resolution must be balanced with the limits imposed by measurement noise.

    Our novel algorithm is defined by combining two main features: (i) discretization refinement guided by gradient-based optimization; (ii) \textcolor{blue}{model complexity refinement guided by statistical model selection criteria.}
    These features are defined as follows:
    \begin{itemize}
    \item[(i)] We introduce an approach that begins with the definition of a loss function based on the maximum a posteriori (MAP) estimate, which integrates prior information with the likelihood derived from the data.
    An initial estimate of the temperature-dependent thermal conductivity is obtained through a gradient-based optimization method, which also serves to initialize the uncertainty quantification stage.
    To improve estimation accuracy while controlling computational effort, we apply a sequential mesh refinement, guided by the behavior of the loss function.
    Spatial and temporal discretizations are progressively refined until a stopping criterion -- based on the measurement noise -- is met, ensuring that the resolution is sufficient to capture relevant thermal dynamics without incurring unnecessary computational cost.
    \item[(ii)] \textcolor{blue}{
    For model complexity selection we consider standard information criteria, namely the Bayesian Information Criterion (BIC) and the Deviance Information Criterion (DIC) \cite{gelman1995bayesian}.
    The BIC is based on the maximized likelihood, while the DIC relies on Markov Chain Monte Carlo (MCMC) sampling of the posterior distribution and thus incorporates uncertainty information.
    While criteria like the BIC are computationally cheap compared to sampling-based approaches, they rely only on point estimates and may therefore not quantify uncertainties adequately.
    Therefore, in our uncertainty quantification framework we rely on the sampling-based DIC.
    In cases where there is less specific interest in uncertainty quantification, the BIC can be attractive.
    }
    \end{itemize}
    
    This paper is structured as follows:
    In \autoref{sec-heat-conduction-problem}, we introduce the heat conduction problem used as our case study, including the physical model, mathematical formulation, and numerical implementation.
    In \autoref{sec-probabilistic-preliminaries}, we present the Bayesian methodology, the gradient-based MAP estimation, and the MCMC-based uncertainty quantification.
    In \autoref{sec-algorithm-methodology}, we detail the methodology used for algorithm development, including mesh refinement, model complexity control, and stopping criteria.
    \autoref{sec-creating-algorithm} presents the application of this framework to synthetic data.
    Finally, in \autoref{sec-application-algorithm}, we demonstrate the algorithm using real experimental measurements, and in \autoref{sec-conclusions}, we summarize the key findings and contributions of the work, as well as directions for future research.

%% file: heat_conduction_problem.tex
\section{Heat conduction problem}
\label{sec-heat-conduction-problem}

\begin{figure}
    \centering
    \begin{subfigure}{0.35\linewidth}
        \centering
        \includegraphics[width=\linewidth]{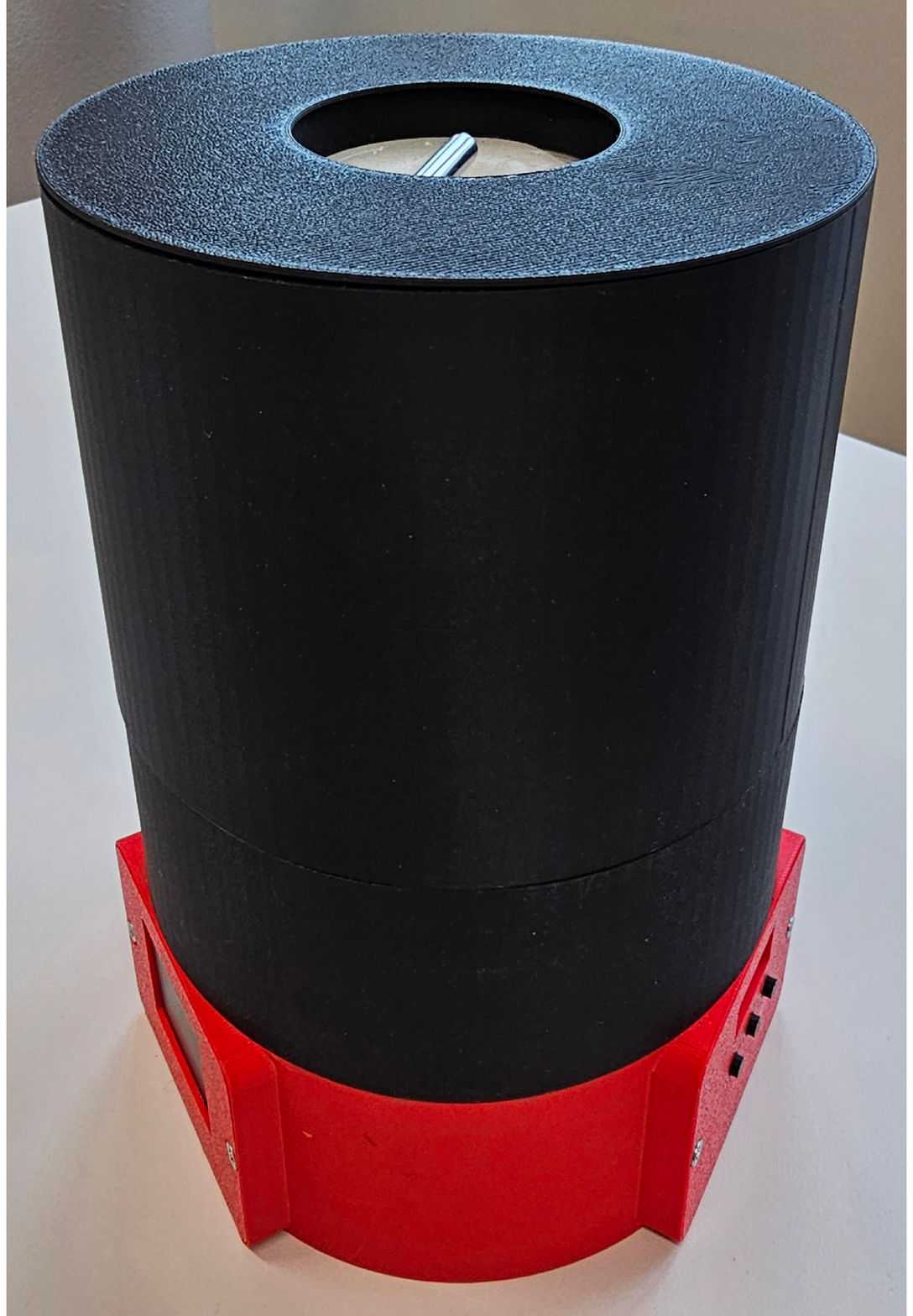}
        \caption{}
        \label{fig-setup}
    \end{subfigure}
    \hfill
    \begin{subfigure}{0.55\linewidth}
        \centering
        \includegraphics[width=\linewidth]{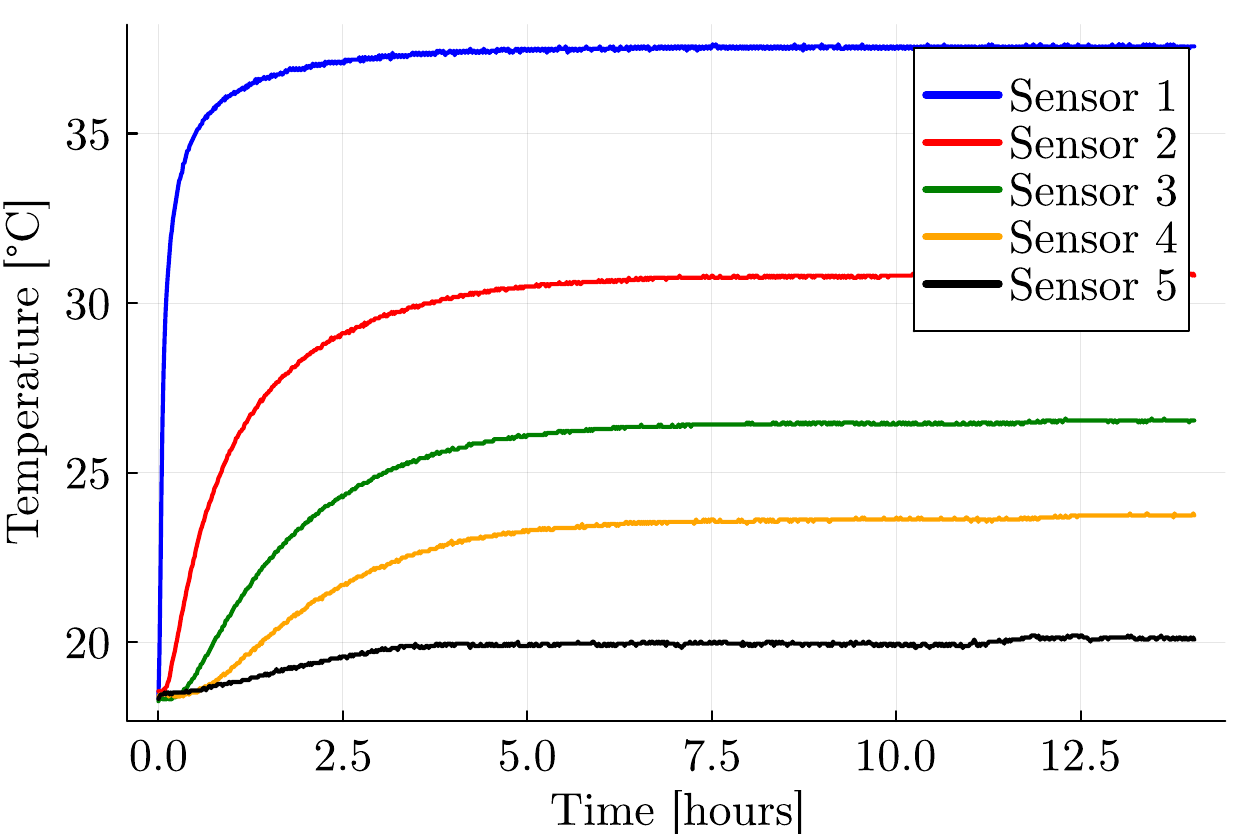}
        \caption{}
        \label{fig-temperature-measurements}
    \end{subfigure}
    \caption{(\subref{fig-setup}) Experimental heat conduction setup;
    (\subref{fig-temperature-measurements}) Typical temperature measurements acquired with the five sensors.}
    \label{fig-setup-and-measurements}
\end{figure}

As an illustrative example, we consider the heat conduction problem described by Rinkens et al. \cite{rinkens2025squeezethermal}.
The setup (\autoref{fig-setup}) consists of a rod made of paraffin wax that is heated at the bottom and cooled at the top.
The lateral surface is covered with an insulating material.
Technical details about this setup are discussed in \autoref{sec-application-algorithm}.

We aim at estimating the temperature-dependent thermal conductivity of the paraffin wax \citep{thermtestinstruments}.
For that, we use a total of 4 sensors, located along the rod's axis, to acquire transient temperature measurements.
The first sensor is close to the heat source and the fourth sensor is close to the top surface of the rod.
Additionally, a fifth sensor is placed above the rod to measure the ambient temperature. 
Typical temperature measurements acquired by the five sensors are shown in \autoref{fig-temperature-measurements}.

\subsection{Physical system}

\begin{figure}
    \centering
    \includegraphics[width=0.9\linewidth]{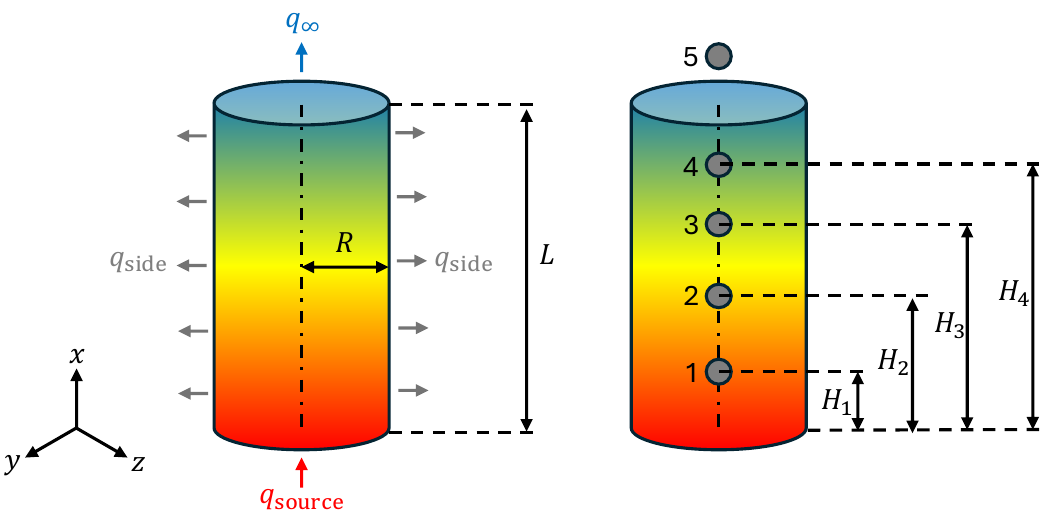}
    \caption{Illustration of the physical system.}
    \label{fig-physical-system}
\end{figure}

An illustration of the physical system is shown in \autoref{fig-physical-system}.
The rod has radius $R = \SI{28.6}{\mm}$ and length $L = \SI{93.0}{\mm}$.
The temperature sensors are denoted by the numbers 1 to 5, and their distances with respect to the bottom of the rod are respectively equal to $H_1 = \SI{5.0}{\mm}$, $H_2 = \SI{25.8}{\mm}$, $H_3 = \SI{45.0}{\mm}$ and $H_4 = \SI{66.5}{\mm}$.
Additionally, the coordinate system $x$ has its origin at the bottom of the rod.

The initial temperature of the rod is $T_0$.
The heat the rod receives at the bottom is denoted by $q_{\rm source}$, and the heat that is lost at the top to the ambient medium is denoted by $q_\infty$.
Additionally, the heat lost to the ambient medium at the lateral surface is denoted by $q_{\rm side}$.
The quantities $q_{\rm source}$, $q_\infty$ and $q_{\rm side}$ are modeled using Newton's law of cooling, which describes the rate of heat exchange as proportional to the temperature differences between the rod and its surroundings.
Specifically, $q_{\rm source}$ is modeled by a heat transfer coefficient $h_{\rm source}$ and a constant source temperature $T_{\rm source}$, while the heat loss to the ambient, $q_\infty$, is modeled by a heat transfer coefficient $h_\infty$ and a time-varying ambient temperature $T_\infty (t)$.
Similarly, the lateral heat exchange $q_{\rm side}$ follows Newton's law with a heat transfer coefficient $h_{\rm side}$ and the same ambient temperature $T_\infty (t)$.

The relevant material properties of the slab are its specific heat $c_p$, density $\rho$, and unknown temperature-dependent thermal conductivity $k(T)$.
In transient heat conduction problems, the temperature dependence of density, specific heat and heat transfer coefficients can also play an important role \citep{huang1991, huang1995, sawaf1995, mota2010, ramos2022}, but this work is focused on the constitutive model, so we consider these parameters to be temperature-independent and precisely known.
Hence, of the physical properties listed above, the thermal conductivity $k(T)$ has a special status, as this is the property that needs to be estimated.

\subsection{Mathematical formulation}

We model the temperature evolution by considering the energy balance in the rod \citep{ozisik1993}, which yields
\begin{equation}
    \rho c_p \dfrac{\partial T(\bm{x},t)}{\partial t} = \nabla \cdot [k(T) \nabla T(\bm{x},t)],
    \label{eq-gov-full}
\end{equation}
where $\bm{x} = (x,y,z)$ and $t$ are the position vector and time, respectively.

\textcolor{blue}{
Because the rod radius is significantly smaller than its length, radial heat diffusion occurs on a characteristic time scale $t_r \sim R^2/\alpha$, whereas axial diffusion occurs on $t_x \sim L^2/\alpha$, where $\alpha = k / (\rho c_p)$ denotes the thermal diffusivity \cite{ozisik1993}.
Their ratio is therefore $t_r/t_x = (R/L)^2 \approx 0.094$ for the present geometry, implying that radial temperature gradients relax approximately 10 times faster than axial gradients.
We therefore approximate the temperature distribution in a cross-section as uniform.
The impact of this modeling assumption is further assessed in \ref{appendix-changes-loss-model}, where a compact sensitivity study is conducted to evaluate how such simplifications and changes in the loss model influence the inferred conductivity.
}

Integration of \autoref{eq-gov-full} over the cross-section then yields the simplified heat equation
\begin{equation}
    \rho c_p \dfrac{\partial T(x,t)}{\partial t} = \dfrac{\partial}{\partial x} \left[ k(T) \dfrac{\partial T(x,t)}{\partial x} \right] +
    \dfrac{2 h_{\rm side}}{R} [T_\infty(t) - T(x,t)].
    \label{eq-gov}
\end{equation}
Note that the temperature field, $T(x,t)$, now depends only on $x$ and $t$.

This differential equation is complemented by the boundary conditions
\begin{subequations}
\label{eq-bc}
    \begin{align}
        k(T) \dfrac{\partial T(0,t)}{\partial x} &= h_\text{source} [T(0,t) - T_\text{source}],
        \\
        k(T) \dfrac{\partial T(L,t)}{\partial x} &= h_\infty [T_\infty(t) - T(L,t)],
    \end{align}
\end{subequations}
and the initial condition $T(x,0) = T_\infty(0)$.
This system of equations models the forward problem, and its solution provides values of $T(x,t)$ given the physical properties of paraffin wax, heat transfer coefficients, and the prescribed temperatures of the ambient medium and the heating element.

\subsection{Numerical simulation}

To evaluate the forward model defined by the initial boundary value problem described above, we use a numerical simulation based on discretization techniques.
We discretize the spatial domain using linear finite elements \citep{lewis1996finite}, while the temporal domain is discretized using the backward Euler method \citep{butcher2016numerical}.
In addition, we define a parameterized model class for the temperature-dependent thermal conductivity, which is integrated into the discretized system.
These strategies used in the numerical simulation are specified below.

\subsubsection{Spatial discretization}

The total number of finite elements is denoted by $n_e$.
For simplicity, we consider that these elements equally divide the spatial domain $[0, L]$, so the element length is $\Delta x = L / n_e$.
The temperature field is then approximated as
\begin{equation}
    T(x,t) \approx \sum_{i=1}^{n_e+1} N_i(x) T_i(t) = \bm{N}(x) \bm{T}(t),
\end{equation}
where $\bm{N}(x) = [N_1(x), \dots, N_{n_e + 1}(x)]$ is the row vector of $n_e + 1$ linear basis functions constructed over the uniform spatial mesh, and $\bm{T}(t) = [T_1(t), \dots, T_{n_e + 1}(t)]^T$ is the column vector of nodal temperatures.

The discretization of the spatial derivatives in \autoref{eq-gov} yields
\begin{equation}
    C \dfrac{d \bm{T}(t)}{dt} = - \left[ K(\bm{T}) + \beta C \right] \bm{T}(t) + \bm{s},
    \label{eq-gov-spatial-discretization}
\end{equation}
where $C$, $K$ and $\bm{s}$ are respectively the capacitance matrix, the conductance matrix and the source vector \citep{lewis1996finite}, and $\beta = 2 h_{\rm side} / (R \rho c_p)$.
Regarding the conductance matrix, the conductivity value used for a specific element is obtained by considering the average temperature of its nodes.

\subsubsection{Temporal discretization}

Similarly to the spatial discretization, we consider uniformly sized time steps, denoted by $\Delta t$.
So $t_m = m \Delta t$, where $m = 1, 2, \dots, n_t$ is the time step index.
We then denote the nodal temperatures at time step $m$ as $\bm{T}(t_m) = \bm{T}_{m}$.
Similarly, the conductance matrix evaluated at the same time step is denoted by $K(\bm{T}_m) = K_m$.

To discretize the temporal domain in \autoref{eq-gov-spatial-discretization} with backward Euler, we replace $\bm{T}(t)$ with $\bm{T}_{m+1}$, and the temporal derivative $d\bm{T}(t) / dt$ is approximated as $(\bm{T}_{m+1} - \bm{T}_m) / \Delta t$, which yields
\begin{equation}
    \left[ \left( \dfrac{1}{\Delta t} + \beta \right) C + K_m \right] \bm{T}_{m+1} = \dfrac{1}{\Delta t} C \bm{T}_m + \bm{s}.
    \label{eq-FEM-linear-system}
\end{equation}
The nodal temperatures at time $m + 1$ are then obtained by solving this linear system.

\subsubsection{Relation between the number of elements and time steps}

While the backward Euler method is unconditionally stable in the classical numerical sense, physical accuracy considerations impose a lower bound on the time step \citep{szabo2008discretization}.
This lower bound is required to preserve the qualitative properties
of the real physical solution.

This constraint leads to a direct relation between the number of finite elements and the number of time steps in the simulation.
In particular, a fixed number of time steps $n_t$ corresponds to a minimum number of finite elements $n_e$.
This relation is given by
\begin{equation}
    n_e > \underline{n}_e(n_t) = \sqrt{\dfrac{n_t L^2 \rho c_p}{6 k_{\rm min} t_{\rm total}}},
    \label{eq-relation-ne-nt}
\end{equation}
where $k_{\rm min}$ is the minimum conductivity value in the finite elements and $t_{\rm total}$ is the total simulation time.
This defines an approximated lower bound for $n_e$, which is sufficiently accurate for our purposes in this work.

\textcolor{blue}{
In practice, the exact value of $k_{\rm min}$ is not known a priori, since it is part of the parameters being inferred.
In our adaptive algorithm, this value is updated at each iteration using the current estimate of the conductivity profile across the finite elements.
This ensures that the corresponding lower bound on the number of elements $n_e$ is always consistent with the latest parameter estimates.
}

\subsubsection{Model class for the conductivity}
\label{sec-model-class-for-the-conductivity}

To discretize the temperature-dependent thermal conductivity $k(T)$, we aim at selecting a model class that properly captures its variation with the temperature.
This choice serves both to approximate the underlying physical behavior and to enable the estimation of the conductivity function within the inference problem.

The model class we choose to represent the conductivity as a function of temperature is piecewise linear functions.
This model class is particularly convenient due to its simplicity and flexibility: the function is defined by linearly interpolating between a fixed set of points, and the model complexity can be increased by adding more segments.

\begin{figure}
    \centering
    \begin{subfigure}{0.49\linewidth}
        \centering
        \includegraphics[width=\linewidth]{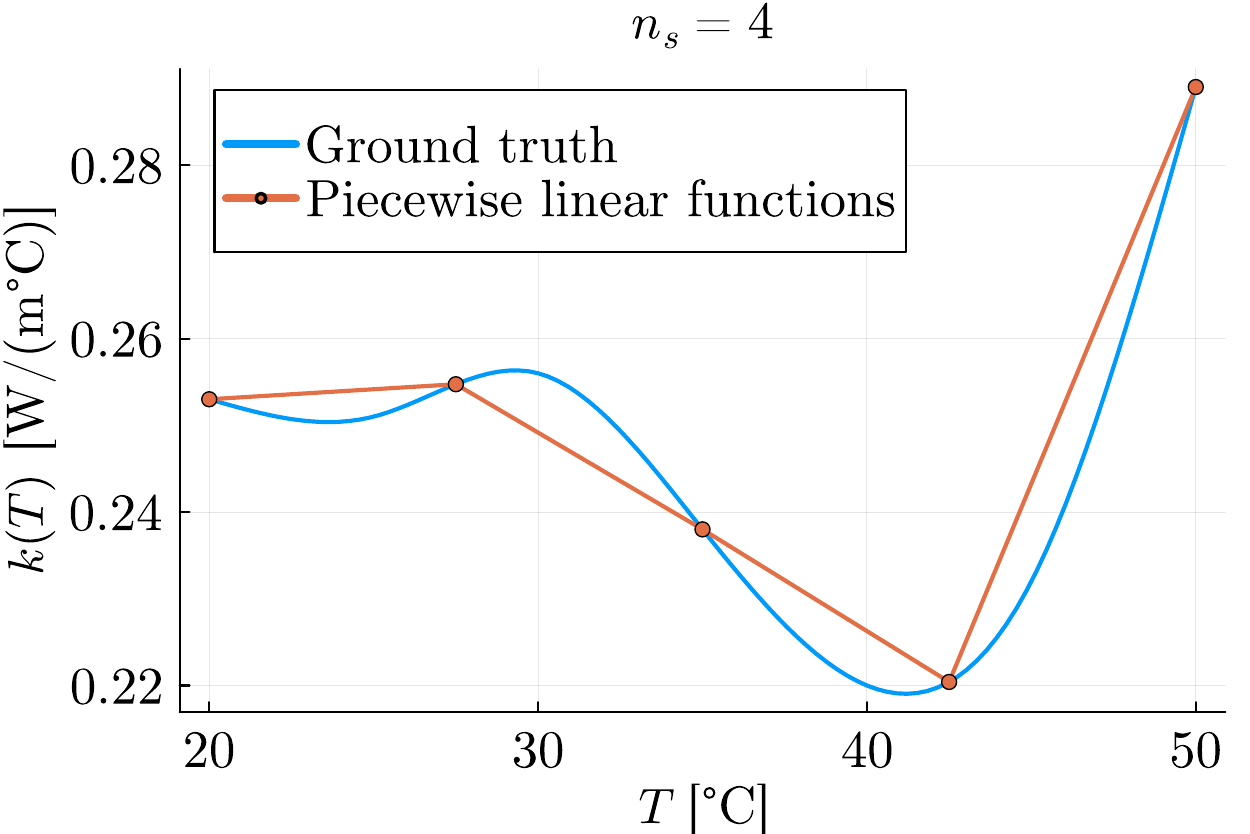}
        \caption{}
        \label{fig-piecewise-ns4}
    \end{subfigure}
    \hfill
    \begin{subfigure}{0.49\linewidth}
        \centering
        \includegraphics[width=\linewidth]{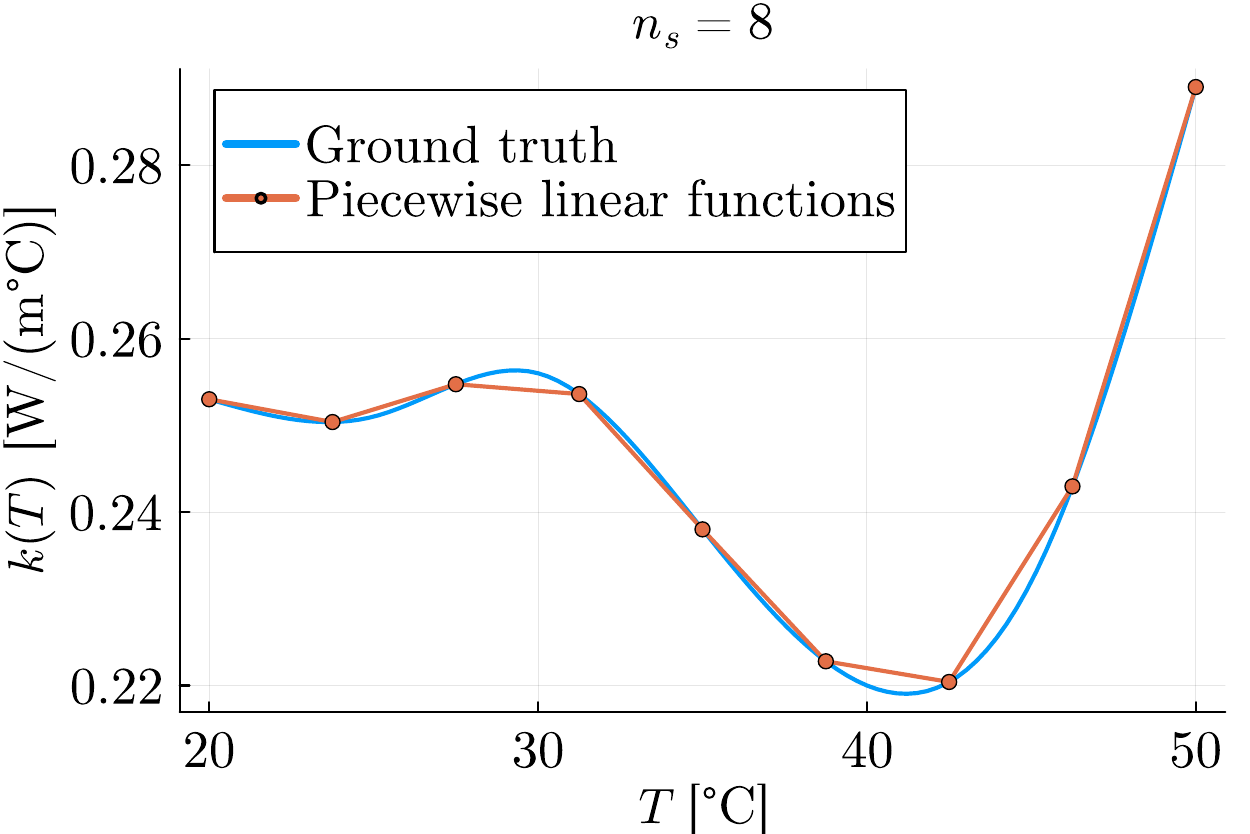}
        \caption{}
        \label{fig-piecewise-ns8}
    \end{subfigure}
    \caption{Discretization of a ground-truth for the conductivity function with piecewise linear functions with (\subref{fig-piecewise-ns4}) $n_s = 4$ segments and (\subref{fig-piecewise-ns8}) $n_s = 8$ segments.}
    \label{fig-piecewise-ground-truth}
\end{figure}

To define the piecewise linear functions, we first define the lower and upper bounds of the temperature domain using the minimum and maximum values observed in the temperature measurements.
For a given number of segments, denoted by $n_s$, we then select $n_s + 1$ interpolation points that are equally spaced within this temperature range.
The conductivity values at these points serve as the model parameters, and the conductivity at any other temperature is obtained by linear interpolation.
This procedure yields a continuous piecewise linear representation of $k(T)$ that is consistent with the temperature range explored in the simulation.
\autoref{fig-piecewise-ground-truth} illustrates the discretization of a ground-truth for the conductivity function with piecewise linear functions, where two different values for the number of segments $n_s$ are considered.
In \autoref{sec-algorithm-methodology} we show how to optimize the number of segments in conjunction with the uncertainty quantification framework.

The piecewise linear conductivity function is evaluated at each element based on the average nodal temperature, as described in the spatial discretization section.
Moreover, we treat the conductivity explicitly in time, that is, we assume that the conductivity values at time step $m + 1$ are calculated with the temperatures obtained at time step $m$.
\textcolor{blue}{
In \ref{appendix-implicit-conductivity} we present a study in which an implicit scheme for the conductivity is considered.
The difference observed between the solutions obtained with the explicit and implicit schemes are much smaller than the typical measurement noise, and the implicit scheme is more computationally expensive.
Hence, we employ the explicit scheme in this work.
}

%% file: probabilistic_preliminaries.tex
\section{Probabilistic preliminaries}
\label{sec-probabilistic-preliminaries}

In this section, we present the main components of our probabilistic methodology, which is applicable to a broad class of inverse problems.
The framework described in this paper does not rely on the specific physical or mathematical details of the application.
Instead, it defines general tools for parameter estimation and uncertainty quantification.
These tools form the foundation of the algorithm introduced in \autoref{sec-algorithm-methodology}, where they are adapted to the particular structure of the thermal conductivity inference problem.
We begin with the Bayesian formulation, which allows prior knowledge to be combined with experimental data.
Next, we define a loss function and describe its minimization using a gradient-based algorithm.
Finally, we introduce the use of Markov chain Monte Carlo (MCMC) methods to explore the posterior distribution and quantify uncertainty in the inferred parameters.

\subsection{Bayesian perspective}

In many practical applications, prior information on the parameters to be estimated is available.
This information can be obtained, for example, from previous experiments, the literature, or experts' opinions.
Bayesian frameworks are useful for such cases, since prior information can be formally taken into account during the estimation of the parameters of interest.

The estimation of parameters within the Bayesian framework can be summarized in the following steps \citep{gelman1995bayesian}:
We start by defining $\bm{p}$ as the vector with the parameters to be estimated.
Based on the information available for $\bm{p}$, we select a prior distribution $P(\bm{p})$, which represents the statistical model of the information available for the parameters.
Then we select a likelihood function $P(\bm{d} \mid \bm{p})$.
This likelihood captures how the data vector $\bm{d}$ is related to the parameters $\bm{p}$.
Finally, we explore the posterior distribution $P(\bm{p} \mid \bm{d})$, which is the conditional probability distribution of the unknown parameters given the data.

In this way, Bayesian estimation amounts to obtaining the posterior distribution.
This posterior is obtained using Bayes' theorem \citep{gelman1995bayesian}, which states
\begin{equation}
    P(\bm{p} \mid \bm{d}) = \dfrac{P(\bm{p}) P(\bm{d} \mid \bm{p})}{P(\bm{d})}.
\end{equation}
In the above equation, $P(\bm{d})$ is called the evidence and plays the role of a normalizing constant.
The evaluation of $P(\bm{d})$ is usually computationally expensive \citep{ozisik2018, kaipio2006}, so we decide not to evaluate it.
Instead, we represent Bayes' theorem as \citep{ozisik2018, kaipio2006}
\begin{equation}
    P(\bm{p} \mid \bm{d}) \propto P(\bm{p}) P(\bm{d} \mid \bm{p}),
    \label{eq-bayes}
\end{equation}
which shows that the posterior $P(\bm{p} \mid \bm{d})$ depends not only on the likelihood $P(\bm{d} \mid \bm{p})$, but also on the prior $P(\bm{p})$.
Hence, even if the prior information is only qualitatively available, it must be mathematically modeled as a statistical distribution \citep{ozisik2018}.

\subsection{Gradient-based MAP estimation}
\label{sec-gradient-based-minimization}

\textcolor{blue}{
To optimize the vector of parameters, we formulate a loss function $S$ according to the maximum a posteriori (MAP) principle \citep{ozisik2018, kaipio2006}.
Maximizing the posterior probability $P(\bm{p} \mid \bm{d})$ is equivalent to minimizing the negative logarithm of the posterior.
So, considering $\bm{d}$ is fixed, \autoref{eq-bayes} leads to the loss function
\begin{equation}
    S(\bm{p})
    = S_\text{prior}(\bm{p}) + S_\text{like}(\bm{p})
    = - \log[P(\bm{p})] - \log[P(\bm{d} \mid \bm{p})].
    \label{eq-loss-function}
\end{equation}
We use the logarithm of the probability densities to prevent numerical overflows from large exponentials and underflows from very small probabilities \citep{ozisik2018, kaipio2006}.
}

To minimize the loss function, gradient-based methods update the parameters iteratively in the direction of the negative gradient of the loss.
For example, the basic gradient descent method \citep{nocedal1999numerical} follows the update rule
\begin{equation}
    \bm{p}_{i+1} = \bm{p}_i - \eta \nabla S(\bm{p}_i),
\end{equation}
where $\eta$ is a positive step size.
In our work, we use a more sophisticated algorithm, namely Newton’s method with a trust region \citep{nocedal1999numerical}.
This approach improves stability by taking advantage of second-order derivative information through the Hessian of the loss function, which allows faster convergence near the optimum.

\subsection{Uncertainty quantification with MCMC}
\label{sec-uncertainty-quantification-with-mcmc}

To extend the point estimates obtained with the gradient-based algorithm, we use MCMC to quantify the uncertainties in the inferred parameters.
In MCMC methods, samples of the posterior distribution are generated by stochastic simulation.
Then, inference on the posterior distribution -- i.e., deriving statistics and expectations -- is performed through inference on these samples \citep{ozisik2018, kaipio2006}.
In this way, MCMC methods can approximate the posterior.

A Markov chain is a stochastic process in which, given the present state, past and future states are independent \citep{gamerman2006}.
If a Markov chain, independently of its initial distribution, reaches a stage that can be represented by a specific distribution $\lambda$, and retains this distribution for all subsequent stages, we say that $\lambda$ is the limit distribution of the chain, and that the chain has reached equilibrium \citep{gamerman2006}.
For the Bayesian approach, the Markov chain is constructed in such a way that its limit distribution coincides with the posterior.

The Metropolis-Hastings algorithm \citep{ozisik2018, kaipio2006} is a classic choice as a method to explore the posterior distribution.
It is easy to implement and widely used for the solution of inverse problems \citep{ozisik2018, kaipio2006}.
In this algorithm, a candidate state is proposed from a specified proposal distribution given the current state, and it is accepted with a probability that depends on the ratio of the unnormalized posterior densities at the candidate and current state (\autoref{eq-bayes}).
It iteratively generates a Markov chain $[\bm{p}^1, \bm{p}^2, \dots, \bm{p}^R]$, where $R$ is the number of steps of the algorithm.

The efficiency of the algorithm is highly sensitive to the choice of step size, as it directly influences how well the proposal distribution explores the parameter space \citep{ozisik2018, kaipio2006, haario2001adaptive}.
To automate the step-size tuning process, thus avoiding the need for manual adjustments, we employ the adaptive Metropolis-Hastings algorithm proposed by Vihola \citep{vihola2012robust}.
This adaptive algorithm dynamically adjusts the proposal distribution by updating its covariance matrix, which improves the sampling procedure efficiency and convergence.
\textcolor{blue}{
In particular, the adaptation mechanism aims at achieving an acceptance ratio close to the theoretical optimal value of approximately 23\% for high-dimensional target distributions \cite{vihola2012robust}.
}

The initial sequence $[\bm{p}^1, \bm{p}^2, \dots, \bm{p}^r]$, $1 < r < R$, which contains all samples before reaching equilibrium, is called the burn-in period \citep{ozisik2018, kaipio2006, gamerman2006}.
If the chain $[\bm{p}^{r+1}, \bm{p}^{r+2}, ..., \bm{p}^R]$ satisfies a convergence criterion, then it is used to represent samples from the limit distribution, and therefore from the posterior.
The convergence criterion we use in this work is the one proposed by Geweke \citep{geweke1991evaluating}.
This criterion evaluates the mean of the samples of the first 10\% ($\mu_{10}$) and of the last 50\% ($\mu_{50}$) of the states in the chain $[\bm{p}^{r+1}, \bm{p}^{r+2}, ..., \bm{p}^R]$.
If the difference between these means is small, then the convergence criterion is satisfied \citep{ozisik2018, geweke1991evaluating}.
We implement this criterion by checking the values of $| (\mu_{10} - \mu_{50}) / \mu_{10} |$ and $| (\mu_{10} - \mu_{50}) / \mu_{50} |$.
If both values are smaller than or equal to $10^{-2}$, then it is satisfied.
\textcolor{blue}{
In \ref{appendix-multiple-chains}, we include a study in which the results obtained from multiple chains with dispersed initializations are analyzed to assess convergence.
}

\textcolor{blue}{
In addition to convergence diagnostics, we evaluate the effective sample size (ESS) \cite{gamerman2006} of the retained samples.
Because successive states in a Markov chain are generally correlated, the number of effectively independent samples is smaller than the total number of posterior samples.
The ESS quantifies this reduction by estimating the equivalent number of independent samples that would provide the same estimator variance.
For all main results, we report the ESS values.
}

%% file: algorithm_methodology.tex
\section{Algorithm methodology}
\label{sec-algorithm-methodology}

In this section, we present the methodology used to develop our algorithm for parameter estimation.
The goal is to maintain a balance between computational efficiency and accuracy.
A key aspect of our methodology is to obtain an appropriate level of simulation fidelity, which involves both the resolution of the numerical discretization and the complexity of the parameterized model.
To achieve this, we introduce a criterion for guiding refinement and model selection based on the measurement errors.

We first describe the optimization strategies used to refine the numerical discretization and select the model complexity.
Finally, we show the iteration scheme that integrates these components, detailing the nested optimization structure and the stopping criteria that control the overall refinement process.

\subsection{Measurement errors and Morozov's discrepancy principle}
\label{sec-measurement-errors-and-morozov}

Measurement errors exist in any data acquisition process.
If an estimation process does not account for measurement errors, it may provide misleading results.
On one hand, if the numerical resolution or model complexity is too high relative to the noise level, the estimation process may overfit and become unnecessarily computationally expensive, capturing random fluctuations rather than the actual underlying behavior.
On the other hand, a coarse discretization or an overly simple model may fail to capture essential characteristics of the data, leading to biased estimates.

To achieve an optimal balance, a criterion is needed to determine the appropriate level of refinement in both discretization and model complexity.
Morozov’s discrepancy principle \citep{morozov2012methods, ozisik2018} provides a way to find this level of refinement.
In \autoref{sec-introduction} we show that the data relates to the model prediction as $\bm{d} = \bm{f} +  \bm{\epsilon}_{\rm pred} + \bm{\epsilon}_{\rm meas}$.
We now define the residual error, which is the difference between the data and the model prediction, as $\bm{\epsilon}_{\rm res} = \bm{d} - \bm{f} =  \bm{\epsilon}_{\rm pred} + \bm{\epsilon}_{\rm meas}$.
Morozov's discrepancy principle suggests that the residual error should be comparable in magnitude to the measurement error, i.e.,
\begin{equation}
    \| \bm{\epsilon}_{\rm res} \| \approx \| \bm{\epsilon}_{\rm meas} \|,
    \label{eq-morozov-original}
\end{equation}
where $\| \cdot \|$ is the $L^2$ norm.
This means that, if the residual is significantly larger than the measurement error, the numerical model should be refined to reduce its error.
However, if the discrepancy is sufficiently close to the measurement error, no further refinement for the numerical model is needed.

If we assume that the measurement error follows a multivariate Gaussian distribution as $\bm{\epsilon}_{\rm meas} \sim \mathcal{N}(\bm{\mu}_{\rm meas}, \Sigma_{\rm meas})$, where $\Sigma_{\rm meas} = {\rm diag}(\bm{\sigma}^2_{\rm meas})$, then a pragmatic formulation of the discrepancy principle is to require that the deviation of the residual error from the measurement bias does not exceed a small multiple of the measurement standard deviation \citep{ozisik2018}, that is,
\begin{equation}
    \| \bm{\epsilon}_{\rm res} - \bm{\mu}_{\rm meas} \| \leq (1 + \gamma) \| \bm{\sigma}_{\rm meas} \|.
    \label{eq-morozov-gamma}
\end{equation}
This corresponds to enforcing the residual error to remain within one measurement error standard deviation, up to a small positive tolerance $\gamma$.

\textcolor{blue}{
The threshold in \autoref{eq-morozov-gamma} depends on the measurement noise and the number of data points through the norm $|| \bm{\sigma}_{\rm meas} ||$, which scales with the collective variance of all measurements.
A larger dataset or higher noise variance increases the magnitude of the threshold, while the tolerance parameter $\gamma$ provides an adjustable margin to control how closely the residuals are required to match the expected measurement error.
This ensures that the criterion adapts to both the data size and noise characteristics when guiding discretization and model complexity refinement.
By using this criterion, we can obtain an appropriate simulation fidelity, preventing overfitting while maintaining sufficient resolution.
}

\subsection{Information criteria}
\label{sec-information-criteria}

\textcolor{blue}{
Selecting an appropriate model requires balancing goodness of fit with complexity.
Information criteria provide quantitative measures to compare models with different numbers of parameters while penalizing over-parameterization.
We consider two fundamentally different information criteria, namely the point-estimate Bayesian Information Criterion (BIC) and the sampling-based Deviance Information Criterion (DIC) \cite{gelman1995bayesian}.
}

\textcolor{blue}{
The BIC evaluates models based on the maximized likelihood, while introducing a penalty that increases with the number of parameters.
It is defined as \cite{gelman1995bayesian}
\begin{equation}
    \mathrm{BIC} = -2 \log(\hat{L}) + n_p \ln (n_d),
\end{equation}
where $\hat{L}=\max_{\bm{p}}P(\bm{p} \mid \bm{d})$ is the maximized value of the likelihood function of the model, and $n_p$ and $n_d$ are respective the dimensions of $\bm{p}$ and $\bm{d}$.
The penalty term $n_p \ln (n_d)$ discourages complex models, favoring those that achieve a good fit with fewer parameters.
Lower BIC values indicate a more favorable trade-off between accuracy and complexity.
}

\textcolor{blue}{
BIC is based on a point estimate of the parameters and does not explicitly account for Markov chain samples.
To incorporate the information from these samples in the model selection, we also consider the DIC, which is defined as \cite{gelman1995bayesian}
\begin{equation}
    \mathrm{DIC} = -2 \log P(\bm{d} \mid \bar{\bm{p}}) + 2p_D,
\end{equation}
where $\bar{\bm{p}}$ is the posterior mean.
Next to that, $p_D$ is the so-called effective number of parameters defined as $p_D = 2V$, where $V$ is the variance of the log of the likelihood function values obtained with the samples of the posterior.
Similarly to BIC, smaller DIC values are preferred.
}

\textcolor{blue}{
There are alternative Bayesian model comparison methods that also use samples from the posterior distribution, such as the Widely Applicable Information Criterion (WAIC) and Leave-one-out cross-validation (LOO-CV) \cite{gelman1995bayesian}.
These can be more robust and informative than the DIC, but their computation is more demanding.
For the sake of computational efficiency and methodological simplicity, we therefore restrict our analysis to the DIC in this paper.
}

\subsection{Prediction fidelity}
\label{sec-prediction-fidelity}

We define prediction fidelity as the combination of discretization fidelity and model fidelity, both of which directly affect the prediction error $\bm{\epsilon}_{\rm pred}$.
The former refers to the resolution of the numerical discretization, and the latter to the complexity of the parameterized model.

\subsubsection{Discretization fidelity}
\label{sec-discretization-fidelity}

A coarse discretization may fail to provide sufficiently accurate estimates, and an excessively fine discretization increases computational cost without necessarily improving accuracy.
An adaptive approach can be used to refine the discretization, ensuring a balance between numerical accuracy and computational efficiency.

If we consider the heat conduction problem of \autoref{sec-heat-conduction-problem}, the discretization fidelity is determined by the number of finite elements used to discretize the spatial domain $n_e$, and the number of time steps used to discretize the temporal domain $n_t$.
A coarse spatial discretization, defined by a small value for $n_e$, may fail to capture temperature gradients accurately, while an insufficient number of time steps $n_t$ can lead to poor resolution of transient effects.
Additionally, an excessively refined discretization in both spatial and temporal domains increases computational cost without necessarily improving the accuracy of the estimates.

In \autoref{sec-gradient-based-minimization} we define the loss function as only depending on the vector of parameters, denoted $S(\bm{p})$.
Now we extend this definition to explicitly include dependence on the numerical discretization parameters used in the simulation.
In particular, since the accuracy of the model predictions is influenced by the number of spatial elements $n_e$ and time steps $n_t$, we write the loss function as $S(\bm{p}; n_e, n_t)$.
This highlights that the loss function also depends on the discretization choices that define the simulation fidelity.

\subsubsection{Model fidelity}
\label{sec-model-fidelity}

A model that is too simple may not capture essential behavior, while an overly complex model can introduce unnecessary parameters and increase the estimation uncertainties.
Hence, similarly to the discretization fidelity, we can use an adaptive approach to select a model fidelity that offers a good balance between estimation accuracy and uncertainties.

For the heat conduction problem described in \autoref{sec-heat-conduction-problem}, the model fidelity is determined by the number of segments $n_s$ used in the piecewise linear functions to represent the temperature-dependent thermal conductivity.
A small number of segments may oversimplify the functional dependence, while too many segments increase the number of degrees of freedom, possibly leading to overfitting and larger uncertainty in the inferred parameters.

In analogy to the discretization fidelity, we introduce the dependence of the loss function on the model fidelity.
We now write the loss function as $S(\bm{p}; n_e, n_t, n_s)$, which highlights that the loss function also depends on the choices that define the model fidelities.

\subsection{Optimization for prediction fidelity}

\textcolor{blue}{
To refine both discretization and the model, we use different strategies that are later combined into a nested scheme.
We first describe the strategy used for the discretization refinement, followed by the strategy used for the model refinement.
We then integrate both procedures into a unified iterative approach.
}

\subsubsection{\textcolor{blue}{Discretization optimization}}

Starting from an initial coarse discretization, we iteratively refine it, tracking the impact on the loss function.
The refinement evolves as follows:
\begin{enumerate}[label=Step \arabic*:, left=1.25em, labelwidth=2cm, labelsep=0.5em, align=left]
    \item We select an initial discretization and perform gradient-based optimization.
    \item We refine the discretization and perform gradient-based optimization again.
    \item We compare the values of the loss function, and proceed with the refinement (Step 2) until a stopping criterion is satisfied.
\end{enumerate}

There are multiple ways to refine the discretization, depending on the problem considered.
In the case of the heat conduction problem shown in \autoref{sec-heat-conduction-problem}, the discretization involves both spatial and temporal meshes.
One possible refinement strategy is to first increase the resolution of the spatial mesh while keeping the temporal discretization fixed, then refine the temporal mesh while maintaining the spatial resolution.
By comparing the resulting loss function values, we can choose the refinement that leads to the greatest improvement.
A detailed analysis of this approach is presented in \autoref{sec-creating-algorithm}.

We consider two stopping criteria for the discretization fidelity.
The first stopping criterion is based on Morozov's discrepancy principle, as indicated in \autoref{sec-measurement-errors-and-morozov}.
However, relying only on this criterion may lead to unnecessary mesh refinement, particularly when the discrepancy remains large due to limited model complexity rather than discretization error.
To address this, a second stopping criterion is introduced to detect when mesh refinement has reached its limits and further improvement requires increasing the model complexity instead.
This second criterion involves monitoring the variation of the loss function between consecutive refinement steps.
If the loss function does not change significantly, it indicates that further mesh refinement is unlikely to yield significant improvements, and the remaining discrepancy may be reduced more effectively by refining the model complexity.

\subsubsection{\textcolor{blue}{Model optimization}}

\textcolor{blue}{
While gradient-based optimization is effective for refining discretization fidelity, we use another strategy more suitable to refine the model complexity.
Increasing the model complexity often improves the fit to the data, but can also lead to overfitting and increase the uncertainties in the estimates \citep{piironen2017comparison}.
For this reason, we rely on the evaluation of the BIC and the DIC.
By monitoring both metrics, we can identify when to properly stop the model complexity refinement.
}

\textcolor{blue}{
The refinement of the model fidelity involves the following steps:
\begin{enumerate}[label=Step \arabic*:, left=1.25em, labelwidth=2cm, labelsep=0.5em, align=left]
    \item We start with a simple model complexity.
    \item Given the discretization, we evaluate the BIC.
    \item We run the MCMC method and evaluate the DIC with samples from the posterior.
    \item If none of the stopping criteria is satisfied, we incrementally increase the model complexity and go back to Step 2.
\end{enumerate}
}

\textcolor{blue}{
We consider two stopping criteria for the model fidelity.
Similarly to the discretization fidelity, the first stopping criterion is based on Morozov's discrepancy principle.
If this condition is not satisfied, a second criterion is used based on the analysis of the information criteria, where the BIC and DIC are evaluated to identify an appropriate balance between model fit and complexity.
}

\subsubsection{Nested optimization}

The optimization of both discretization fidelity and model fidelity can be combined into an iterative scheme.
In this process, we first optimize the discretization fidelity while keeping the model complexity fixed, and then we optimize the model fidelity after the discretization has been refined.
This approach ensures that the computational cost is mainly focused on refining the discretization, which is often the more computationally demanding task.
Once the discretization is sufficiently refined, the model complexity is increased to improve the representation of the ground truth.

\textcolor{blue}{
The nested optimization involves the following steps:
\begin{enumerate}[label=Step \arabic*:, left=1.25em, labelwidth=2cm, labelsep=0.5em, align=left]
    \item We select the initial model complexity and discretization.
    \item We refine the discretization keeping the model complexity fixed, until a stopping criterion for the discretization is reached.
    \item If a stopping criterion for the model complexity is not reached, we increase the model complexity and go back to Step 2.
\end{enumerate}
}
    
In \autoref{sec-creating-algorithm}, we elaborate on how to define the nested optimization scheme for the heat conduction problem indicated in \autoref{sec-heat-conduction-problem}.

%% file: creating_an_algorithm_with_synthetic_data.tex
\section{Creating an algorithm with synthetic data}
\label{sec-creating-algorithm}

    In this section, we use the heat conduction problem as a case study to create a concrete algorithm.
    The goal is to estimate and quantify the uncertainties of a temperature-dependent thermal conductivity, while simultaneously determining appropriate spatial and temporal meshes for the numerical discretization.

    By using synthetic data generated from a known ground truth, we create a controlled environment in which we can evaluate the performance of the algorithm.
    This approach allows us to test the refinement strategies used for the discretization and model fidelities, and define suitable stopping criteria before applying the framework to real experimental data.
    Once the methodology is finalized using synthetic data, we proceed with a real set of data, as detailed in \autoref{sec-application-algorithm}.

    \subsection{Generating synthetic data}
    \label{sec-generating-synthetic-data}

        In the context of the heat conduction problem, the synthetic data translates to simulated temperature measurements.
        These simulated temperature measurements are obtained by solving the direct problem for a given ground truth and then adding randomly generated noise to this solution.

\begin{figure}
            \centering
            \includegraphics[width=0.5\linewidth]{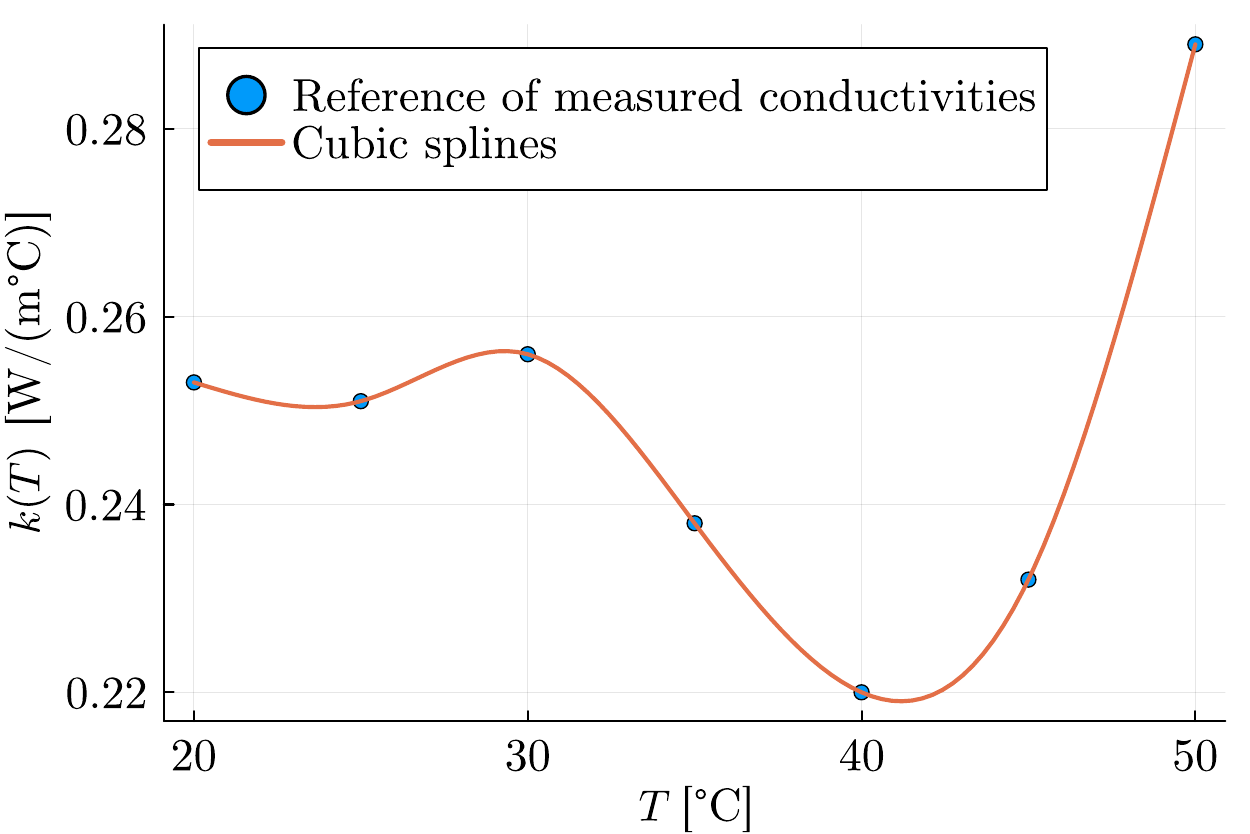}
            \caption{Thermal conductivity as a function of temperature.
            The blue circles represent the references of measured conductivities \citep{thermtestinstruments}.
            The orange curve represents the cubic splines interpolating these points.}
            \label{fig-ground-truth}
        \end{figure}

        We define the ground truth for the thermal conductivity as a function of temperature based on measured values of the conductivity of paraffin wax \citep{thermtestinstruments}.
        We consider a total of seven discrete conductivity values at specific temperature points, as shown in \autoref{fig-ground-truth}.
        To create a smooth and continuous representation of the conductivity function, we interpolate these measured points using cubic B-splines.
        The resulting spline function serves as the ground truth for the conductivity in our algorithm development.

        Once the ground truth is defined, the next step is to specify the time instances at which these measurements will be recorded.
        We set the simulation time to 12 hours, which was enough for reaching the steady state condition.
        We consider that each sensor records temperature values every 20 seconds, resulting in a total of 2160 measurements per sensor.
        We then solve the direct problem using the heat conduction model described in \autoref{sec-heat-conduction-problem}, with the modification that the ambient temperature is assumed to be constant over time (i.e., $T_\infty(t) = T_\infty$).
        The values of the physical properties and boundary conditions are shown in \autoref{tab-parameters-simulated-measurements}.
        
\begin{figure}
            \centering
            \includegraphics[width=0.6\linewidth]{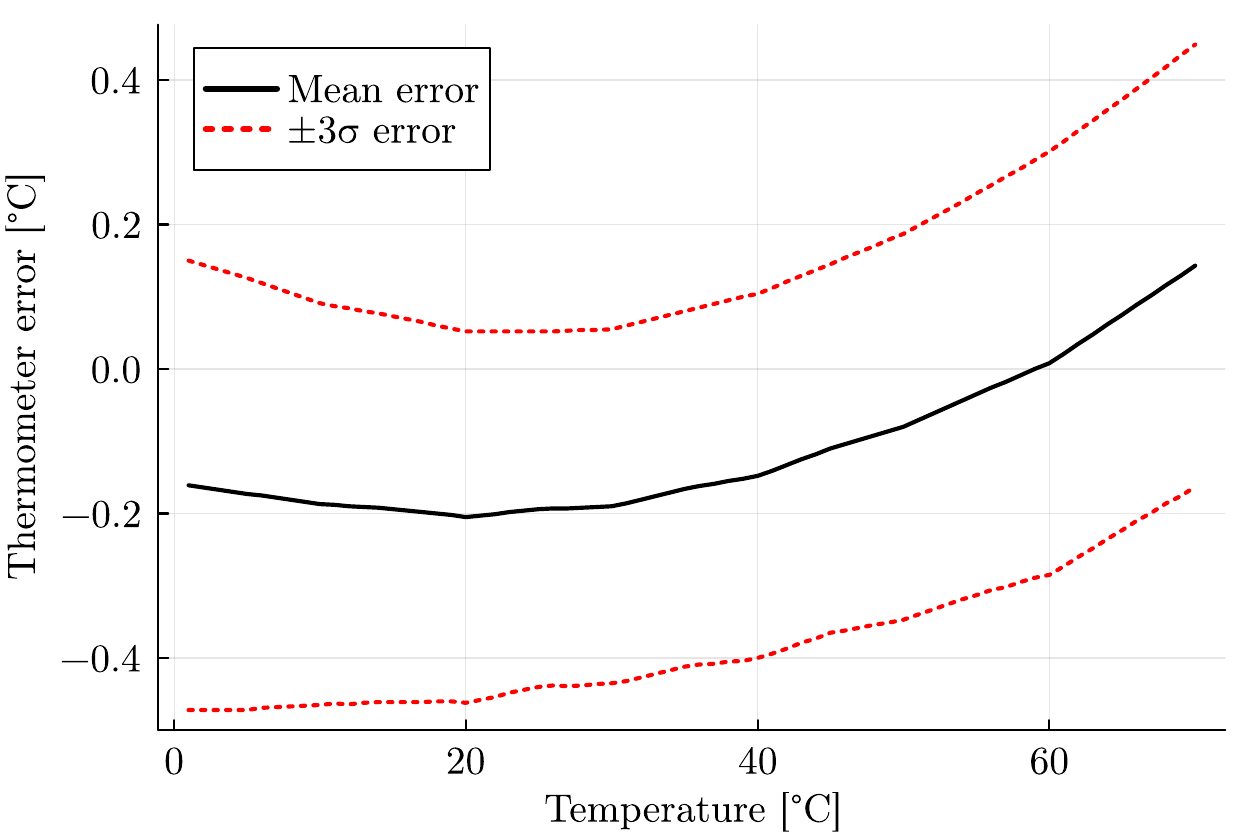}
            \caption{Error mean and standard deviation curves for the temperature sensors \citep{DS18B20}.}
            \label{fig-sensor-error-curve}
        \end{figure}

\begin{figure}
            \centering
            \includegraphics[width=0.6\linewidth]{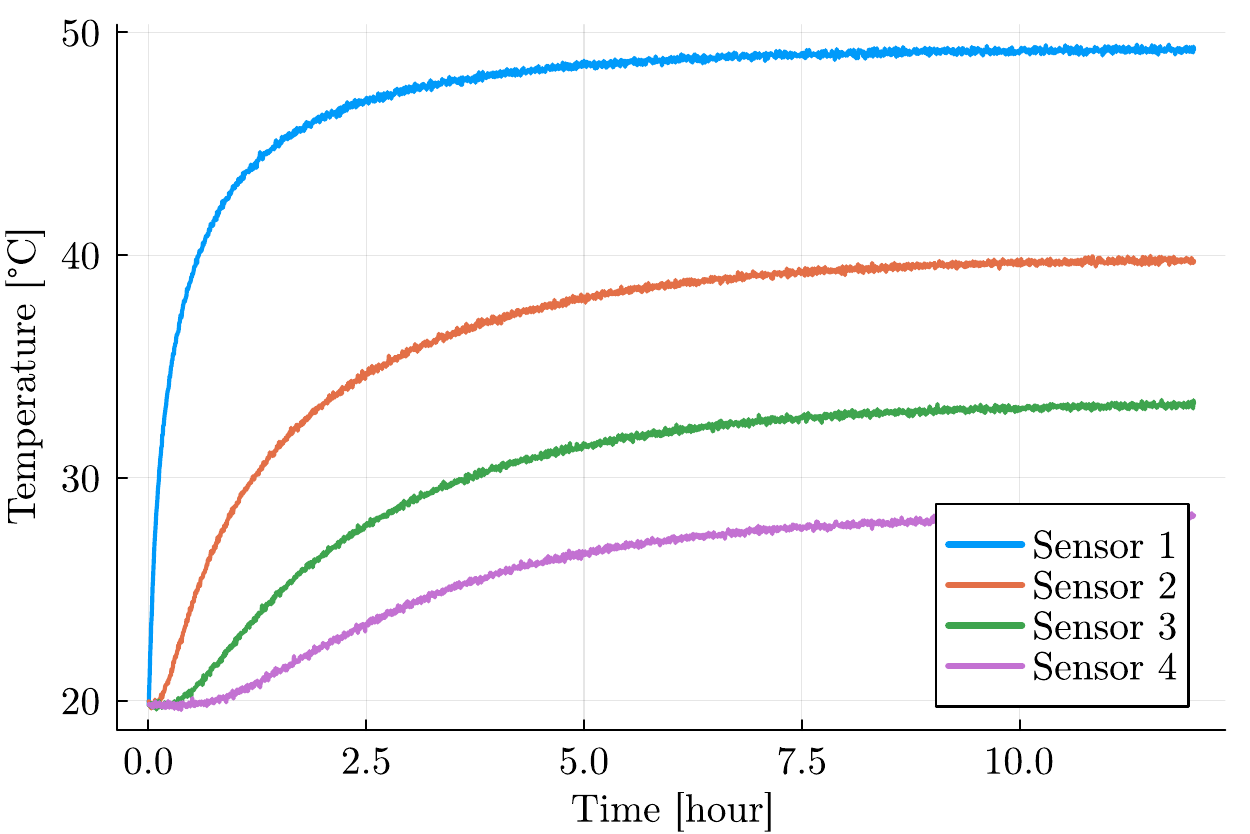}
            \caption{Simulated temperature measurements.}
            \label{fig-simulated-data}
        \end{figure}

        We introduce Gaussian noise to the obtained temperature values to simulate experimental conditions.
        This noise has a mean and standard deviation that are temperature-dependent (\autoref{fig-sensor-error-curve}).
        Their values are based on the specifications of the real sensors used for real data acquisition in \autoref{sec-application-algorithm}, ensuring that the virtual measurements reflect realistic uncertainty levels.
        To estimate the mean and standard deviation at a given temperature, we first select 70 points over the mean and standard deviation curves.
        These points are equally spaced in the temperature range from $\SI{1}{\degreeCelsius}$ to $\SI{70}{\degreeCelsius}$.
        We use a linear interpolation to find the mean and standard deviation at the given temperature.
        The obtained simulated temperature measurements are illustrated in \autoref{fig-simulated-data}.

        \begin{table}
            \centering
            \caption{Values of the physical properties and boundary conditions used to obtain the simulated temperature  \citep{rinkens2025squeezethermal}.}
            \begin{tabular}{lcr}
                \toprule
                Parameter & Value & Unit
                \\
                \midrule
                $\rho$ & $9.0 \times 10^2$ & \si{kg/m^3}
                \\
                $c_p$ & $2.1 \times 10^3$ & \si{J/(kg.\degreeCelsius)}
                \\
                $h_\text{source}$ & $2.5 \times 10^1$ & \si{W/(m^2.\degreeCelsius)}
                \\
                $h_{\rm side}$ & $1.0 \times 10^0$ & \si{W/(m^2.\degreeCelsius)}
                \\
                $h_\infty$ & $1.0 \times 10^1$ & \si{W/(m^2.\degreeCelsius)}
                \\
                $T_\text{source}$ & $5.7 \times 10^1$ & \si{\degreeCelsius}
                \\
                $T_\infty$ & $2.0 \times 10^1$ & \si{\degreeCelsius}
                \\
                $T_0$ & $2.0 \times 10^1$ & \si{\degreeCelsius}
                \\
                \midrule
            \end{tabular}
            \label{tab-parameters-simulated-measurements}
        \end{table}

    \subsection{Bayesian setup}
    \label{sec-bayesian-setup}

        The piecewise linear functions used to represent the conductivity curve are obtained with the linear interpolation of a set of discrete points $(T_i, k_i)$, $i = 1, ..., n_s + 1$.
        As mentioned in \autoref{sec-model-class-for-the-conductivity}, the values $T_i$ are chosen to be equally spaced in the temperature domain.
        So these values are known in advance, with $T_1$ and $T_{n_s + 1}$ equal to the minimum and maximum measured temperature, respectively.
        Hence, only the corresponding conductivity values $k_i$ need to be estimated, so we define the vector of parameters as $\bm{p} = [k_1, ..., k_{n_s + 1}]^T$.
        Additionally, we define the vector with data $\bm{d}$ as the vertical concatenation of vectors $\bm{d}_s$, $s = 1, ..., 4$, which contain the (simulated) temperature measurements of the $s$-th sensor (see \autoref{fig-simulated-data}).
        Hence, $\bm{d}$ has $ n_d = 4 \times 2160 = 8640$ elements.

        \textcolor{blue}{
        We use a multivariate normal distribution as the prior for the conductivity values $k_i$, defined as
        \begin{equation}
            P(\bm{p}) =
            \mathcal{N}(\bm{p} \mid \bm{\mu}_k, \Sigma_k) =
            (2 \pi)^{-(n_s+1)/2} |\Sigma_k|^{-1/2}
            \exp \left[ -\dfrac{1}{2} \left( \bm{p} - \bm{\mu}_k \right)^T
            \Sigma_k^{-1} \left( \bm{p} - \bm{\mu}_k \right) \right].
        \end{equation}
        The mean vector $\bm{\mu}_k$ has all its elements equal to
        \SI{0.3}{W/(m.\degreeCelsius)}, a value chosen based on typical conductivities reported for paraffin wax \citep{rinkens2025squeezethermal}.
        To impose smooth correlation between conductivity values as a function of temperature, the covariance matrix $\Sigma_k$ is constructed using the squared exponential kernel \cite{rasmussen2003gaussian},
        \begin{equation}
            \Sigma_k(T_i,T_j) =
            \sigma_k^2 \exp \left[
            -\dfrac{1}{2} \left( \dfrac{T_i - T_j}{l} \right)^2
            \right],
        \end{equation}
        where $\sigma_k = \SI{0.03}{W/(m.\degreeCelsius)}$ is the prior standard deviation and $l$ is the length scale controlling how rapidly correlations decay with temperature difference.
        In principle, $l$ can be estimated as a
        hyperparameter from the data (e.g., via marginal likelihood maximization or a hierarchical Bayesian formulation) \cite{rasmussen2003gaussian}.
        Since this is not the objective of the present work, we keep $l$ fixed.
        After manual tuning to obtain a reasonable smoothness over the observed temperature range, we set $l = (T_{\max} - T_{\min})/3$, where $T_{\max}$ and $T_{\min}$ are the maximum and minimum measured temperatures, respectively.
        This choice enforces moderate smoothness without over constraining the prior.
        }
        
        We select the likelihood function as the same multivariate normal used to model the measurement errors, i.e.,
        \begin{align}
            P(\bm{d} \mid \bm{p}) = P(\bm{\epsilon}_{\rm res}  \mid  \bm{p}) &= \mathcal{N}(\bm{\epsilon}_{\rm res} \mid \bm{\mu}_{\rm meas}, \Sigma_{\rm meas})
            \label{eq-likelihood}\\
            &= (2 \pi)^{-n_d/2} |\Sigma|^{-1/2}_{\rm meas} \exp \left[ -\dfrac{1}{2} (\bm{\epsilon}_{\rm res} - \bm{\mu}_{\rm meas})^T \Sigma^{-1}_{\rm meas} (\bm{\epsilon}_{\rm res} - \bm{\mu}_{\rm meas}) \right].\nonumber
        \end{align}
        Values of $\bm{\mu}_{\rm meas}$ and $\Sigma_{\rm meas} = \text{diag} (\bm{\sigma}_{\rm meas}^2)$ are obtained from the data in \autoref{fig-sensor-error-curve}, and we use the same linear interpolation considered for the generation of the synthetic data in \autoref{sec-generating-synthetic-data}.
        \textcolor{blue}{
        In \ref{appendix-measurement-error}, we perform a compact sensitivity analysis study in which we disturb the mean and standard deviation vectors considered for the likelihood function, in order to assess how the measurement error model influences the inferred conductivity.
        }

        \textcolor{blue}{
        We assume that measurement errors are independent across time and sensors, corresponding to a diagonal likelihood covariance.
        This assumption is commonly adopted in inverse problems in heat transfer, where dense time series and multiple sensors are used \cite{ozisik2018, khatoon2023fast}.
        The measurements used in the analysis are approximately 20 seconds apart, which, together with the spatial separation of the sensors, reduces the effect of temporal and spatial correlations.
        While models with correlated noise structures could in principle better capture effects such as ambient fluctuations, electronics, or common model mismatch, incorporating such correlations requires additional probabilistic modeling and leads to a substantial increase in computational cost due to the handling of non-diagonal covariance matrices.
        Since these aspects are orthogonal to the main focus of this work, we retain the independence assumption.
        Nevertheless, extending the framework to include more advanced covariance structures remains an important direction for future research.
        }

        The prior and likelihood represented above are used to define the loss function (\autoref{eq-loss-function}) in \autoref{sec-gradient-based-minimization}.
        There, we define the loss function as being dependent only on $\bm{p}$, and its value $S(\bm{p}; n_e, n_t, n_s)$ is calculated for fixed meshes (spatial and temporal) and model complexity.

    \subsection{\textcolor{blue}{Investigation of discretization optimization}}

        To investigate the \textcolor{blue}{discretization optimization} strategy, we consider the example described in \autoref{sec-discretization-fidelity}, where two refinement stages are compared.
        In finite element analyses, it is common practice to refine the spatial mesh by doubling the number of elements, which improves resolution in a systematic way.
        For consistency, we adopt the same strategy not only for the spatial discretization but also for the temporal discretization and for the model complexity, increasing the number of time steps and the number of segments by a factor of two at each refinement stage.
        The goal is to obtain sufficiently refined spatial and temporal meshes by estimating appropriate values for the number of finite elements and the number of time steps. 
        This procedure involves the following steps:
        \textcolor{blue}{
        \begin{enumerate}[label=Step \arabic*:, left=1.25em, labelwidth=2cm, labelsep=0.5em, align=left]
            \item For a fixed $n_s$, define initial values of $n_e$ and $n_t$.
            \item
                \begin{enumerate} 
                    \item Define $n_e^+ = 2 n_e$ and perform the gradient-based optimization to minimize $S(\bm{p}; n_e^+, n_t, n_s)$.
                    \item Define $n_t^+ = 2 n_t$ and check if $n_e$ satisfies the time-step constraint (\autoref{eq-relation-ne-nt}).
                    If not, set $n_e$ as the lowest integer that satisfies this condition.
                    Then, perform the gradient-based optimization to minimize $S(\bm{p}; n_e, n_t^+, n_s)$.
                \end{enumerate}
            \item If $S(\bm{p}; n_e^+, n_t, n_s) < S(\bm{p}; n_e, n_t^+, n_s)$ set $n_e = n_e^+$.
            Otherwise set $n_t = n_t^+$.
            \item If the stopping criteria for the discretization fidelity are not satisfied, return to Step 2.
        \end{enumerate}
        }
    
        The conductivity values obtained from the gradient-based algorithm are used as an initial guess for the MCMC method, as shown in the next section.
        This means that the goal is not to achieve extremely precise estimates of the conductivity values, but rather to obtain a reasonable approximation that facilitates the convergence of the MCMC method.
        Therefore, we adopt simple stopping criteria for the gradient-based algorithm, using both absolute and a relative tolerances for the conductivity values set to $10^{-2}$.
        Initial results showed that these values offer a practical balance between numerical precision and computational efficiency.
        
        We now turn our attention to the definition of the stopping criteria for the discretization fidelity.
        The first approach consists of using Morozov's discrepancy principle, so we consider that the spatial and temporal meshes are sufficiently refined when the difference between the data and the model prediction is comparable in magnitude to the measurement error.
        We can then define a threshold for the loss function, denoted $S^\text{Morozov}$.
        This threshold is obtained by applying the relation between $\bm{\epsilon}_{\rm res}$ and $\bm{\mu}_{\rm meas}$ given by \autoref{eq-morozov-gamma} (\autoref{sec-measurement-errors-and-morozov}) in the definition of the likelihood function given by \autoref{eq-likelihood} (\autoref{sec-bayesian-setup}):
        \begin{equation}
            S^\text{Morozov}(\bm{p}; n_s) = S_\text{prior}(\bm{p}; n_s) + S^\text{Morozov}_\text{like}.
        \end{equation}
        Next to that, $S^\text{Morozov}_\text{like}$ is obtained by replacing $(\bm{\epsilon}_{\rm res} - \bm{\mu}_{\rm meas})$ in \autoref{eq-likelihood} by $(1 + \gamma) \bm{\sigma}_{\rm meas}$, i.e.,
        \begin{equation}
            S^\text{Morozov}_\text{like} = - \log \left[ \mathcal{N} \bigl( (1 + \gamma) \bm{\sigma}_{\rm meas} \mid \bm{0}, \Sigma_{\rm meas} \bigr) \right].
        \end{equation}
        Our first stopping criterion for the discretization fidelity is then given by $S \leq S^\text{Morozov}$, which yields
        \begin{equation}
            S_\text{like}(\bm{p}; n_e, n_t, n_s) \leq S^\text{Morozov}_\text{like}.
            \label{eq-stopping-criterion-morozov}
        \end{equation}
        If this stopping criterion is satisfied, the mesh refinement process can be stopped, as the numerical discretization is considered adequate relative to the measurement noise.
    
\begin{figure}
            \centering
            \begin{subfigure}{0.49\linewidth}
                \centering
                \includegraphics[width=\linewidth]{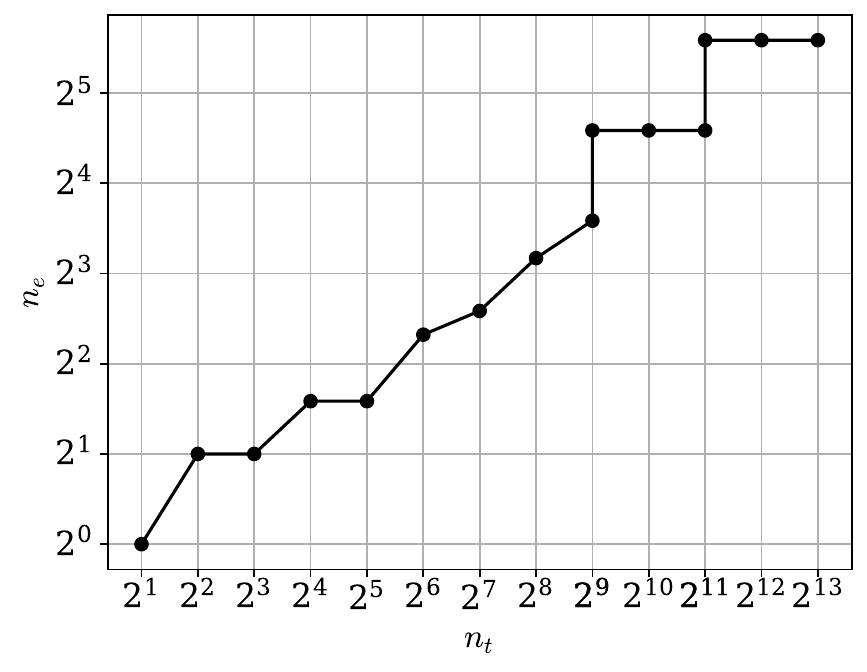}
                \caption{}
                \label{fig-nt-vs-ne}
            \end{subfigure}
            \hfill
            \begin{subfigure}{0.49\linewidth}
                \centering
                \includegraphics[width=\linewidth]{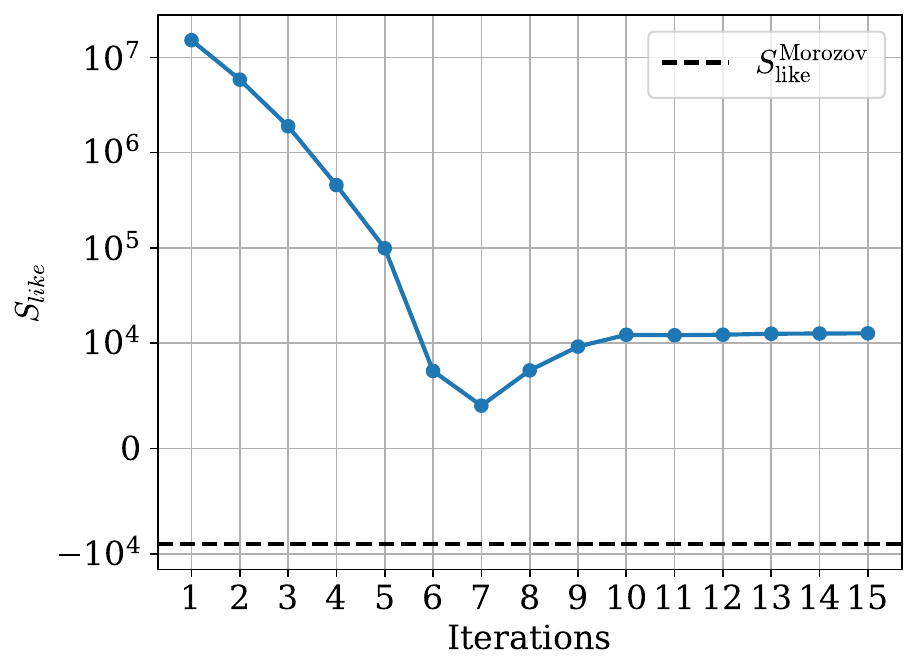}
                \caption{}
                \label{fig-S_like}
            \end{subfigure}
        \caption{Results from the discretization fidelity optimization with $n_s = 1$ (linear conductivity model) and $\gamma = 1\%$.
        (\subref{fig-nt-vs-ne}) Values of $n_e$ and $n_t$.
        The iterations are represented by the points.
        (\subref{fig-S_like}) Values of $S_\text{like}$ obtained for each iteration.}
        \label{fig-results-mesh-optimization-15}
        \end{figure}

        To investigate the behavior of the mesh refinement strategy, we perform a sequence of 15 iterations using a model with a single segment (i.e., $n_s = 1$).
        In this particular case, we set $\gamma = 1\%$, which results in $S_\text{like}^\text{Morozov} = - 9.1 \times 10^3$.
        The obtained results are illustrated in \autoref{fig-results-mesh-optimization-15}.
        \autoref{fig-nt-vs-ne} shows the obtained values of $n_e$ and $n_t$ for each iteration.
        This represents the evolution of both spatial and temporal meshes throughout the optimization.
        We observe that, in some iterations, both $n_e$ and $n_t$ increase simultaneously.
        This occurs because of the relation between these parameters given by \autoref{eq-relation-ne-nt}.
        Next to that, \autoref{fig-S_like} presents the values of $S_\text{like}$ for each iteration.
        The vertical axis is represented in two different scales:
        Above $10^4$, we use a logarithmic scale.
        Below this value, we use a linear scale.
        The threshold $S_{\rm like}^{\rm Morozov}$ is indicated by the black dashed line.
        We observe a consistent decrease in this quantity up to iteration 7.
        Beyond this point, its value starts to increase and eventually stabilizes around $10^4$.
    
        The fact that $S_{\rm like}$ stabilizes at values above the threshold $S_{\rm like}^{\rm Morozov}$ indicates the need of introducing an additional stopping criterion, since further mesh refinement does not necessarily lead to improved data fit.
        In particular, for a fixed model complexity, it may occur that the discrepancy between model predictions and observed data cannot be reduced just by refining the spatial and temporal discretization.
        In such cases, increasing the model complexity -- e.g., by adding more segments in the piecewise linear functions -- is mandated.
        At first, one might argue that the refinement should stop as soon as the loss function begins to increase.
        However, such an increase can occur when using coarse meshes, which may still lead to inaccurate or misleading parameter estimates.
        Therefore, if the stopping criterion defined in \autoref{eq-stopping-criterion-morozov} has not yet been met, it is crucial to continue refining the mesh until the loss function stabilizes.
        We then need to introduce an additional stopping criterion.
        Specifically, we state that, if the values of $S_\text{like}$ stabilize over multiple iterations, without reaching the threshold imposed by Morozov's discrepancy principle, the refinement process can be terminated.
        The criterion we propose is
        \begin{equation}
            \sigma_{\text{rel,} \delta} \leq \phi,
            \label{eq-stopping-criterion-rel-std}
        \end{equation}
        where $\sigma_{\text{rel}, \delta}$ is the relative standard deviation of the last $\delta$ values of $S_\text{like}$, and $\phi$ is a small positive number.
        This criterion ensures that unnecessary refinements are avoided when further improvements become insignificant, even if the error level remains above the measurement noise threshold.
        
        Preliminary results showed that, for the problem under analysis, monitoring the relative standard deviation of the last three values of $S_\text{like}$ and setting $\phi = 5\%$ was sufficient to stabilize the mesh refinement when the criterion imposed by Morozov's principle is not reached.
        Hence, for the results shown in this paper, we consider the fixed values $\delta = 3$ and $\phi = 5\%$.
        If we consider the results shown in \autoref{fig-results-mesh-optimization-15}, this additional stopping criterion would be satisfied after evaluating the standard deviation of values of $S_\text{like}$ of iterations 10--12, thus indicating that the solution of the mesh refinement would correspond to iteration 10.
        This means that we stop at the first of the last three values.

    \subsection{\textcolor{blue}{Investigation of model optimization}}
    \label{sec-investigation-model-optimization}

\begin{figure}
            \centering
            \includegraphics[width=0.7\linewidth]{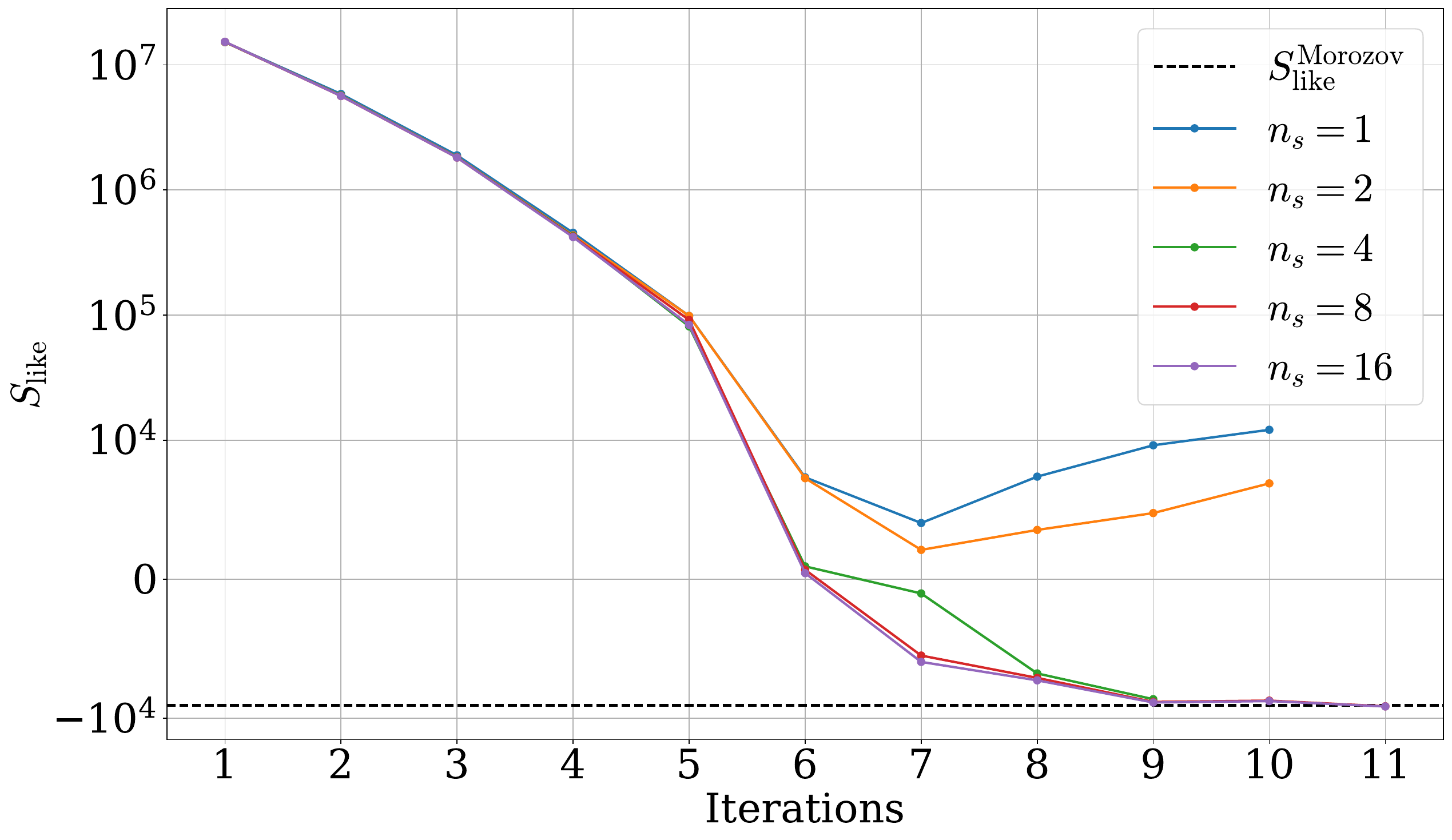}
            \caption{Evolution of $S_{\rm like}$ for different model complexities obtained with the synthetic data.}
            \label{fig-evol-S_like-synthetic}
        \end{figure}

        \autoref{fig-evol-S_like-synthetic} shows the evolution of $S_{\rm like}$ for different model complexities obtained with the synthetic data.
        These values correspond to the iterations up to the point where the stopping criterion given by \autoref{eq-stopping-criterion-rel-std} is satisfied, which means that the last two values of $S_{\rm like}$ used in the calculation of $\sigma_{{\rm rel}, \delta}$ are not shown.
        Similarly to the mesh refinement, we refine the model complexity by doubling the number of segments used in the piecewise linear functions.

        Preliminary analyses indicated that increasing $n_s$ beyond 16 would not lead to a significant improvement in the representation of the conductivity curve, and therefore we limit the refinement to this value.
        In this example, the threshold $S_{\rm like}^{\rm Morozov}$ is crossed twice: first when $n_s = 8$ and again when $n_s = 16$.
        This would indicate that the selected model complexity would be $n_s = 8$.
        However, we must also consider scenarios in which $S_{\rm like}^{\rm Morozov}$ is not reached for any $n_s \leq 16$.
        This can occur, for instance, when $\gamma$ is chosen to be smaller than 1\%.
        In such cases, similar to the behavior observed in the mesh refinement process, where increasing spatial and temporal resolutions does not necessarily minimize $S_{\rm like}$, increasing the model complexity in this case does not necessarily guarantee a better fit.

        \textcolor{blue}{
        For this reason, we introduce an additional stopping criterion for the model complexity refinement, which relies on the information criteria.
        We investigate the model optimization strategy by performing the mesh refinement across a range of model complexities.
        The gradient-based estimates obtained during the mesh optimization are used as initial guesses for the MCMC simulations.
        This choice is justified by the fact that the gradient-based optimization provides point estimates located close to a region with high posterior density.
        Initializing the Markov chains in this region accelerates convergence, thereby improving sampling efficiency and decreasing the number of iterations required to properly represent the posterior.
        }

        \textcolor{blue}{For each model complexity, we run Markov chains with \num{100000} samples to quantify the uncertainty in the estimated conductivity function.
        The burn-in period is set to the first \num{10000} samples, and the remaining \num{90000} samples are used to represent the posterior.
        This configuration was sufficient to satisfy the convergence criterion described in \autoref{sec-uncertainty-quantification-with-mcmc}.
        Nevertheless, there might be cases in which additional robustness may be required.
        In such cases, multiple chains can be initialized from different starting points to assess convergence more rigorously.
        This can be easily implemented within the present framework, as the gradient-based stage can be complemented with perturbed initializations or alternative prior-based draws.
        A representative example comparing diagnostics obtained from multiple chains is presented in \ref{appendix-multiple-chains}.
        }

\begin{figure}
            \centering
            \begin{subfigure}{0.33\linewidth}
                \centering
                \includegraphics[width=\linewidth]{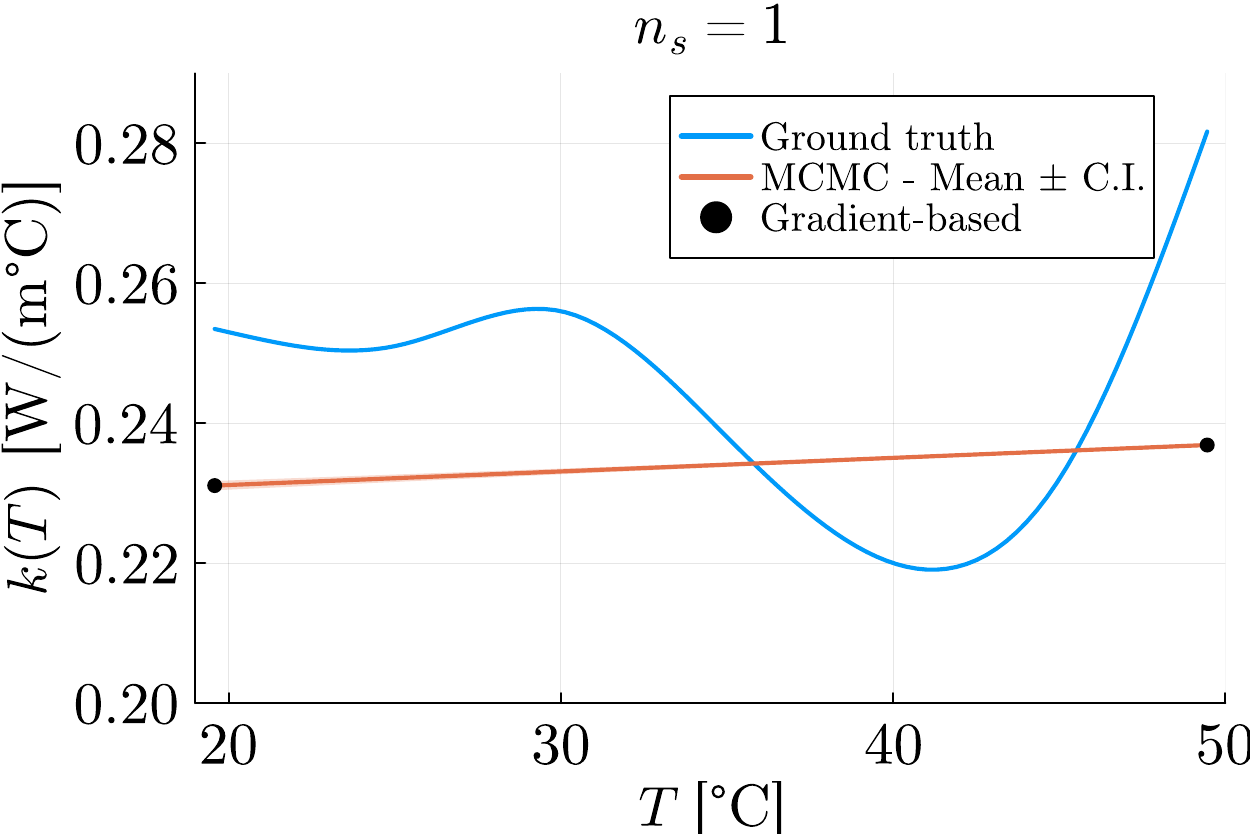}
            \end{subfigure}
            \hfil
            \begin{subfigure}{0.33\linewidth}
                \centering
                \includegraphics[width=\linewidth]{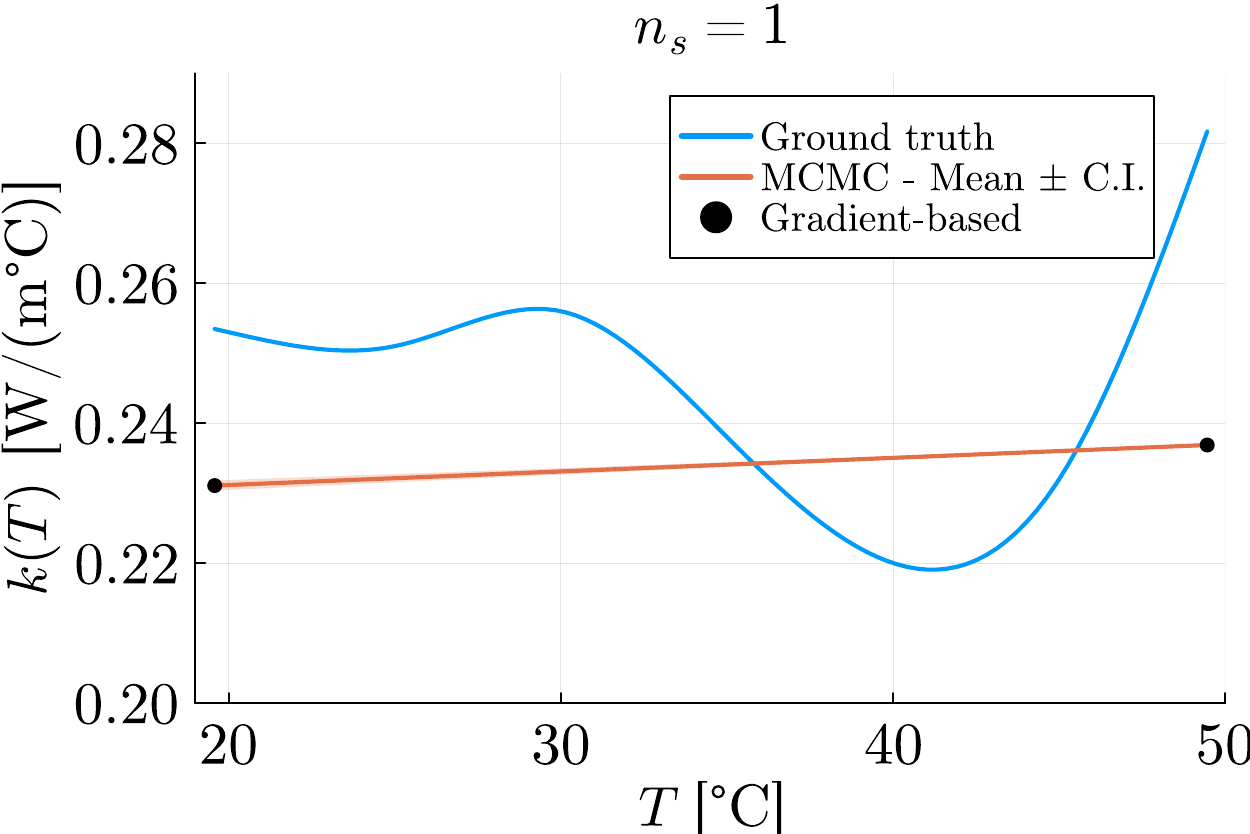}
            \end{subfigure}
            
            \vspace{1em}
            \begin{subfigure}{0.33\linewidth}
                \centering
                \includegraphics[width=\linewidth]{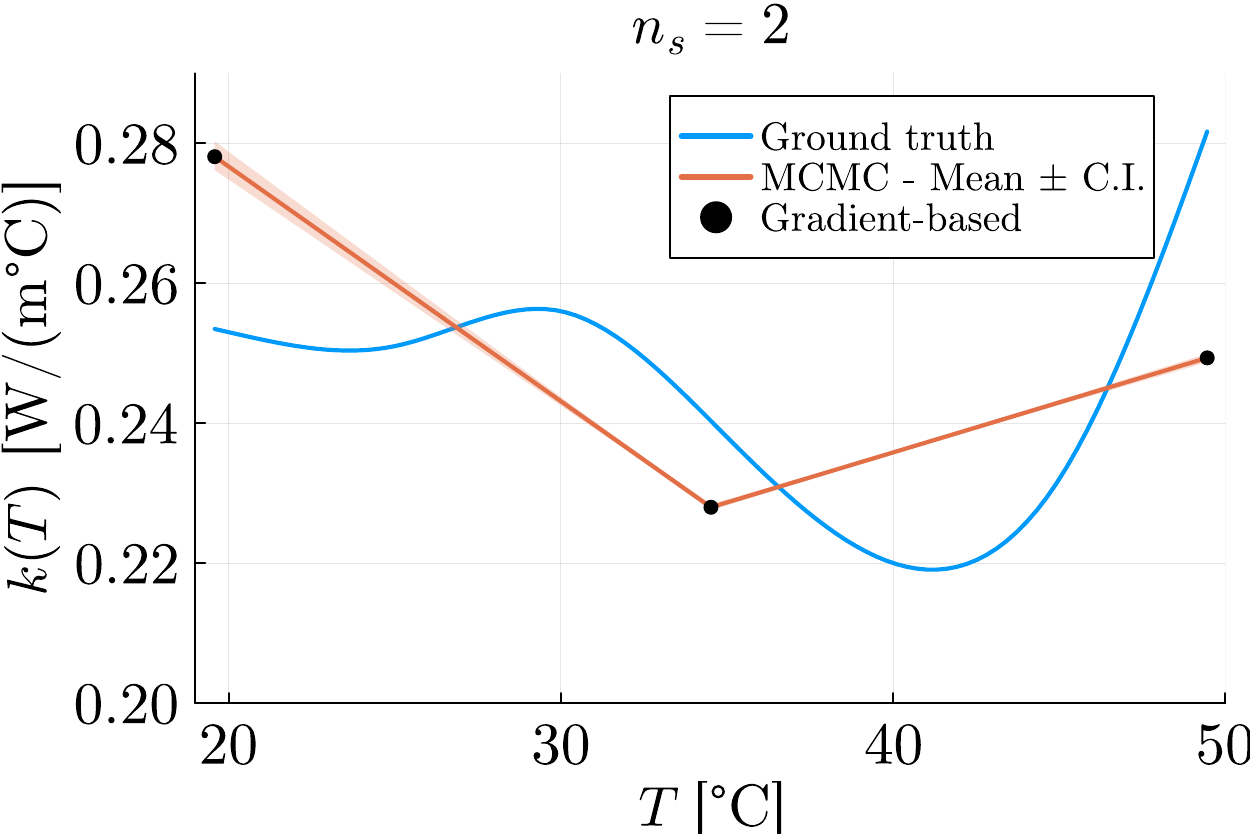}
            \end{subfigure}
            \hfil
            \begin{subfigure}{0.33\linewidth}
                \centering
                \includegraphics[width=\linewidth]{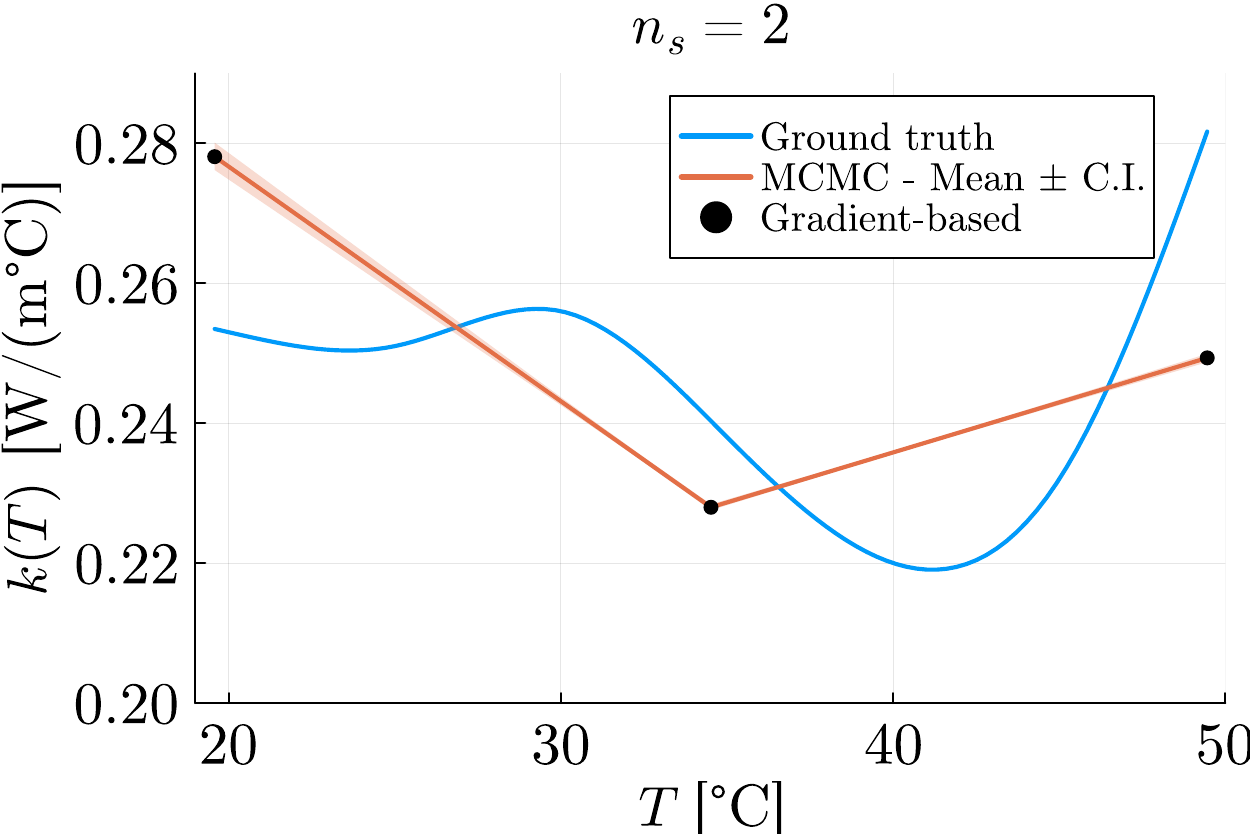}
            \end{subfigure}

            \vspace{1em}
            \begin{subfigure}{0.33\linewidth}
                \centering
                \includegraphics[width=\linewidth]{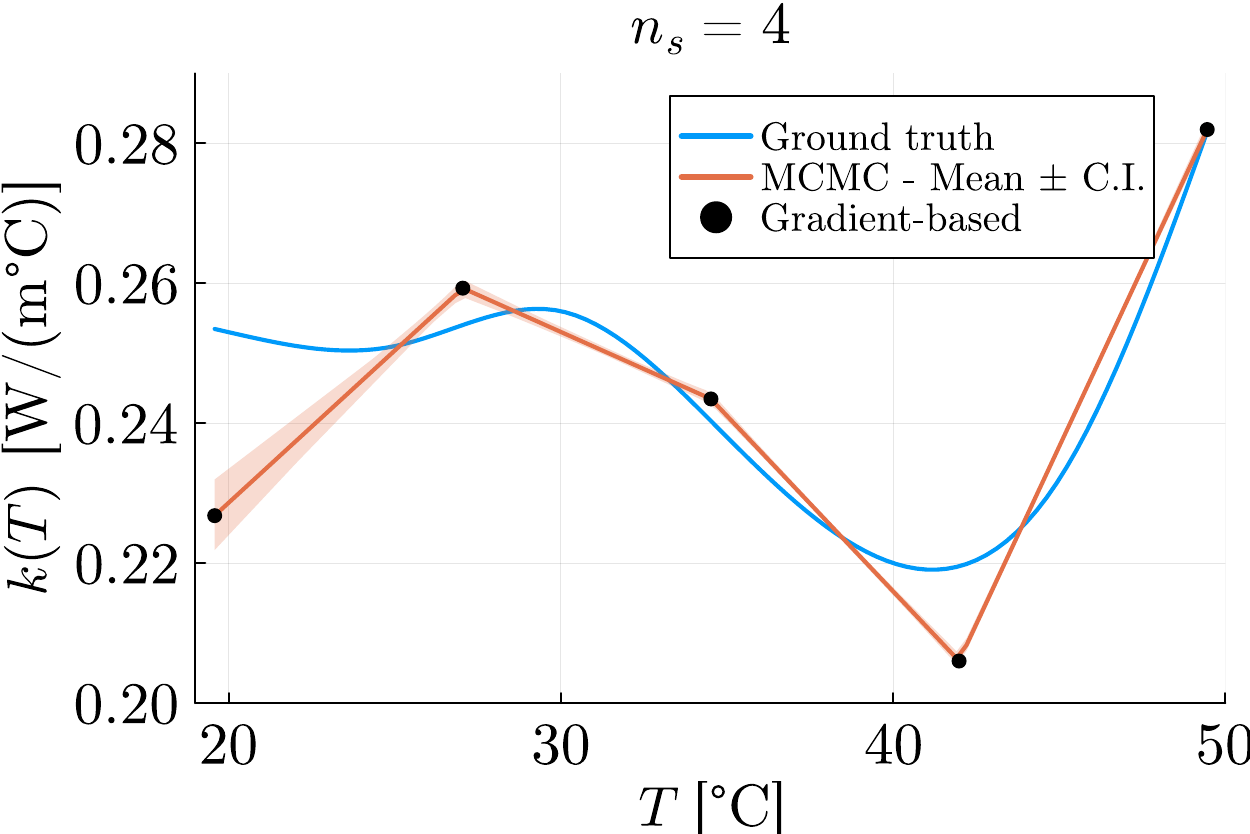}
            \end{subfigure}
            \hfil
            \begin{subfigure}{0.33\linewidth}
                \centering
                \includegraphics[width=\linewidth]{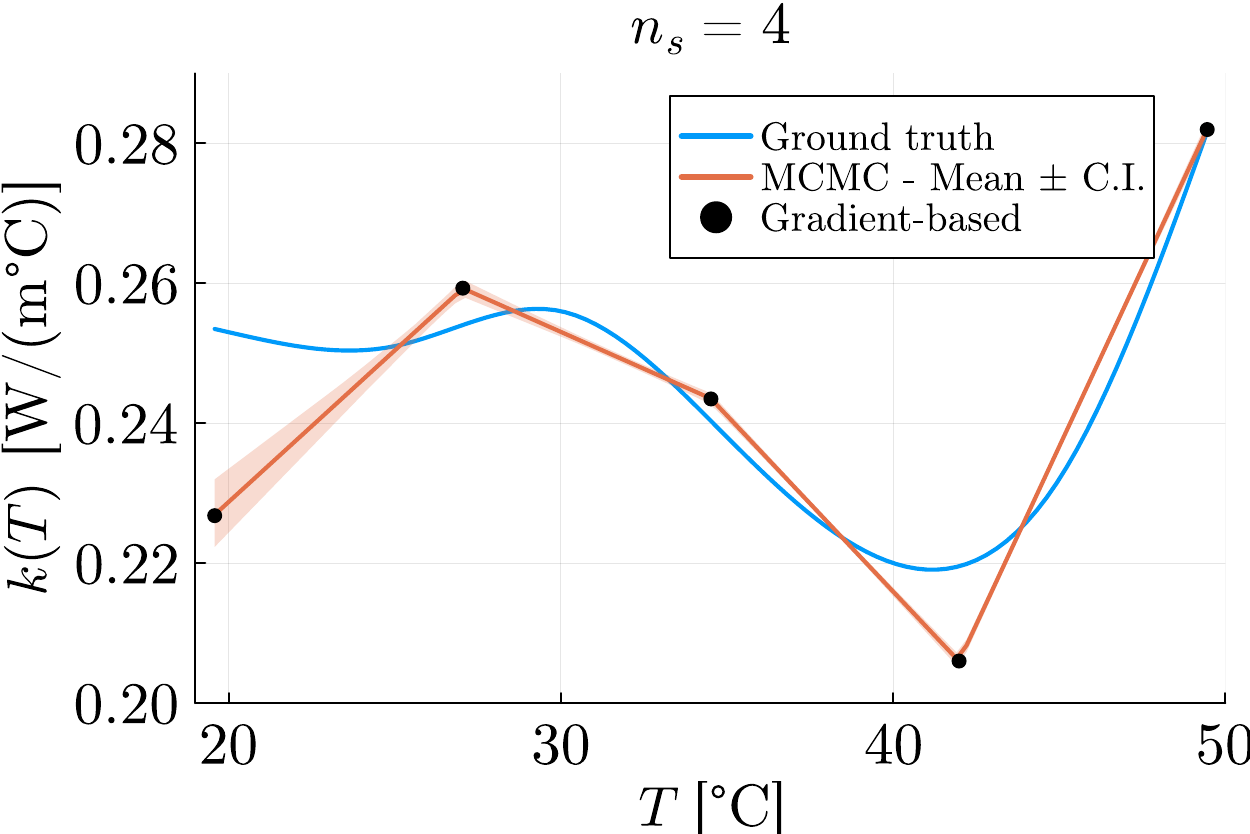}
            \end{subfigure}

            \vspace{1em}
            \begin{subfigure}{0.33\linewidth}
                \centering
                \includegraphics[width=\linewidth]{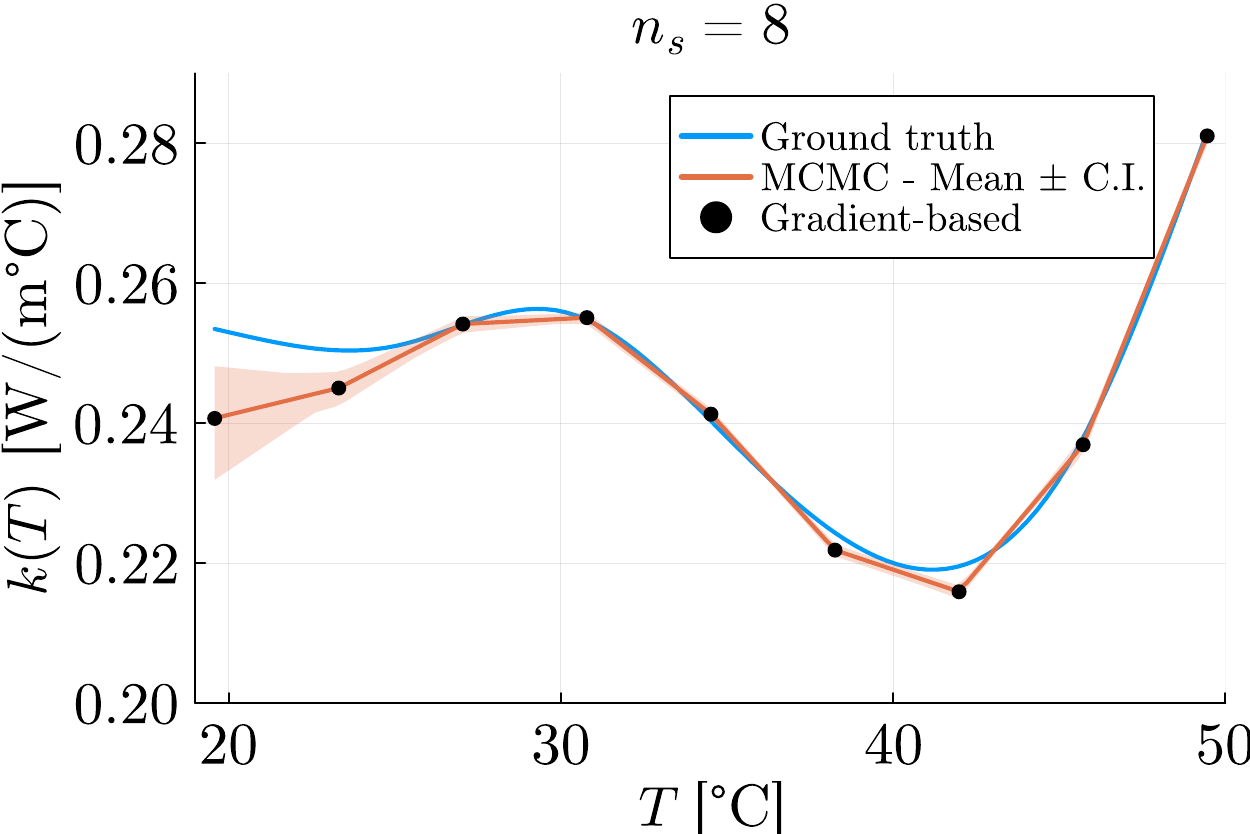}
            \end{subfigure}
            \hfil
            \begin{subfigure}{0.33\linewidth}
                \centering
                \includegraphics[width=\linewidth]{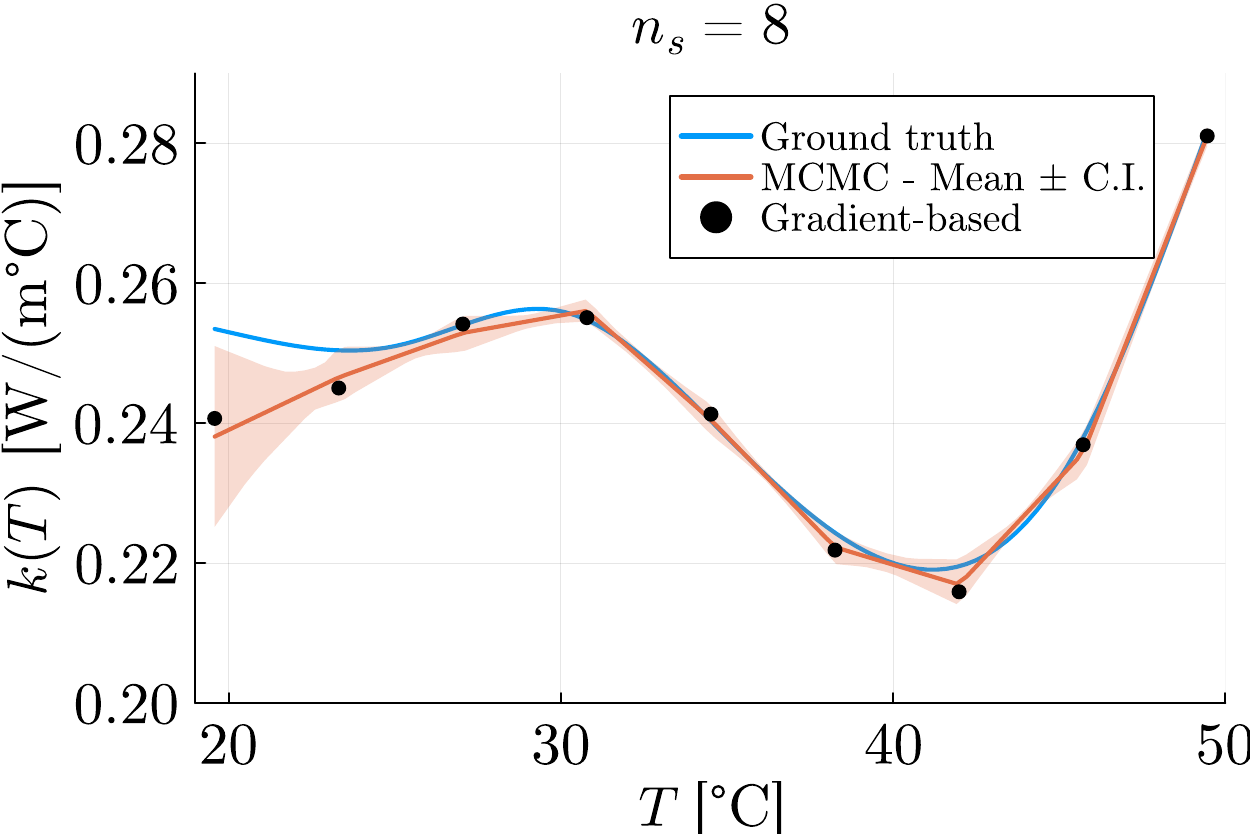}
            \end{subfigure}

            \vspace{1em}
            \begin{subfigure}{0.33\linewidth}
                \centering
                \includegraphics[width=\linewidth]{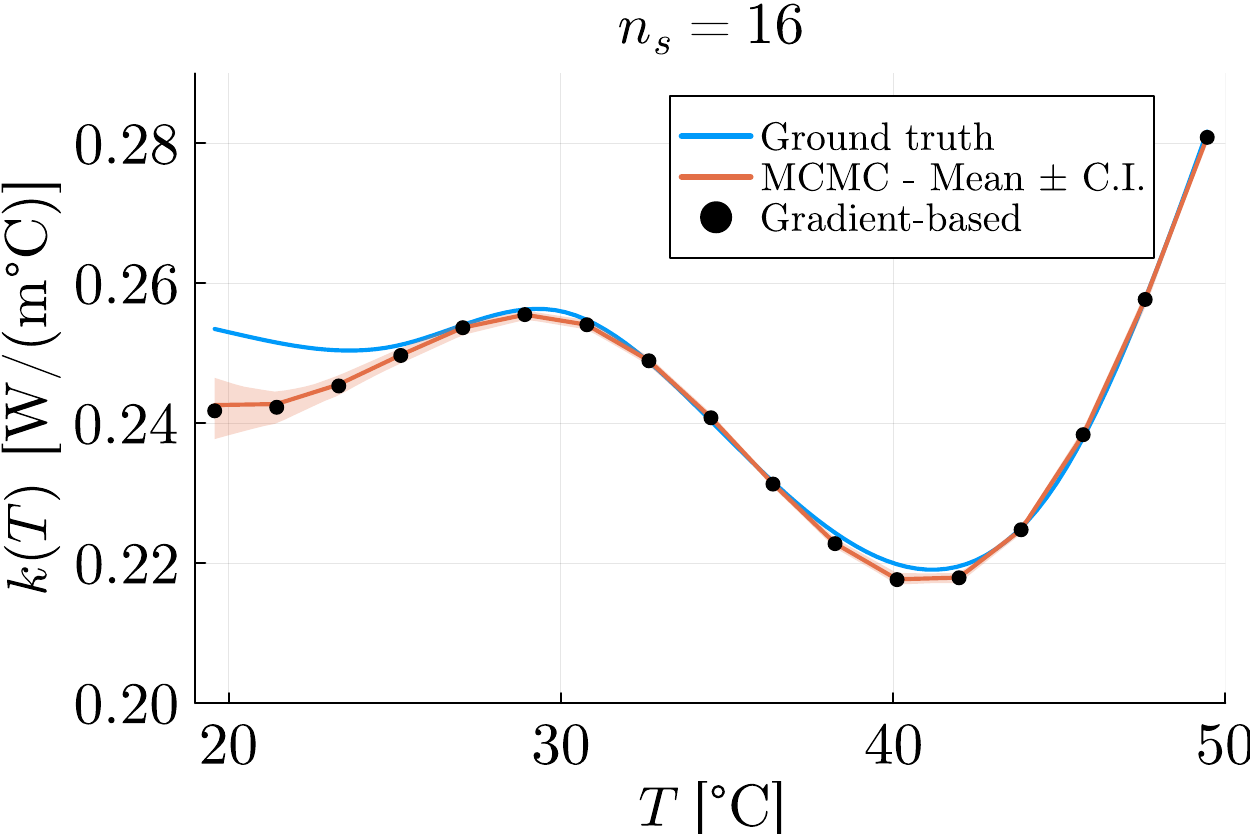}
                \caption{Stronger correlation.}
                \label{fig-UQ-stronger-correlation}
            \end{subfigure}
            \hfil
            \begin{subfigure}{0.33\linewidth}
                \centering
                \includegraphics[width=\linewidth]{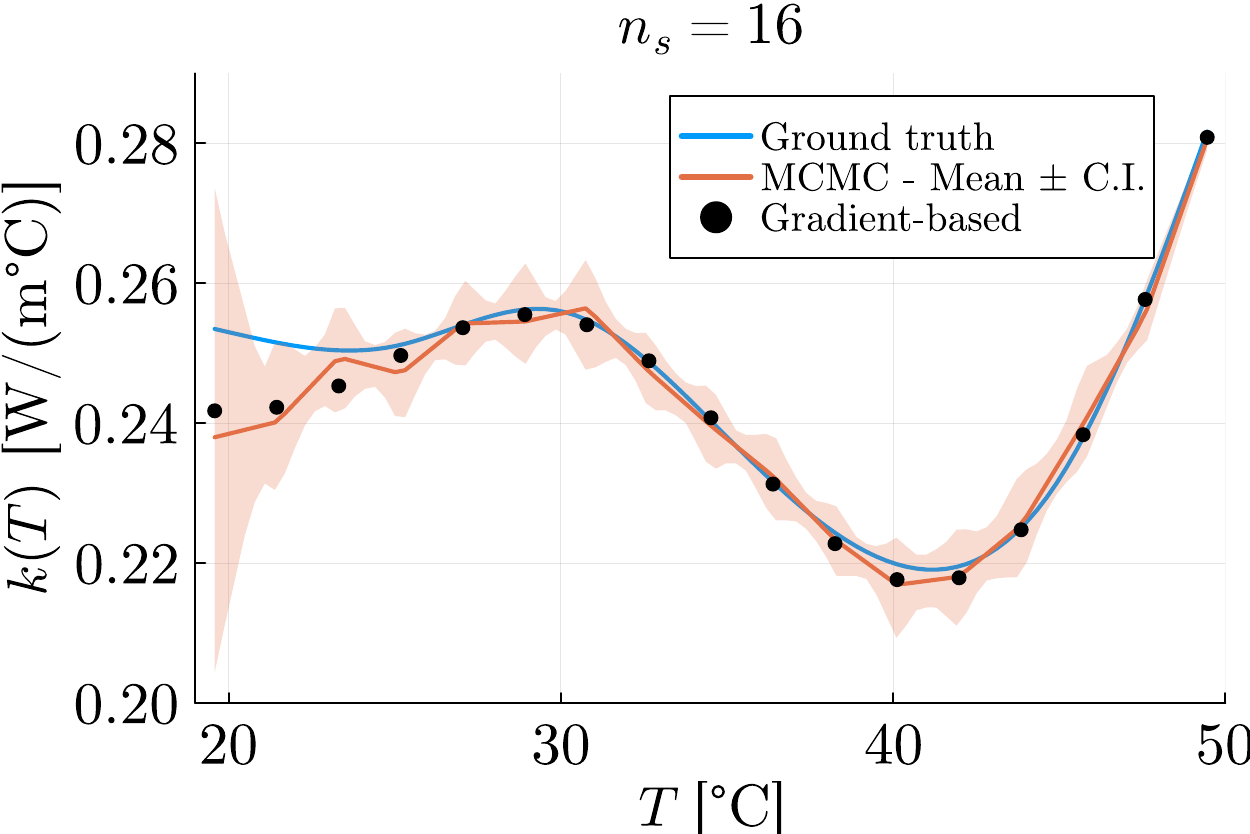}
                \caption{Weaker correlation.}
                \label{fig-UQ-weaker-correlation}
            \end{subfigure}
            \caption{\textcolor{blue}{Estimation of $k(T)$ after the discretization optimization for different model complexities and correlations.}}
            \label{fig-plot-UQ-synthetic}
        \end{figure}

        \textcolor{blue}{
        \autoref{fig-plot-UQ-synthetic} shows a comparison between the ground truth and the posterior estimates of the conductivity for different values of $n_s$ (model complexity).
        The credible intervals correspond to 99\% posterior intervals, computed from the quantiles of the samples by removing 0.5\% from each tail of the distribution at every temperature value.
        The black dots represent the solution obtained from the gradient-based algorithm.
        \autoref{fig-UQ-stronger-correlation} shows this comparison for the case in which the original length scale is used to model the correlation between the conductivity values, while \autoref{fig-UQ-weaker-correlation} presents the same comparison when the length scale is chosen to be 10 times smaller, resulting in a weaker correlation.
        We observe that the stronger correlation stabilizes the uncertainties in the estimated conductivity curve.
        In contrast, when the correlation is weaker, the conductivity values become more independent, allowing the estimates to vary more freely and resulting in wider credible interval bands.
        }

\begin{figure}
            \centering
            \begin{subfigure}{0.49\linewidth}
                \centering
                \includegraphics[width=\linewidth]{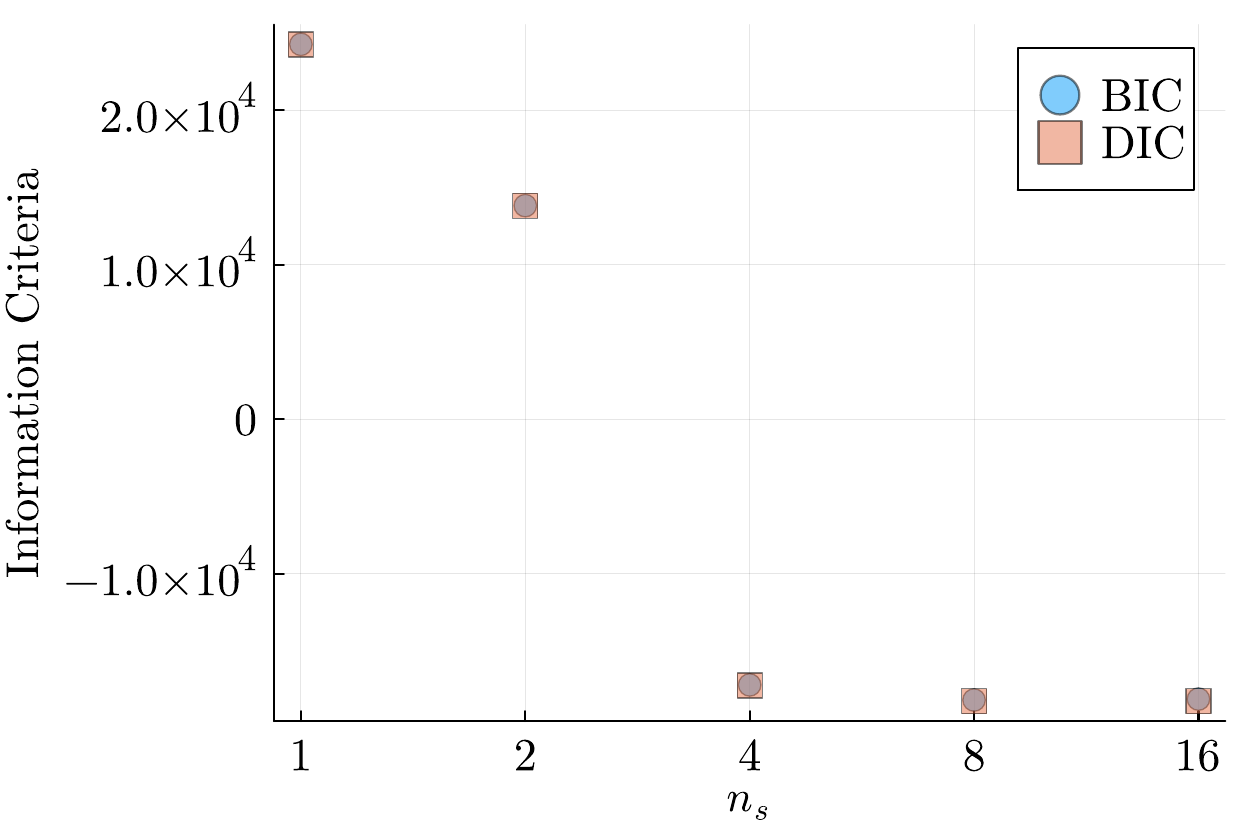}
                \caption{Stronger correlation.}
            \end{subfigure}
            \hfill
            \begin{subfigure}{0.49\linewidth}
                \centering
                \includegraphics[width=\linewidth]{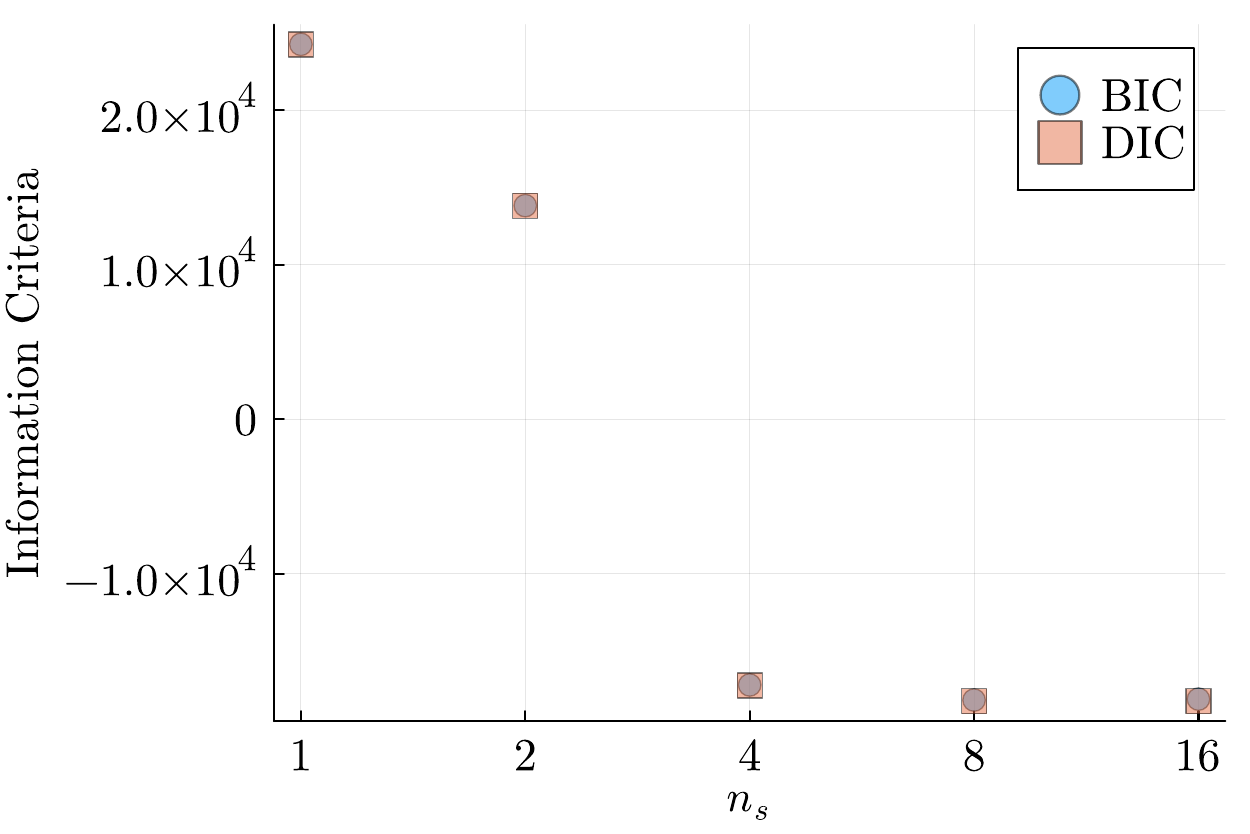}
                \caption{Weaker correlation.}
            \end{subfigure}
            \caption{\textcolor{blue}{Values of BIC and DIC for different model complexities.}}
            \label{fig-BIC-DIC}
        \end{figure}

        \textcolor{blue}{
        \autoref{fig-BIC-DIC} shows the computed BIC and DIC values for the range of model complexities considered.
        The figure presents these quantities for both the stronger correlation (original length scale) and weaker correlation (length scale 10 times smaller).
        We observe that the information criteria obtained for both cases are essentially the same, so we focus on the case with the original length scale.
        Both criteria exhibit a clear decrease when the number of segments is increased to $n_s = 4$, suggesting that the model at this complexity better captures the underlying physics of the system.
        Beyond this point, improvements in model fit become more gradual.
        In particular, increasing the complexity from $n_s = 8$ to $n_s = 16$ leads to a slight increase in the BIC from $-1.818 \times 10^4$ to $-1.812 \times 10^4$, indicating that the additional parameters do not justify the improvement in likelihood.
        In contrast, the DIC shows a minor decrease from $-1.825 \times 10^4$ to $-1.826 \times 10^4$, but this reduction is not substantial enough to select the more complex model.
        }

        \textcolor{blue}{
        Since the information criteria obtained for both correlation length scales are essentially the same, while \autoref{fig-plot-UQ-synthetic} shows a noticeable difference in the resulting uncertainty levels, additional metrics such as the area of the credible intervals could in principle be incorporated into the model selection procedure.
        However, because the present framework focuses on priors that stabilize the uncertainty through the correlation structure, we define the stopping criterion based only on the information criteria.
        To formalize this stopping criterion, we adopt a tolerance-based approach: if the improvements in the information criteria are smaller than 5\% relative to the previous model, we consider the increase in complexity unjustified.
        Applying this threshold, the minor decrease in DIC at $n_s = 16$ is negligible, and therefore the selected model complexity is $n_s = 8$.
        This ensures a model that balances estimation accuracy and computational efficiency while avoiding over-parameterization.
        }

        \textcolor{blue}{
        Since the BIC and DIC values obtained in this study are very close for all model complexities considered, the BIC alone can be used as the metric for the stopping criterion.
        This has the practical advantage that the MCMC simulations do not need to be performed for every candidate model complexity, which significantly reduces the computational cost of the procedure.
        In situations where the user is interested in analyzing the uncertainty associated with each model complexity -- for example, to investigate how the credible intervals change when weaker correlations between the parameters are assumed -- the DIC can be evaluated together with the BIC by performing the corresponding MCMC simulations.
        In this work, we consider the more general case in which the MCMC simulations are carried out for each model complexity, allowing both information criteria and the associated posterior uncertainties to be evaluated at every step.
        }

    \subsection{The algorithm}

        \autoref{alg-myalgo} summarizes the proposed adaptive refinement algorithm, which is used to estimate the temperature-dependent thermal conductivity in the heat conduction problem, along with the corresponding mesh and model parameters, and to perform the uncertainty quantification of the estimated conductivity curve.
        The function that performs the gradient-based optimization used for mesh refinement is \texttt{gradient\_based\_optimization}.
        The mesh refinement continues until the function \texttt{satisfies\_mesh\_stopping\_criteria} determines if a stopping criterion is satisfied.
        This function relies on the two criteria defined by \autoref{eq-stopping-criterion-morozov} and \autoref{eq-stopping-criterion-rel-std}.
        Once the mesh refinement process is complete, the function \texttt{mcmc\_based\_uq} performs the MCMC-based uncertainty quantification.
        \textcolor{blue}{
        This final refinement stage is terminated by the function \texttt{satisfies\_model\_complexity\_stopping\_criteria}, which again considers the stopping criterion given by Morozov's discrepancy principle (\autoref{eq-stopping-criterion-morozov}), and also analyze the information criteria to determine whether increasing the model complexity leads to a justified improvement in the data likelihood.
        }

        For simplicity, the update of $n_e$ based on the constraint related to the number of time steps $n_t$ (\autoref{eq-relation-ne-nt}) is not incorporated in the pseudo-code \autoref{alg-myalgo}.
        In our implementation, this update is applied, if needed, after increasing the number of time steps during the mesh refinement.

        \SetKwComment{Comment}{}{}
        \begin{algorithm}
            \caption{Adaptive refinement algorithm for discretization and model complexity}
            \label{alg-myalgo}
            
            \KwIn{Initial conductivity values $k_1$, $k_2$. 
            Stopping criteria parameters $\gamma$, $\delta$ and $\phi$.}
            \KwOut{Mesh and model parameters $n_e$, $n_t$, $n_s$ and $\bm{p}$. 
            Posterior distribution of conductivity $P(\cdot|\bm{d})$}
            
            first\_iteration $\gets$ \textbf{true}
            \;
            $n_s \gets 1$
            \Comment*[r]{Start conductivity model with one segment}
            $\bm{p}_0 \gets (k_1, k_2)$
            \Comment*[r]{Initial conductivity parameters}
            \Repeat{\textnormal{\texttt{satisfies\_model\_stopping\_criteria}}$(P(\cdot|\bm{d}), S, \gamma)$}{
            
                \If{\textnormal{\textbf{not} first\_iteration}}{
                    $n_s \gets 2\,n_s$
                    \;
                    $\bm{p}_0 \gets \texttt{linear\_interpolation}(\bm{p}, n_s)$
                    \Comment*[r]{Include mid points}
                }
            
                $n_e \gets 1$, $n_t \gets 1$
                \Comment*[r]{Initial spatial and temporal resolution}
            
                \Repeat{\textnormal{\texttt{satisfies\_mesh\_stopping\_criteria}}$(S, \gamma, \delta, \phi)$}{
            
                    $\bm{p}_e, S_e \gets \texttt{gradient\_based\_optimization}(\bm{p}_0, 2n_e, n_t)$\;
                    $\bm{p}_t, S_t \gets \texttt{gradient\_based\_optimization}(\bm{p}_0, n_e, 2n_t)$\;
            
                    \If{$S_e < S_t$}{
                        $n_e \gets 2n_e$\;
                        $\bm{p} \gets \bm{p}_e$\;
                        $S \gets S_e$\;
                    }
                    \Else{
                        $n_t \gets 2n_t$\; \label{alg-line-update-nt}
                        $\bm{p} \gets \bm{p}_t$\;
                        $S \gets S_t$\;
                    }
            
                }
            
                $P(\cdot|\bm{d}) \gets \texttt{mcmc\_based\_uq}(\bm{p}, n_e, n_t, n_s)$\;
            
                first\_iteration $\gets$ \textbf{false}\;
            }
        \end{algorithm}

%% file: application_of_the_algorithm_to_real_data.tex
\section{Application of the algorithm to real data}
\label{sec-application-algorithm}

    In this section, we demonstrate the practical application of the proposed algorithm using real experimental data.
    We begin by describing the experimental setup and measurement process, providing details about the physical conditions and data acquisition methods.
    Next, we present the results obtained by applying our algorithm to infer the temperature-dependent thermal conductivity.
    A discussion of the outcomes highlights the algorithm’s effectiveness in balancing model complexity, discretization fidelity, and uncertainty quantification in relation to the data.

    \subsection{The experiment}

         Our experimental setup (\autoref{fig-setup}) was manufactured using fused deposition modeling (FDM) printing with polylactic acid (PLA).
         The column of paraffin wax is contained within a thin polyvinyl chloride (PVC) tube.
         A brass base plate is heated by a \SI{60}{\W} Peltier element, which is regulated by a control module with a temperature sensor integrated in the base plate.

         The PVC tube is insulated on the outside by a thick layer of styrofoam.
         Throughout the experiment, the temperature of the paraffin is kept below the melting temperature of approximately $58 - \SI{60}{\degreeCelsius}$, such that the material does not exhibit phase changes.
         The five temperature sensors used to acquire data are of model DS18B20 \citep{DS18B20}, and they were calibrated prior to assembly.
         The temperatures of all sensors are recorded approximately for 14 hours, with a total of \num{32057} measurements per sensor.
         \textcolor{blue}{This corresponds to a measurement frequency of approximately \SI{0.64}{Hz}.}
         The registered ambient temperature is used as input for the model.
         The experiment was reproduced four times, but with minor variations in the ambient temperature ($\pm \SI{2}{\degreeCelsius}$).
         The measurement with the smallest variation in ambient temperature is used in the remainder of this work, as this is convenient for the interpretation of the inference results.
         \textcolor{blue}{In \ref{appendix-testing-different-datasets} we test the fitted model on two different datasets to show that the predicted conductivity is not an artifact of a single ambient trajectory}.

\begin{figure}
            \centering
            \begin{subfigure}{0.47\linewidth}
                \centering
                \includegraphics[width=\linewidth]{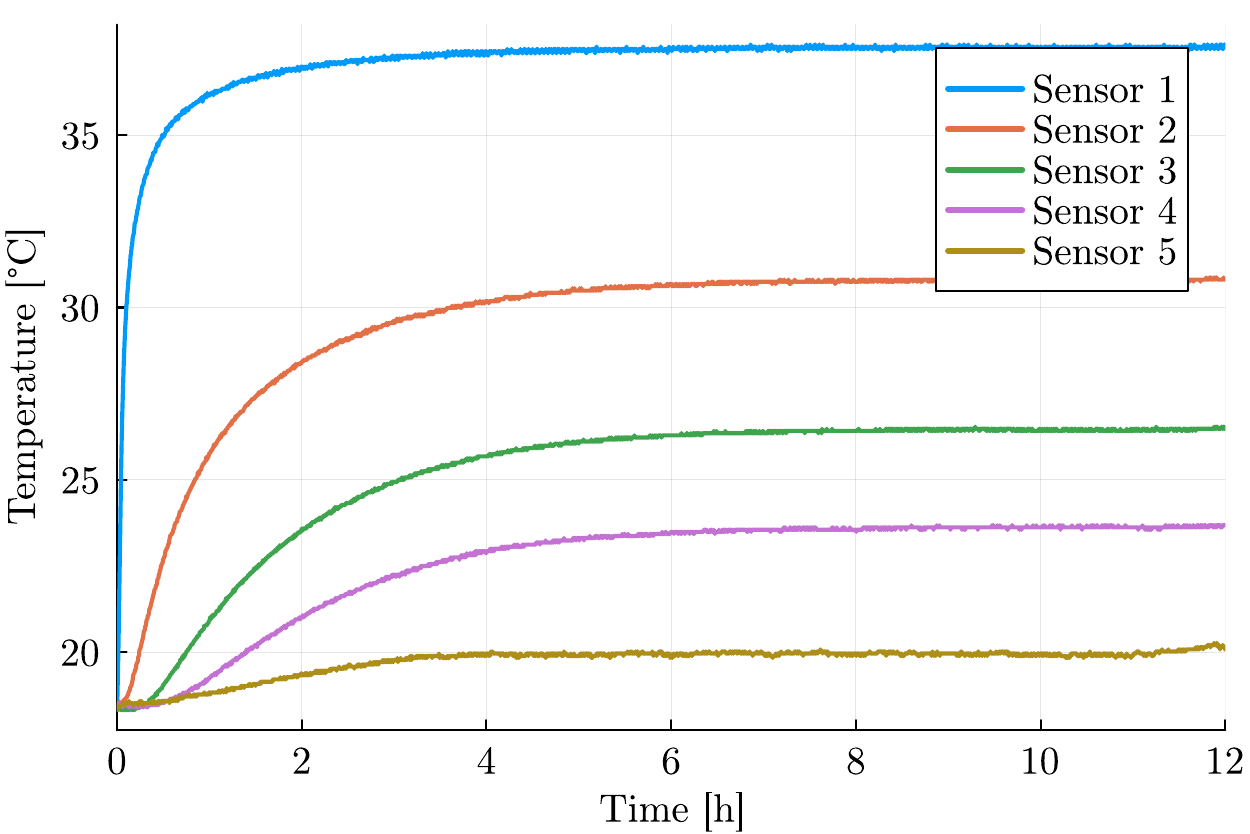}
                \caption{Full regime.}
                \label{fig-meas-real-data-full}
            \end{subfigure}
            \hfill
            \begin{subfigure}{0.47\linewidth}
                \centering
                \includegraphics[width=\linewidth]{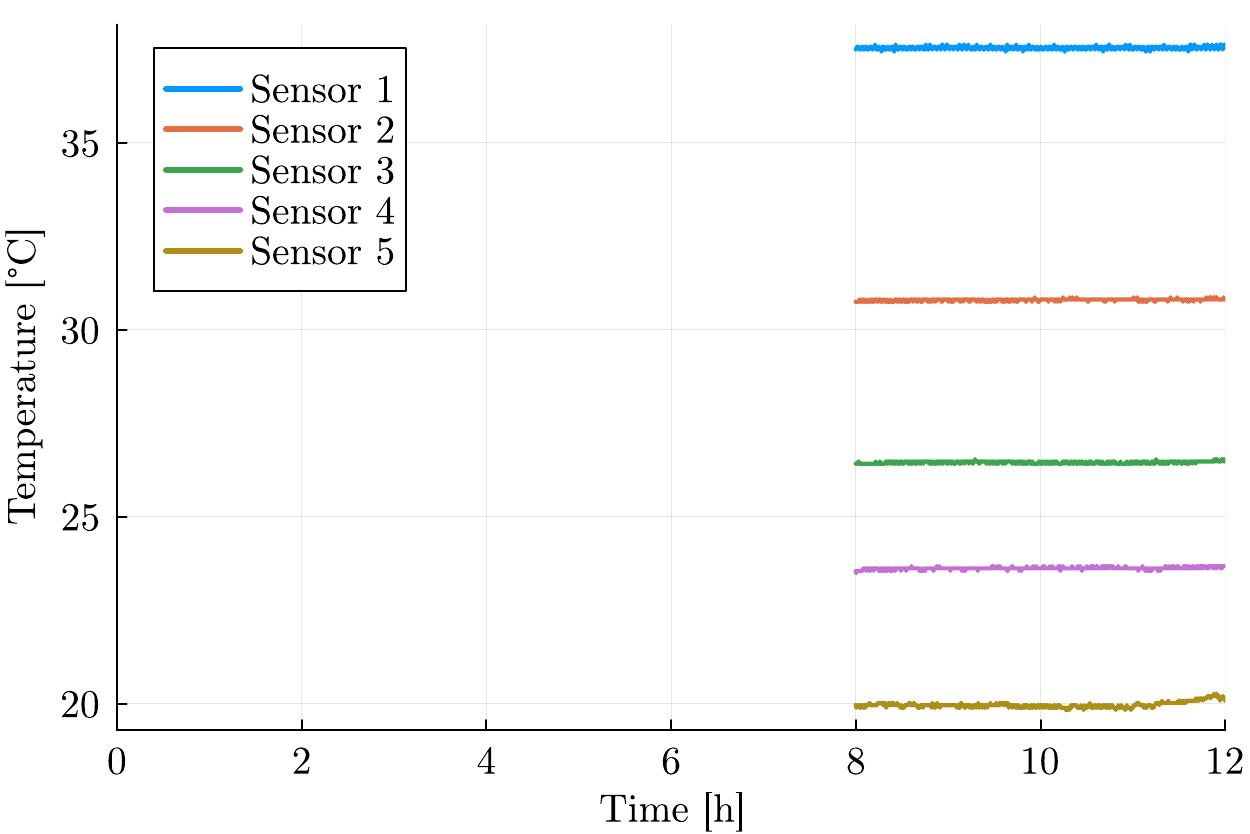}
                \caption{Steady state.}
                \label{fig-meas-real-data-steady}
            \end{subfigure}
            \caption{Datasets used for the estimation of the conductivity.}
         \end{figure}

         We extract an experimental dataset by selecting 2109 (approximately) equally spaced measurements from each sensor, with the final measurement taken 12 hours after the start of the experiment (\autoref{fig-meas-real-data-full}).
        This dataset provides sufficient resolution to capture both the transient and steady state regimes.
        To evaluate the performance of the algorithm when data cover only the steady state regime, we also construct an additional subset that consists of the last quarter of the full dataset, with 528 measurements from each sensor (\autoref{fig-meas-real-data-steady}).

    \subsection{Results and discussions}

    \subsubsection{Estimation of contextual parameters}

        \textcolor{blue}{
        Since the true values of the physical properties and boundary conditions are unknown in the case with real data, in principle one should estimate all parameters simultaneously with the thermal conductivity and quantify their uncertainties.
        We study this approach in \ref{appendix-estimate-all-parameters}, which conveys that the additional parametric uncertainty translates into increased uncertainty in the conductivity.
        }

        \textcolor{blue}{
        In the context of studying the relation between model complexity and discretization parameters, jointly estimating all parameters complicates the interpretation, as the additional uncertainty excludes the consideration of conductivity models with significant number of parameters.
        Therefore, we consider a simplified calibration procedure in which we first calibrate the physical properties and boundary conditions.
        This calibration is performed using a gradient-based optimization algorithm under the assumption of a thermal conductivity that is not temperature-dependent.
        Although this procedure yields an estimate for a constant thermal conductivity, this value is not used in the subsequent uncertainty quantification.
        Only the remaining calibrated parameters are retained, after which the temperature-dependent conductivity is estimated using the proposed framework.
        While this simplified approach enables the detailed study of our adaptive algorithm, we acknowledge that it affects the inferred results.
        }
        
        The prior distributions used for the calibration are summarized in \autoref{tab-calibration-priors}.
        \textcolor{blue}{These are also used in the analysis in \ref{appendix-estimate-all-parameters}.}
        The truncated priors have lower and upper bounds respectively equal to 0.1 and 10 times the mean value. Next to that, their means and standard deviations pertain to the distributions before truncation.
        Once the calibration is completed, the resulting estimates are kept fixed for all subsequent simulations.
        The calibrated values obtained from this step are also reported in \autoref{tab-calibration-priors}.

        \begin{table}
            \centering
            \caption{Truncated normal priors used for calibration, and resulting calibrated values.}
            \begin{tabular}{lcccr}
                \toprule
                Parameter & Mean & Std. Dev. & Calibrated Value & Unit
                \\
                \midrule
                $k$ & $3.00 \times 10^{-1}$ & $3.00 \times 10^{-2}$ & $2.70 \times 10^{-1}$ & \si{W/(m.\degreeCelsius)}
                \\
                $\rho$ & $9.00 \times 10^2$ & $9.00 \times 10^1$ & $7.35 \times 10^2$ & \si{kg/m^3}
                \\
                $c_p$ & $2.50 \times 10^3$ & $2.50 \times 10^2$ & $2.48 \times 10^3$ & \si{J/(kg.\degreeCelsius)}
                \\
                $h_{\rm source}$ & $1.00 \times 10^2$ & $5.00 \times 10^1$ & $1.38 \times 10^2$ & \si{W/(m^2.\degreeCelsius)}
                \\
                $h_{\rm side}$ & $1.00 \times 10^0$ & $5.00 \times 10^{-1}$ & $2.00 \times 10^0$ & \si{W/(m^2.\degreeCelsius)}
                \\
                $h_{\infty}$ & $1.00 \times 10^1$ & $5.00 \times 10^0$ & $1.48 \times 10^1$ & \si{W/(m^2.\degreeCelsius)}
                \\
                $T_{\rm source}$ & $4.00 \times 10^1$ & $2.00 \times 10^{-1}$ & $4.08 \times 10^1$ & \si{\degreeCelsius}
                \\
                \midrule
            \end{tabular}
            \label{tab-calibration-priors}
        \end{table}

        \subsubsection{Description of running the algorithm \& its results}
\begin{figure}
            \centering
            \begin{subfigure}{0.49\linewidth}
                \centering
                \includegraphics[width=\linewidth]{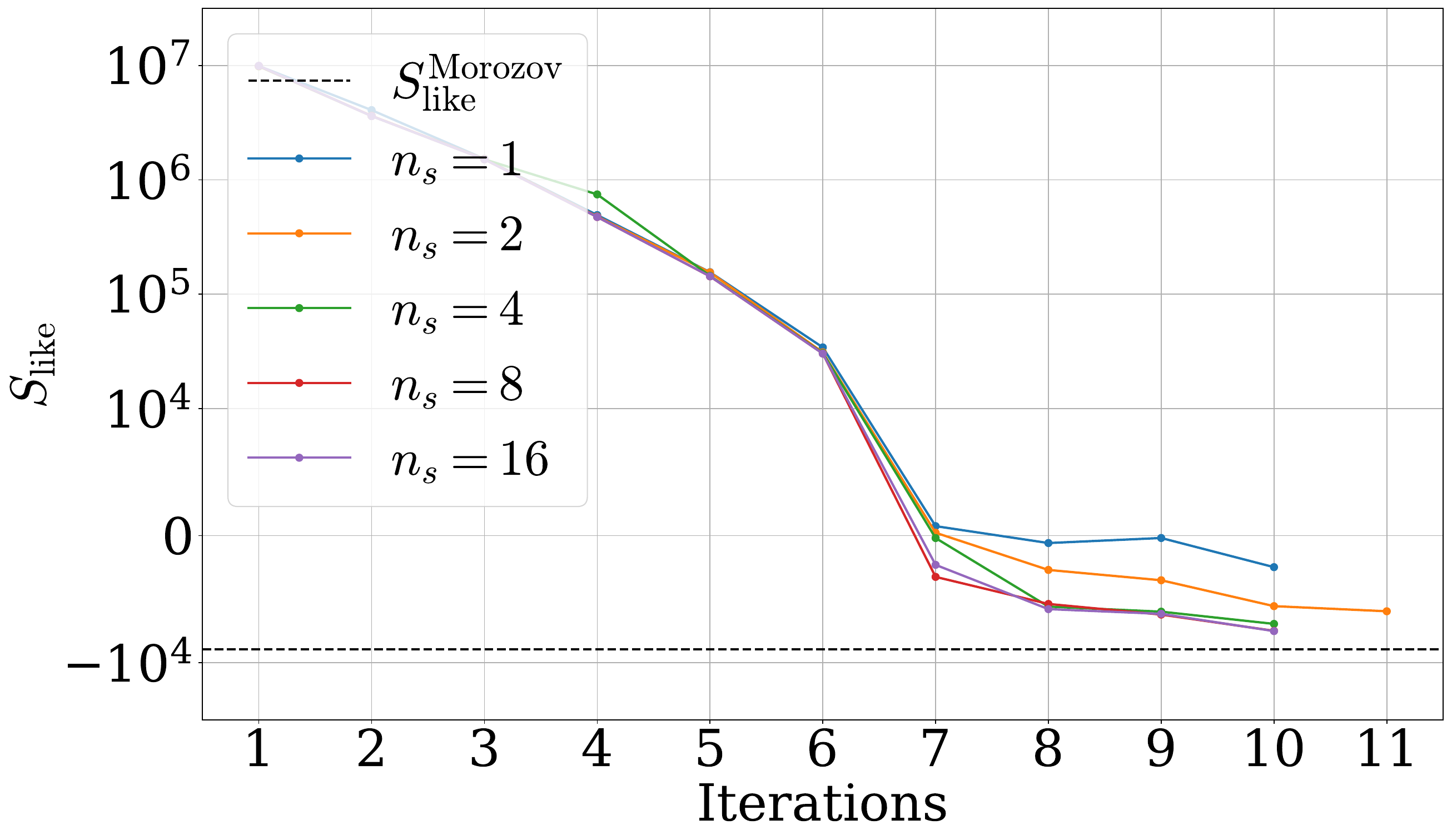}
            \end{subfigure}
            \hfill
            \begin{subfigure}{0.49\linewidth}
                \centering
                \includegraphics[width=\linewidth]{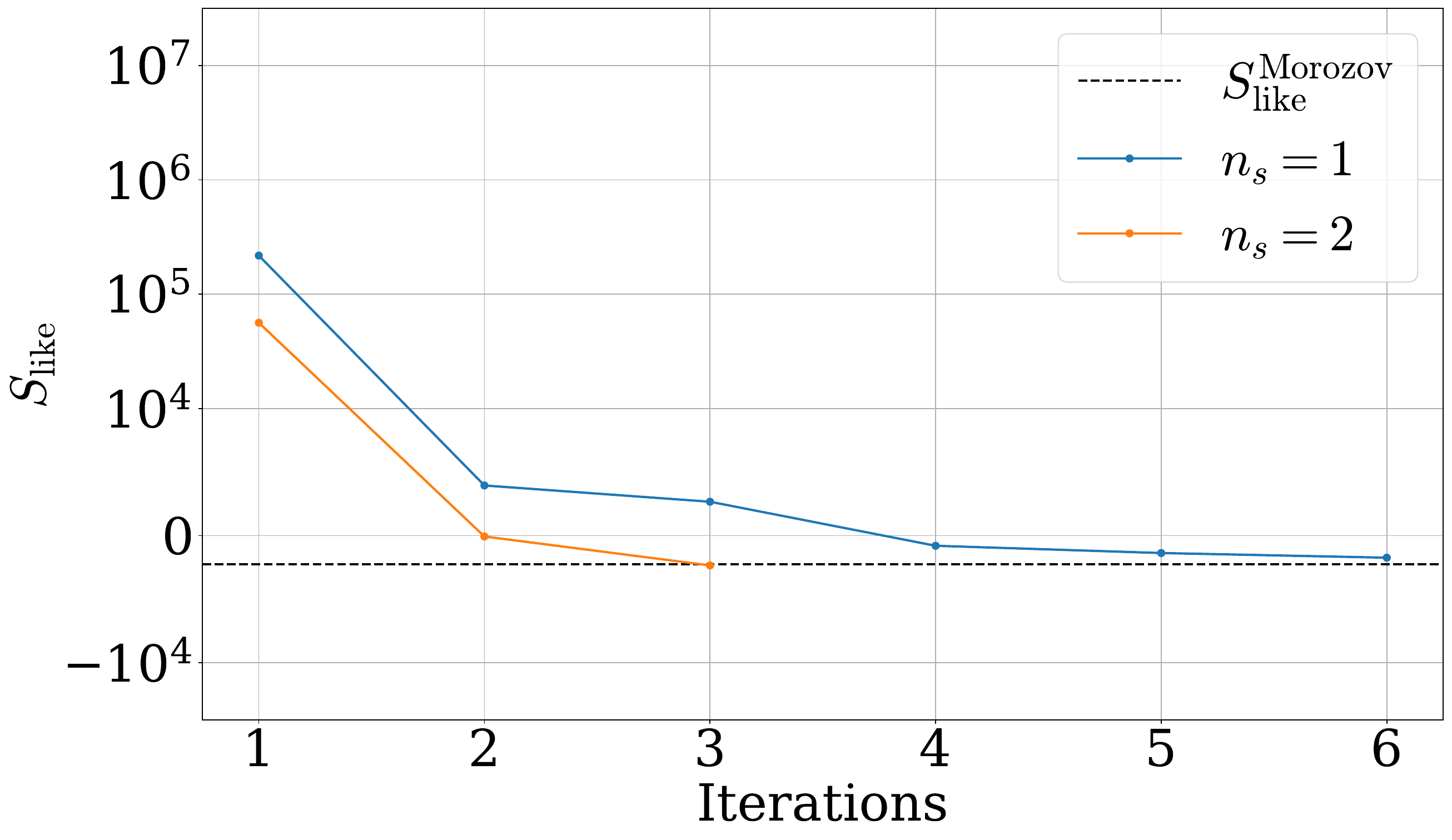}
            \end{subfigure}

            \begin{subfigure}{0.45\linewidth}
                \centering
                \includegraphics[width=\linewidth]{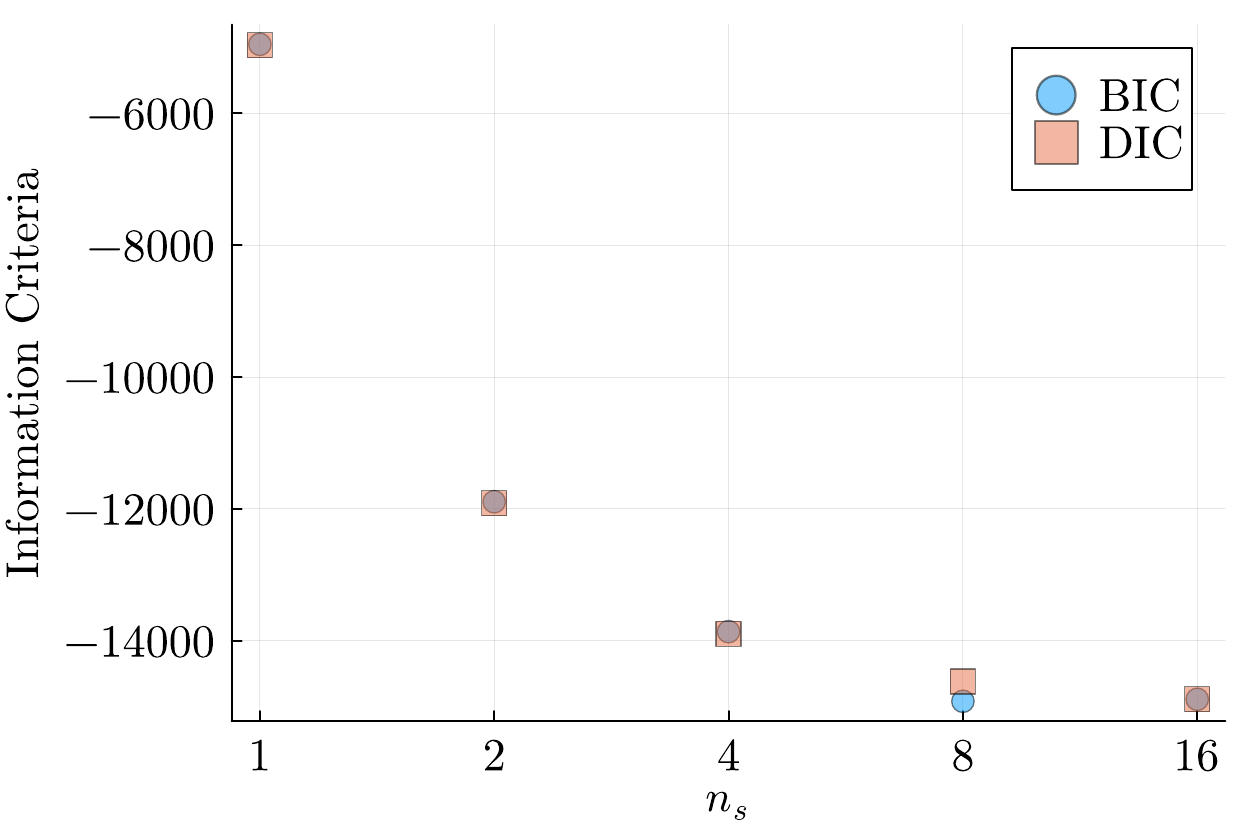}
                \caption{Full regime.}
                \label{fig-evol-S_like-full-regime}
            \end{subfigure}
            \hfill
            \begin{subfigure}{0.45\linewidth}
                \centering
                \includegraphics[width=\linewidth]{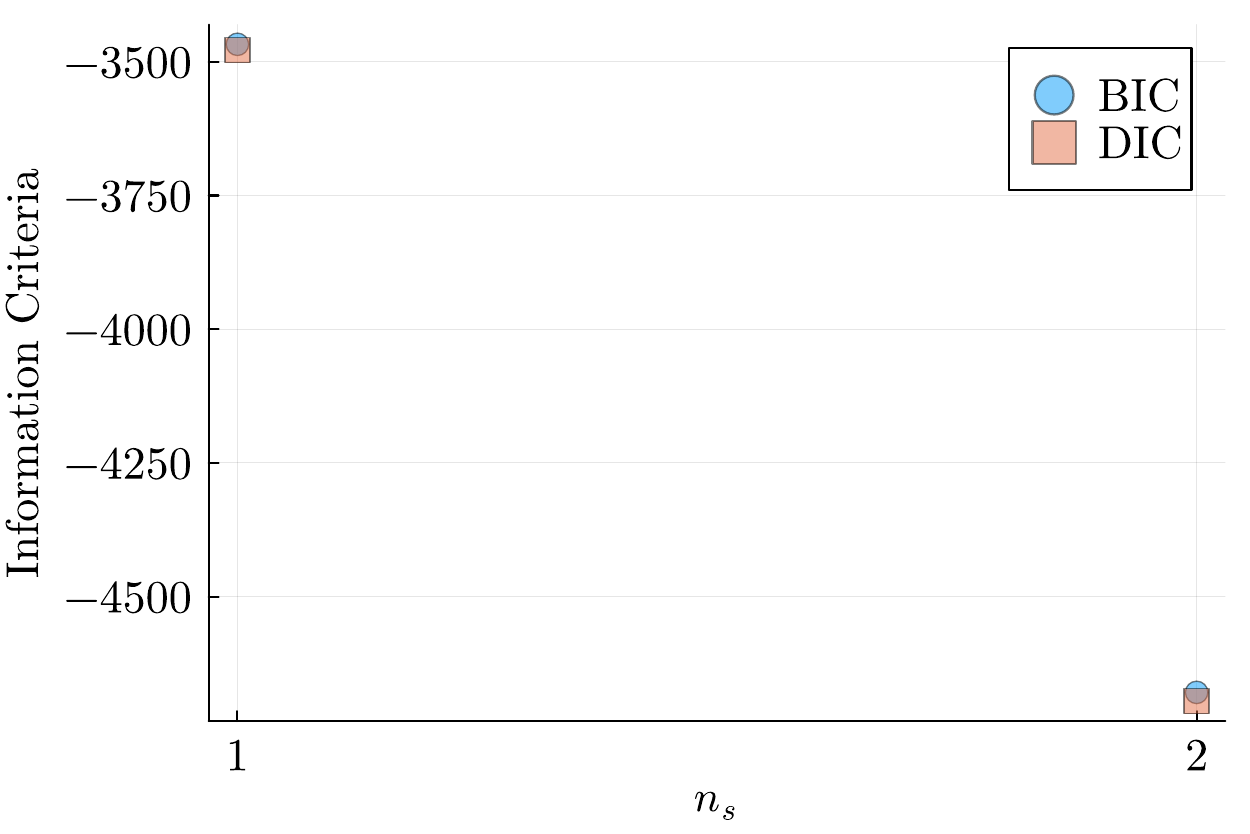}
                \caption{Steady state.}
                \label{fig-evol-S_like-steady-state}
            \end{subfigure}
            \caption{\textcolor{blue}{Evolution of $S_{\rm like}$ throughout the mesh adaptation iterations and information criteria for different model complexities.}}
            \label{fig-evol-S_like}
        \end{figure}

        \autoref{fig-evol-S_like} illustrates the evolution of $S_\text{like}$ throughout the mesh adaptation iterations for different model complexities, and the corresponding values of the information criteria, for both the full dataset and the steady-state subset.
        Each $S_\text{like}$ curve corresponds to a different model complexity, i.e., the number of segments $n_s$ considered for the piecewise linear functions.
        This figure shows how the gradient-based optimization minimizes $S_\text{like}$ as the spatial and temporal meshes are refined.
        The black dashed lines indicate the threshold $S_\text{like}^\text{Morozov}$.

\begin{figure}
            \centering
            \begin{subfigure}{0.47\linewidth}
                \centering
                \includegraphics[width=\linewidth]{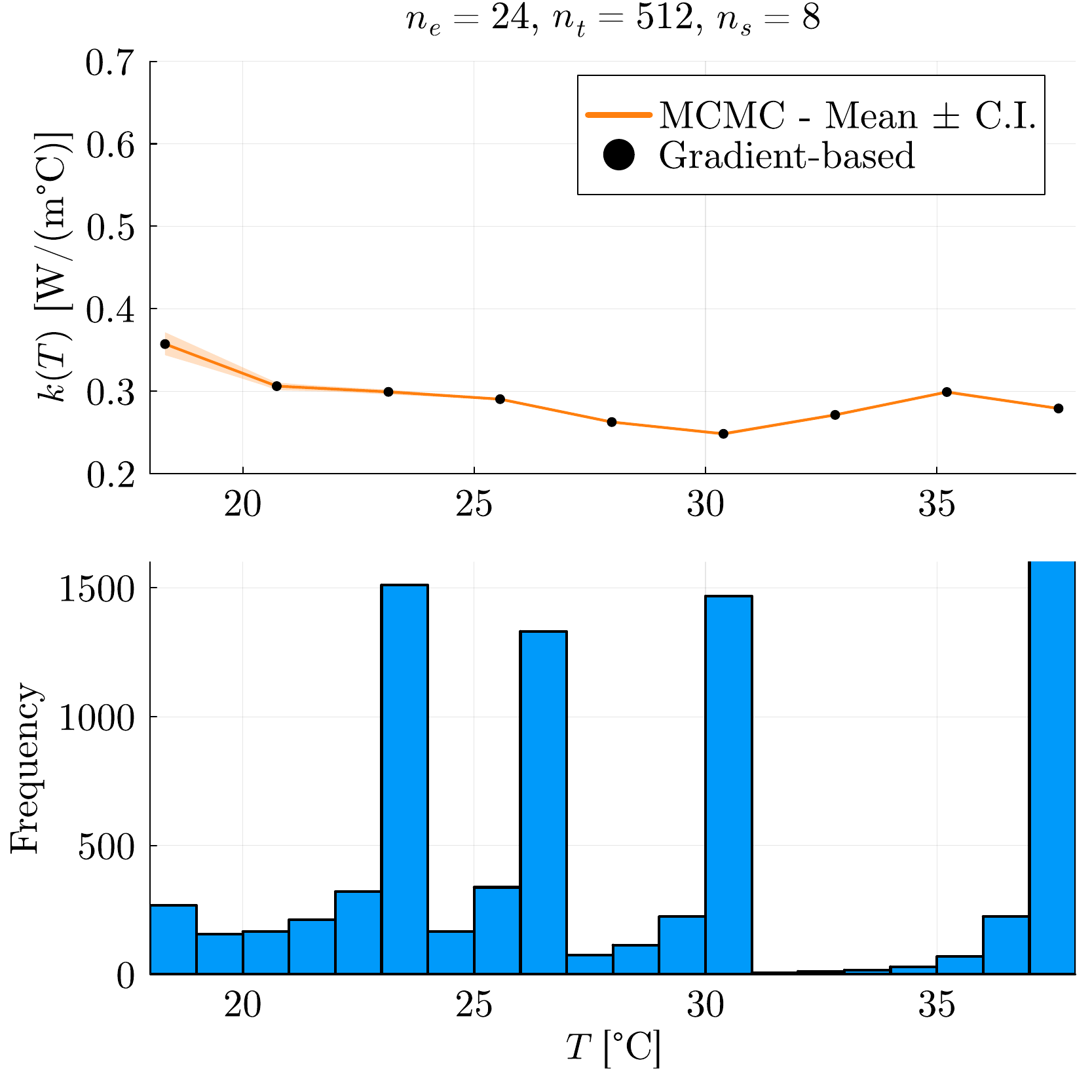}
                \caption{Full regime.}
                \label{fig-results-full-regime}
            \end{subfigure}
            \hfill
            \begin{subfigure}{0.47\linewidth}
                \centering
                \includegraphics[width=\linewidth]{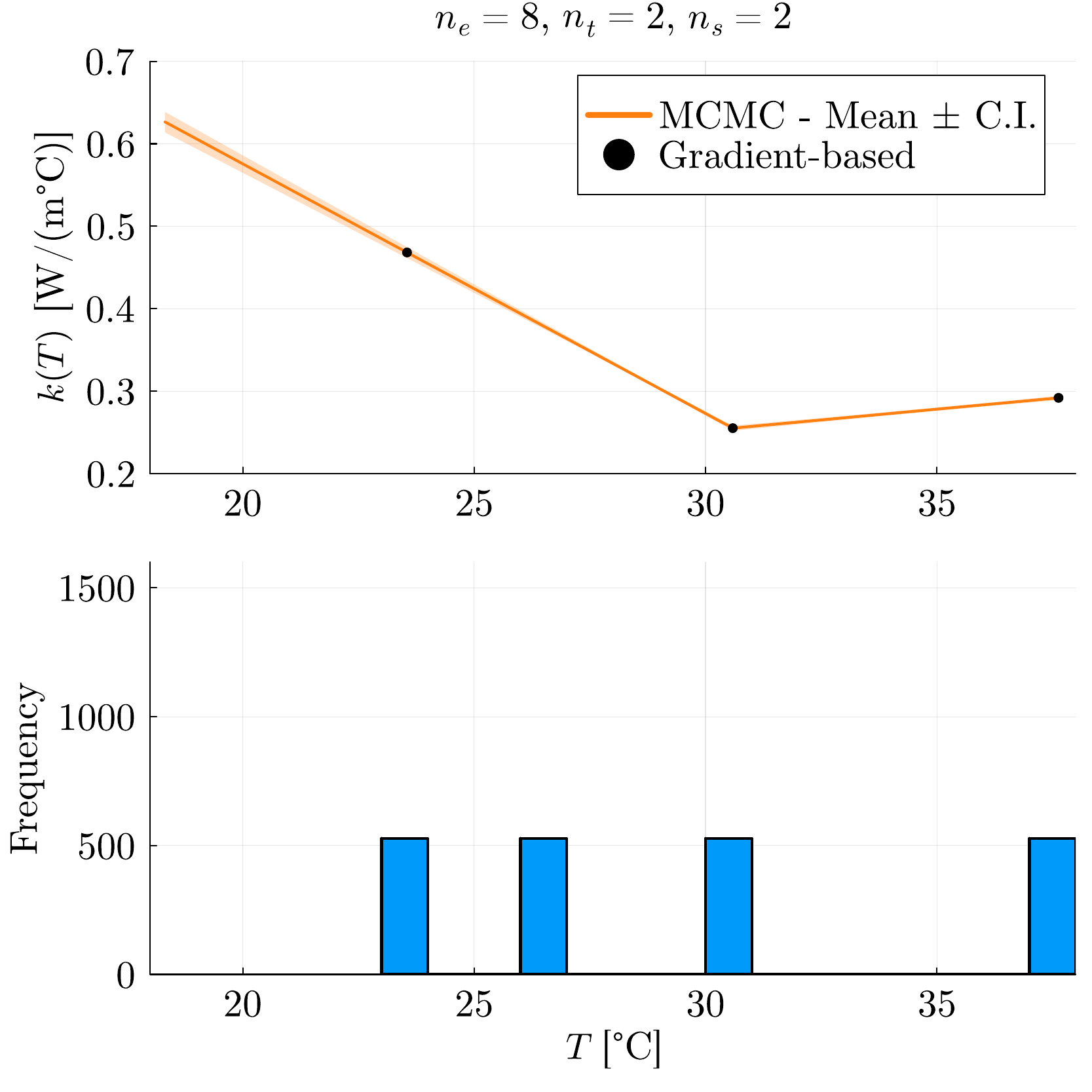}
                \caption{Steady state.}
                \label{fig-results-steady-state}
            \end{subfigure}
            \caption{Results obtained with different datasets for the real experimental data case.}
            \label{fig-results}
        \end{figure}

        \textcolor{blue}{
        For the case using the full dataset (\autoref{fig-evol-S_like-full-regime}), the algorithm selected the model with $n_s = 8$.
        In this case, values of $S_{\rm like}$ did not reach the threshold $S_{\rm like}^{\rm Morozov}$.
        However, the algorithm identified that further refinement would not be required, since no considerable improvement in the information criteria values was obtained when increasing the complexity to $n_s = 16$.
        Next to that, the algorithm selected $n_e = 24$ and $n_t = 512$, as shown in \autoref{fig-results-full-regime}.
        This choice reflects the need for high spatial and temporal resolution to capture transient dynamics, as well as a more complex conductivity model to represent the full temperature range.
        The corresponding histograms indicate that the estimation incorporates data covering the entire temperature range, with peaks concentrated around the steady state values.
        For the selected model complexity, the acceptance rate of the MCMC samples was 24\%, which is consistent with the targeted optimum.
        The effective sample sizes corresponding to the nine inferred conductivity values all lie between \num{1000} and \num{2900}.
        }
        
        \textcolor{blue}{
        For the case with the data covering only the steady state regime (\autoref{fig-evol-S_like-steady-state}), the threshold $S_{\rm like}^{\rm Morozov}$ is reached when $n_s = 2$, and the algorithm immediately selects this model complexity.
        In this case, the stopping criterion based on $S_{\rm like}^{\rm Morozov}$ is sufficient to determine the optimal balance between model accuracy and complexity.
        Therefore, the model selection based on the values of the information criteria is not actually required, since no additional model complexities are considered for comparison beyond $n_s = 2$.
        Next to that, the algorithm returned $n_e = 8$ and $n_t = 2$ (\autoref{fig-results-steady-state}).
        This outcome reflects coarser spatial and temporal meshes and a simpler model complexity, as fine resolution is not required when only steady state data are available.
        For the selected configuration, the MCMC acceptance rate was 24\%, and the effective sample sizes associated with the three inferred conductivity values were all approximately 8200.
        }

        \subsubsection{Computational cost analysis}

        \begin{table}
            \centering
            \caption{\textcolor{blue}{Computational units for each mesh refinement iteration and MCMC obtained with the real dataset.}}
            \begin{adjustbox}{max width=\linewidth}
            \begin{tabular}{cccccccc}
            \toprule
            & \multicolumn{5}{c}{Full regime} & \multicolumn{2}{c}{Steady state}
            \\
            \cmidrule(lr){2-6} \cmidrule(lr){7-8}
            Iterations & $n_s = 1$ & $n_s = 2$ & $n_s = 4$ & $n_s = 8$ & $n_s = 16$ & $n_s = 1$ & $n_s = 2$
            \\
            \midrule
            1  & $9.2 \times 10^1$ & $8.0 \times 10^1$ & $8.0 \times 10^1$ & $7.2 \times 10^1$ & $1.8 \times 10^2$ & $1.0 \times 10^2$ & $1.0 \times 10^2$
            \\
            2  & $2.4 \times 10^2$ & $2.1 \times 10^2$ & $2.2 \times 10^2$ & $1.4 \times 10^2$ & $4.7 \times 10^2$ & $5.9 \times 10^2$ & $7.0 \times 10^2$
            \\
            3  & $9.9 \times 10^2$ & $9.9 \times 10^2$ & $9.9 \times 10^2$ & $1.2 \times 10^3$ & $2.9 \times 10^3$ & $2.1 \times 10^3$ & $2.6 \times 10^3$
            \\
            4  & $2.8 \times 10^3$ & $2.1 \times 10^3$ & $1.9 \times 10^3$ & $2.6 \times 10^3$ & $5.8 \times 10^3$ & $8.4 \times 10^3$ & --
            \\
            5  & $8.4 \times 10^3$ & $7.8 \times 10^3$ & $3.7 \times 10^3$ & $8.9 \times 10^3$ & $2.7 \times 10^4$ & $1.7 \times 10^4$ & --
            \\
            6  & $1.7 \times 10^4$ & $1.7 \times 10^4$ & $1.6 \times 10^4$ & $1.6 \times 10^4$ & $6.0 \times 10^4$ & $6.8 \times 10^4$ & --
            \\
            \cdashline{7-7}[2pt/2pt]
            7  & $6.8 \times 10^4$ & $5.9 \times 10^4$ & $5.9 \times 10^4$ & $8.8 \times 10^4$ & $2.5 \times 10^5$ & $1.4 \times 10^5$ & --
            \\
            8  & $3.0 \times 10^5$ & $3.0 \times 10^5$ & $2.9 \times 10^5$ & $3.9 \times 10^5$ & $8.4 \times 10^5$ & $2.7 \times 10^5$ & --
            \\
            9  & $1.3 \times 10^6$ & $1.1 \times 10^6$ & $9.5 \times 10^5$ & $8.8 \times 10^5$ & $3.1 \times 10^6$ & -- & --
            \\
            10 & $4.3 \times 10^6$ & $4.3 \times 10^6$ & $4.0 \times 10^6$ & $4.0 \times 10^6$ & $1.1 \times 10^7$ & -- & --
            \\
            \cdashline{2-2}[2pt/2pt]
            \cdashline{4-4}[2pt/2pt]
            \cdashline{5-5}[2pt/2pt]
            \cdashline{6-6}[2pt/2pt]
            11 & $8.7 \times 10^6$ & $8.7 \times 10^6$ & $1.4 \times 10^7$ & $1.6 \times 10^7$ & $4.4 \times 10^7$ & -- & --
            \\
            \cdashline{3-3}[2pt/2pt]
            12 & $1.7 \times 10^7$ & $1.7 \times 10^7$ & $5.9 \times 10^7$ & $3.2 \times 10^7$ & $8.8 \times 10^7$ & -- & --
            \\
            13 & -- & $6.5 \times 10^7$ & -- & -- & -- & -- & --
            \\
            MCMC & $1.3 \times 10^{10}$ & $2.6 \times 10^{10}$ & $2.6 \times 10^{10}$ & $2.9 \times 10^{10}$ & $2.9 \times 10^{10}$ & $2.0 \times 10^{8}$ & $1.3 \times 10^{7}$
            \\
            \midrule
            Total per $n_s$ & $1.3 \times 10^{10}$ & $2.6 \times 10^{10}$ & $2.6 \times 10^{10}$ & $3.0 \times 10^{10}$ & $3.0 \times 10^{10}$ & $2.1 \times 10^{8}$ & $1.3 \times 10^{7}$
            \\
            \midrule
            Total over all $n_s$ &  &  &  &  & $1.6 \times 10^{11}$ &  & $2.2 \times 10^{8}$
            \\
            \bottomrule
            \end{tabular}
            \end{adjustbox}
            \label{tab-comp-units-steady-state}
        \end{table}
        
        \autoref{tab-comp-units-steady-state} gives an overview of the computational effort related to both the full regime and steady-state datasets.
        Assuming the use of a sparse direct matrix solver, the computational effort of a single iteration scales as $\mathcal{O}(n_{\rm eval} n_e^2 n_t)$, where $n_{\rm eval}$ is the number of model evaluations for a given spatial and temporal discretization $(n_e, n_t)$ and model complexity $n_s$.
        We refer to the case in which the product $n_{\rm eval} n_e^2 n_t = 1$ as one computational unit, and we list the computational units per algorithm step in \autoref{tab-comp-units-steady-state}.

        Mesh refinement stops either because Morozov’s discrepancy principle is satisfied or because $S_{\rm like}$ stabilizes (cf. \autoref{eq-stopping-criterion-rel-std}).
        In the latter case, the selected mesh refinement is the third-to-last one.
        In the table, this is indicated with a dashed line below the iteration of the selected mesh.
        
        We observe that the full regime dataset can make use of more model refinement steps, and more mesh refinement steps per model complexity, compared to the steady-state dataset.
        The reason for this is that the transient behavior requires a higher resolution in space and time, as well as a more refined conductivity parametrization on account of the wider temperature range being spanned.

        \autoref{tab-comp-units-steady-state} also reports the total computational units per model complexity and for the entire algorithm run.
        When using the full dataset, the total number of computational units is roughly 700 times larger than for the steady-state dataset.
        This ratio is consistent with the observed computational times, as the CPU time for the full dataset is also approximately 700 times higher than for the steady-state case.
        There are two key contributing factors to this difference.
        First, the full regime dataset requires finer meshes and time steps, thereby involving computational more demanding finite element simulations.
        Second, this case also requires one additional model complexity evaluation.
        Note that, if the method were applied to a three-dimensional problem, the number of computational units would exponentially increase, reflecting the higher cost of solving three-dimensional finite element problems with refined meshes and multiple model evaluations.

        \textcolor{blue}{
        Finally, a relevant distinction arises when considering only the use of the BIC for model selection.
        In particular, if only the BIC is employed, the MCMC sampling is not required for all candidate model complexities.
        As a result, the substantial computational effort associated with the MCMC sampling reported in \autoref{tab-comp-units-steady-state} would be avoided for all but the selected model.
        As result, the total computational cost would be significantly reduced, with MCMC sampling performed only once for the chosen model complexity to enable uncertainty quantification.
        }

%% file: conclusions_and_recommendations.tex
\section{Conclusions and recommendations}
\label{sec-conclusions}

    In this work, we developed a Bayesian framework for model calibration in heat conduction, where discretization fidelity and model complexity are jointly refined.
    By integrating gradient-based optimization with uncertainty quantification (UQ), we define an adaptive algorithm that balances accuracy of the estimates with confidence in the inferred parameters.
    The use of Morozov’s discrepancy principle as a stopping criterion for discretization refinement ensures that numerical errors were controlled relative to the measurement noise, while \textcolor{blue}{the analysis of the information criteria} guided the refinement of model complexity to prevent overfitting.
    This dual refinement strategy provides a systematic approach to determining both the spatial and temporal resolution of the numerical model as well as the functional complexity of the inferred constitutive relation.

    Application to synthetic data demonstrated that the algorithm reliably detects when mesh refinement should be stopped and when additional model complexity is needed.
    \textcolor{blue}{The results revealed the trade-off between improving data fit and model refinement, indicating that the proposed methodology improves computational efficiency while preserving robustness in parameter inference.}
    The methodology was then applied to real experimental data, and the algorithm successfully adapted to different datasets: when the full dataset covering both transient and steady state regimes was used, it converged to refined meshes and higher model complexity.
    In the case of steady state data only, it produced coarser meshes and a simpler conductivity representation.
    These results confirm that the algorithm adapts naturally to the information content of the data, and avoids unnecessary refinement and computational effort.

    Overall, the proposed approach provides a general strategy for inverse problems governed by PDEs, where the challenges of discretization error, model complexity, and data noise must be simultaneously considered.
    While this study focused on the estimation of a temperature-dependent thermal conductivity, the methodology is directly applicable to other physical systems that can use different types of models and discretization methods.

    Advanced adaptive mesh refinement techniques have been developed over the past decades.
    \textcolor{blue}{
    In the context of Bayesian techniques, where models are calibrated based on specific quantities of interest (i.e., measurement points), in particular goal-oriented error estimation and adaptivity techniques \citep{oden2001goal, oden2002estimation, becker2001optimal} are anticipated to be effective.
    }
    While such techniques represent an important and active area of research, they are orthogonal to the goals of this paper.
    Our focus is on the coordinated refinement of discretization and model complexity guided by the measurement noise, rather than on the development of advanced mesh adaptation algorithms.
    Integrating these mesh refinement strategies into our Bayesian framework does present an interesting direction for future extensions.

    Looking ahead, several recommendations arise.
    First, the algorithm could be extended to account for more general error structures, such as correlated measurement errors or modeling biases that go beyond simple Gaussian assumptions.
    Second, extending the framework to multiphysics problems or three-dimensional domains would demonstrate scalability to more realistic engineering applications.
    \textcolor{blue}{In this context, the computational cost of MCMC may increase substantially, making surrogate models essential to approximate the forward solution or the log-likelihood, while parallelization and GPU acceleration can be leveraged to further reduce the expense of the simulations.}
    Finally, while this work relied on Markov chain Monte Carlo for uncertainty quantification, alternative sampling or variational inference techniques could be explored to further improve computational efficiency.

    In conclusion, the framework presented here provides an approach to discretization-optimized Bayesian model calibration, ensuring that the trade-offs between accuracy, uncertainty, and computational cost are explicitly managed.
    By using uncertainty quantification as a component of the refinement process, the methodology not only delivers reliable estimates of material properties but also provides a systematic way to prevent overfitting and maintain consistency with the information content of the data.

%% file: acknowledgments.tex
\section*{Acknowledgments}

This research was conducted as part of the DAMOCLES project within the EMDAIR program of the Eindhoven Artificial Intelligence Systems Institute (EAISI).

%% file: data_availability.tex
\section*{Data availability statement}

Data and scripts that enable reproduction can be found at \url{https://doi.org/10.5281/zenodo.17642255}.

%% file: apendices.tex
\appendix

\section{Changes in loss model}
\label{appendix-changes-loss-model}

    \textcolor{blue}{
    To assess the potential impact of model errors on the inferred conductivity, we perform a compact sensitivity analysis focused on the heat exchange boundary conditions.
    Specifically, we increase the heat transfer coefficients in Newton's law of cooling by 2\% at both the top and bottom surfaces of the rod.
    Additionally, the lateral heat exchange term $2 h_{\rm side} / R$ is also increased by 2\%.
    These perturbations represent plausible variations in the boundary and heat loss models that could arise from modeling simplifications or uncertainties.
    }

\begin{figure}
        \centering
        \includegraphics[width=0.5\linewidth]{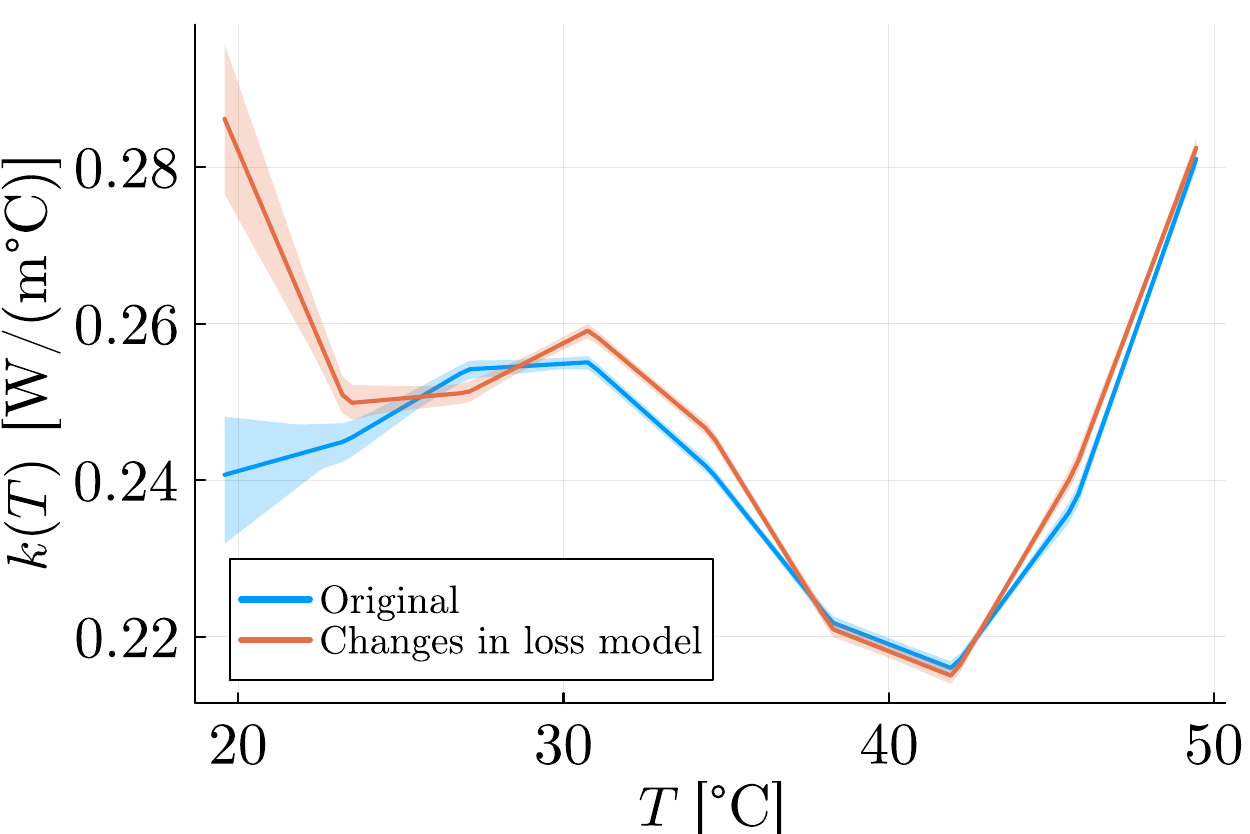}
        \caption{Comparison between original inferred conductivity curve and the one obtained after changing the loss model.}
        \label{fig-uq-changes-loss-model}
    \end{figure}

    \textcolor{blue}{
    The study is conducted using the synthetic data case, where the boundary conditions and physical properties are fully known.
    \autoref{fig-uq-changes-loss-model} compares the original inferred conductivity curve with the curve obtained under these modified boundary conditions and heat loss terms.
    Overall, the two curves show good agreement, indicating that the inference is relatively robust to small variations in the boundary modeling.
    However, a clear mismatch is observed at temperatures close to the initial condition, corresponding to early times in the simulation.
    This behavior is consistent with the one-dimensional approximation of the rod, which neglects radial temperature gradients that relax rapidly at the beginning of the transient.
    The mismatch highlights that, although the 1D model accurately represents the overall axial heat conduction, small deviations due to radial temperature gradients can influence the inferred conductivity near the initial temperature.
    }

    \textcolor{blue}{
    We note that model discrepancy terms, which account for structural errors in the forward model, can be treated explicitly within a Bayesian calibration framework \cite{kennedy2001bayesian, ling2014selection}.
    In any case, this approach does not influence the behavior of the algorithm and do not change the main conclusions drawn in this work.
    }

\section{Comparison between explicit and implicit integration schemes for the conductivity}
\label{appendix-implicit-conductivity}

    \textcolor{blue}{
    In this appendix, we compare the results obtained with explicit and implicit updating schemes for the temp\-era\-ture-dependent conductivity.
    For the implicit scheme, the conductivity at time step $m + 1$ is evaluated using the temperature at time step $m + 1$, which is unknown a priori.
    To address this, we implement a Picard iteration \cite{evans2022partial}, with a maximum number of 50 iterations and a relative tolerance of $10^{-9}$.
    }
    
    \textcolor{blue}{
    The comparison is carried out for the synthetic data described in \autoref{sec-investigation-model-optimization}, with the obtained optimized model complexity $n_s = 8$.
    We denote by $\bm{f}_{\rm ex}$ the solution obtained with the explicit scheme and by $\bm{f}_{\rm im}$ the solution obtained with the implicit scheme.
    These solutions represent vectors of temperature predictions at the four sensors.
    The then define the relative difference as $|| \bm{f}_{\rm ex} - \bm{f}_{\rm im} || / ||\bm{f}_{\rm im} ||$.
    }

\begin{figure}
        \centering
        \includegraphics[width=0.5\linewidth]{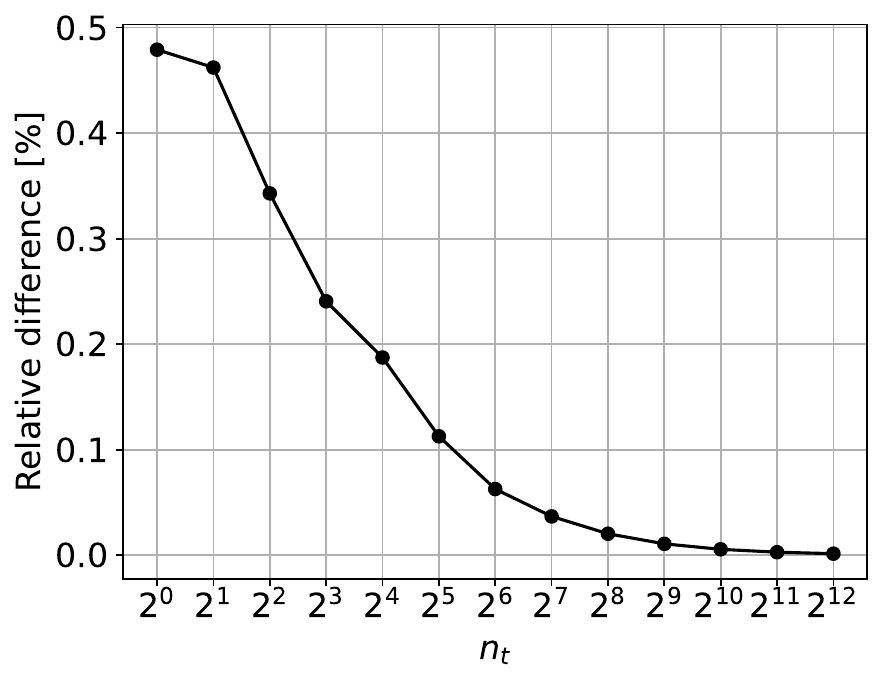}
        \caption{\textcolor{blue}{Relative difference between the solutions obtained with the explicit and implicit schemes for different number of time steps.}}
        \label{fig-nt-vs-rel-diff}
    \end{figure}

    \textcolor{blue}{
    \autoref{fig-nt-vs-rel-diff} shows the relative difference for different time steps.
    We note that the relative difference is already very small for small numbers of time steps.
    In particular, when $n_t = 2^{10} = 1024$, the relative difference is $6.5 \times 10^{-3} \%$, and the computational time required for the implicit scheme is approximately 4.5 times larger than that of the explicit scheme.
    So we use the explicit treatment of the conductivity in the main part of the paper.
    }
    
\section{Case study with multiple Markov chains}
\label{appendix-multiple-chains}

    \textcolor{blue}{
    In order to assess the robustness of the MCMC sampler, we study the impact of using multiple Markov chains for the case with the steady-state dataset.
    This configuration is selected because it requires a simpler conductivity parametrization and coarser spatial and temporal discretizations, which reduce the computational cost.
    This is particularly relevant when running several chains, as the total cost scales linearly with the number of chains.
    In this study, we generate three independent Markov chains initialized from different starting points.
    The initial guesses are sampled from the prior distributions, rather than taken from the gradient-based optimization, in order to promote dispersion in the initial states and to test robustness with respect to initialization.
    }

\begin{figure}
        \centering
        \includegraphics[width=0.5\linewidth]{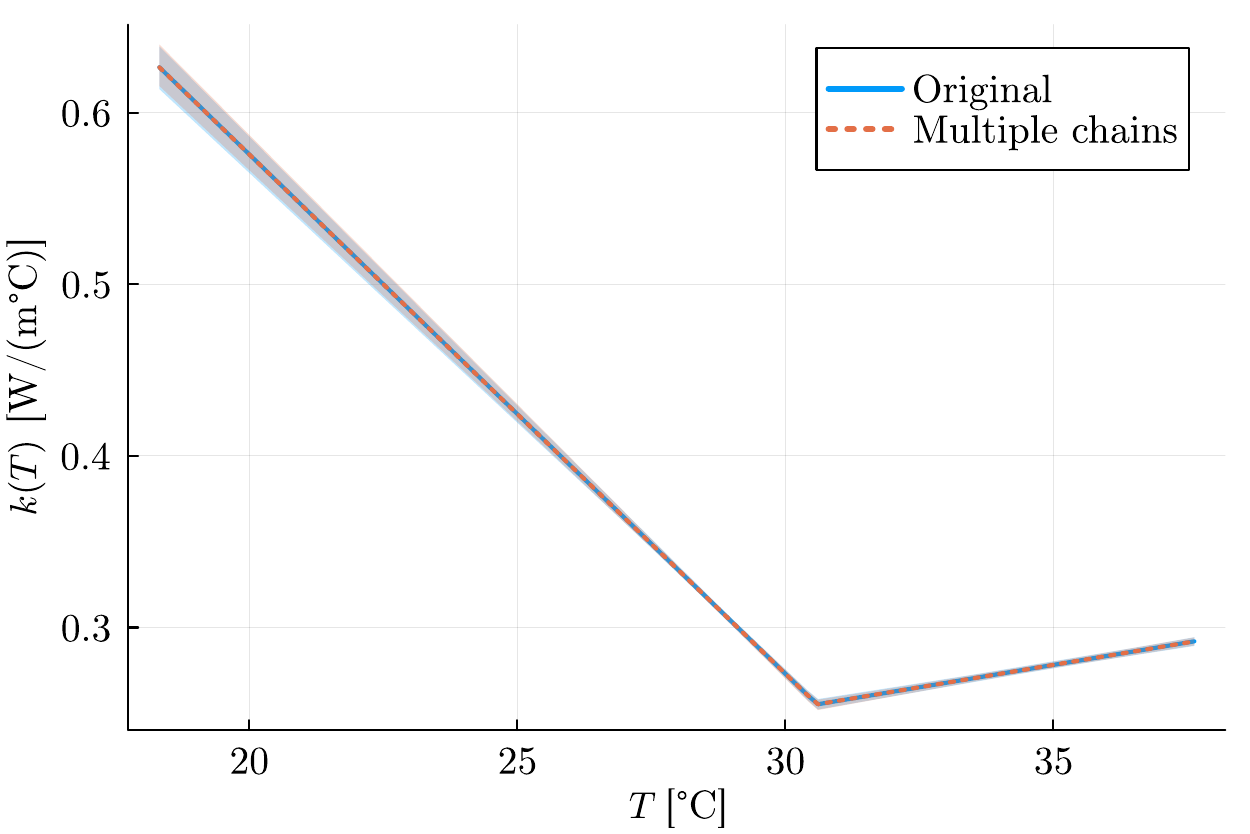}
        \caption{Comparison between the original inferred conductivity curve and the one obtained with multiple Markov chains.}
        \label{fig-uq-multiple-chains}
    \end{figure}

    \textcolor{blue}{
    Each chain is run for \num{100000} samples.
    Since the initialization is no longer informed by the gradient-based solution, more samples are required to reach equilibrium.
    Consequently, the burn-in period is set to the first \num{50000} samples, and the remaining \num{50000} samples of each chain are combined to represent the posterior distribution.
    \autoref{fig-uq-multiple-chains} compares the conductivity curve obtained in the original single-chain setting with the one inferred using the three chains.
    The posterior means and credible intervals are essentially the same in both cases.
    This indicates that, for the present problem, a single chain initialized at the gradient-based estimate is sufficient to obtain a reliable characterization of the posterior distribution of the conductivity.
    }

\section{Changing the measurement error bias and standard deviation}
\label{appendix-measurement-error}

\begin{figure}
        \centering
        \begin{subfigure}{0.49\linewidth}
            \centering
            \includegraphics[width=\linewidth]{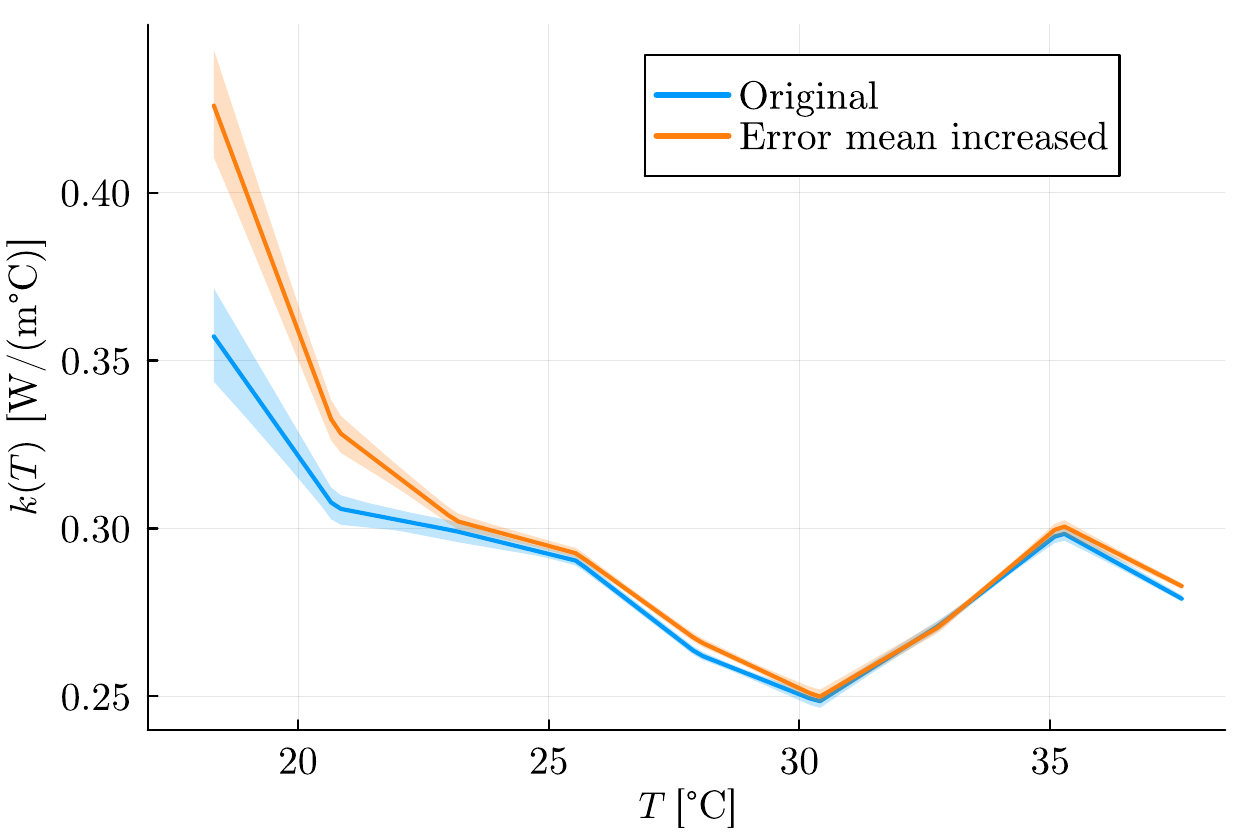}
            \caption{Error mean increased.}
        \end{subfigure}
        \hfill
        \begin{subfigure}{0.49\linewidth}
            \centering
            \includegraphics[width=\linewidth]{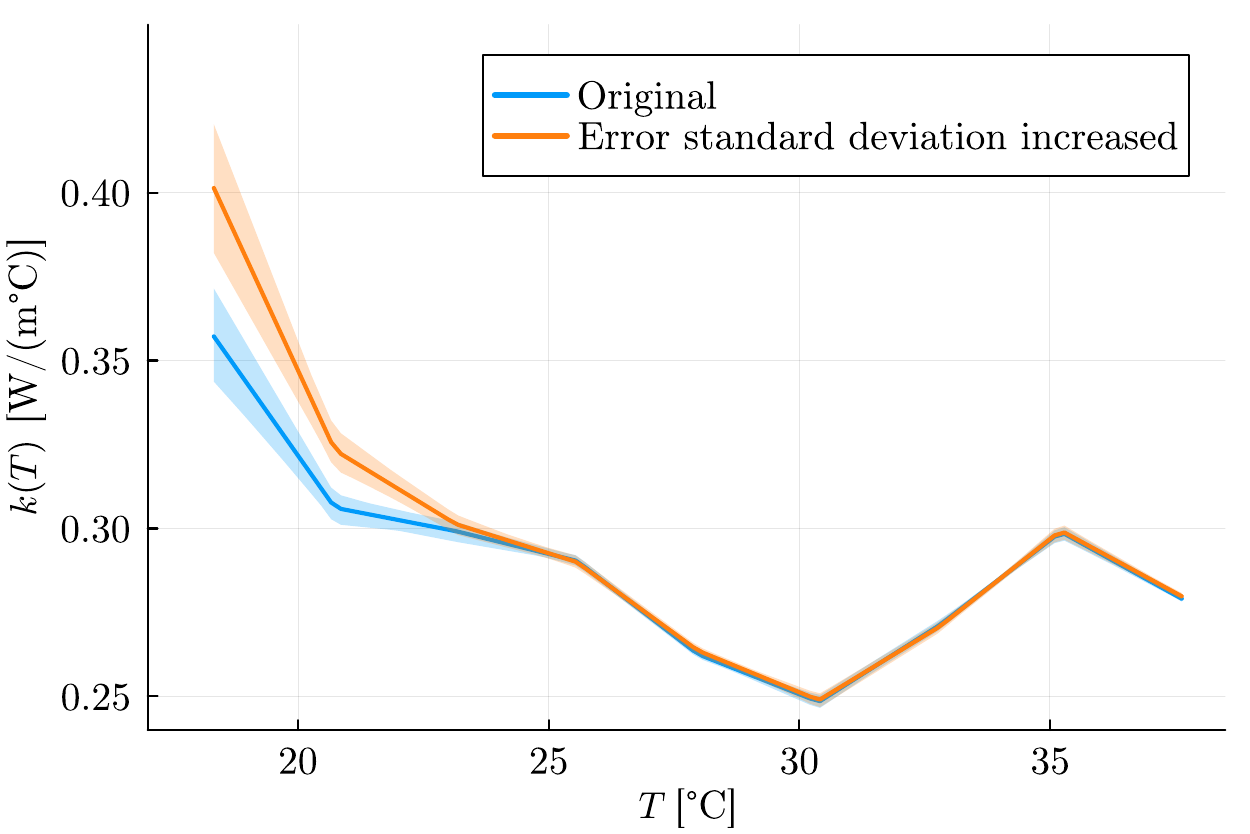}
            \caption{Error standard deviation increased.}
        \end{subfigure}
        \caption{Comparison between the inferred conductivity curves obtained with different measurement error models.}
        \label{fig-comparison-meas-error}
    \end{figure}

    \textcolor{blue}{
    To assess how the measurement error model influences the inferred conductivity, we conduct a compact sensitivity analysis in which the likelihood parameters are perturbed.
    \autoref{fig-comparison-meas-error} compares the posterior conductivity curves obtained under the original error model and under two modified scenarios:
    (i) a 10\% increase in the mean of the measurement error and (ii) a 10\% increase in its standard deviation.
    These perturbations are applied independently, yielding separate inference runs for each case.
    The objective is to evaluate the changes in both the posterior mean estimate and the associated credible intervals with respect to moderate misspecification of the likelihood.
    Note that the model bias and the standard deviation of the measurement error can be treated as hyperparameters within the Bayesian framework, so they are estimated jointly with the conductivity curve.
    }

    \textcolor{blue}{
    The results indicate that the inferred conductivity curves are not considerably sensitive to these perturbations.
    Across most of the temperature range, the posterior means and credible bands remain nearly unchanged relative to the baseline case.
    Noticeable deviations occur only in regions where the temperature is near the initial condition, which is consistent with the reduced data availability in this regime.
    For all three cases, the algorithm selected the same discretization and model refinement parameters, namely $n_e = 24$, $n_t = 512$, and $n_s = 8$, indicating that moderate changes in the measurement error model do not affect the mesh and model refinement decisions.
    Overall, this analysis suggests that the proposed framework yields stable conductivity estimates under reasonable variations of the assumed measurement noise characteristics.
    }
    
\section{Testing with different datasets}
\label{appendix-testing-different-datasets}

    \textcolor{blue}{
    We assess the robustness of the inferred temperature-dependent thermal conductivity by considering multiple independent realizations of the experiment. 
    We denote by Dataset 1 the dataset used in \autoref{sec-application-algorithm} to infer the conductivity. 
    Additionally, we consider two independent runs of the same experiment, denoted by Dataset 2 and Dataset 3.
    All three datasets correspond to the same geometry, instrumentation, and heating, but differ in their measured ambient temperature trajectories and experimental realizations.
    }
    
\begin{figure}
        \centering
        \begin{subfigure}{0.49\linewidth}
            \centering
            \includegraphics[width=\linewidth]{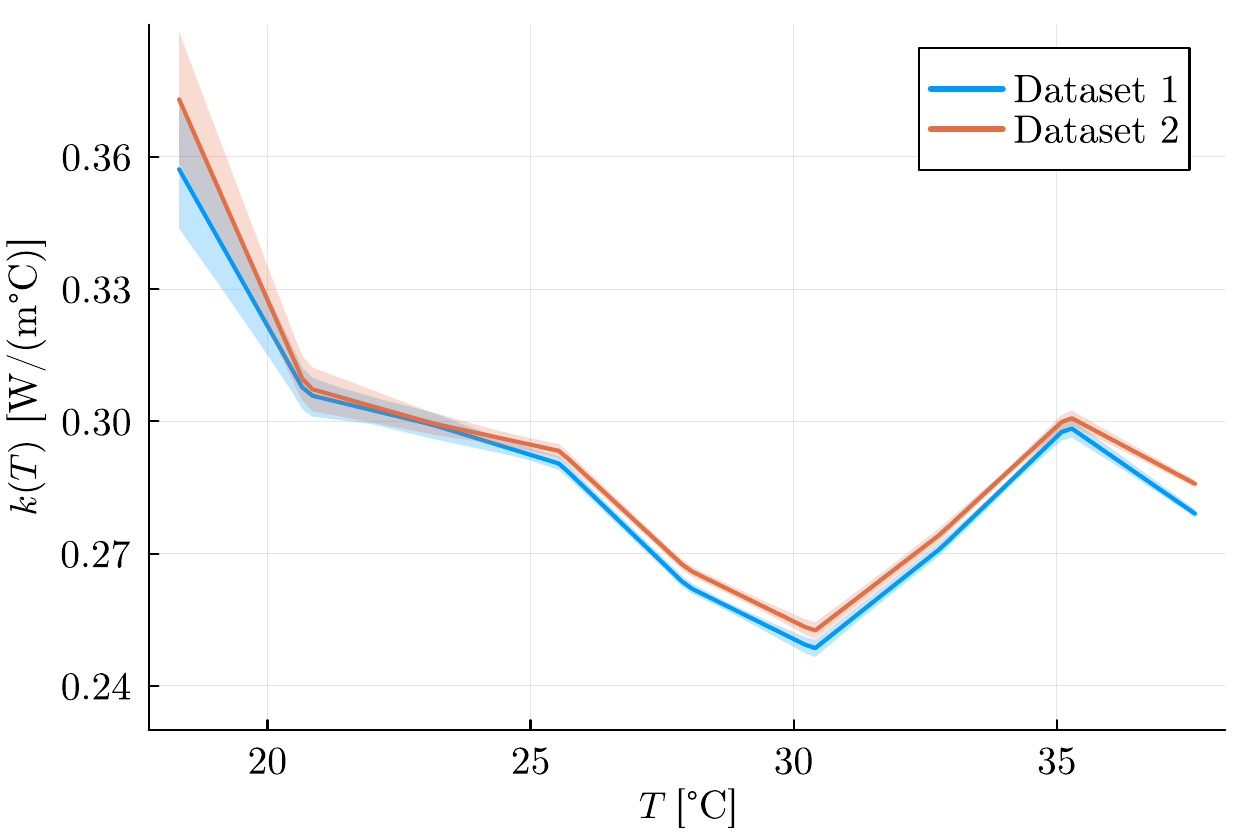}
            \caption{Datasets 1 and 2}
        \end{subfigure}
        \hfill
        \begin{subfigure}{0.49\linewidth}
            \centering
            \includegraphics[width=\linewidth]{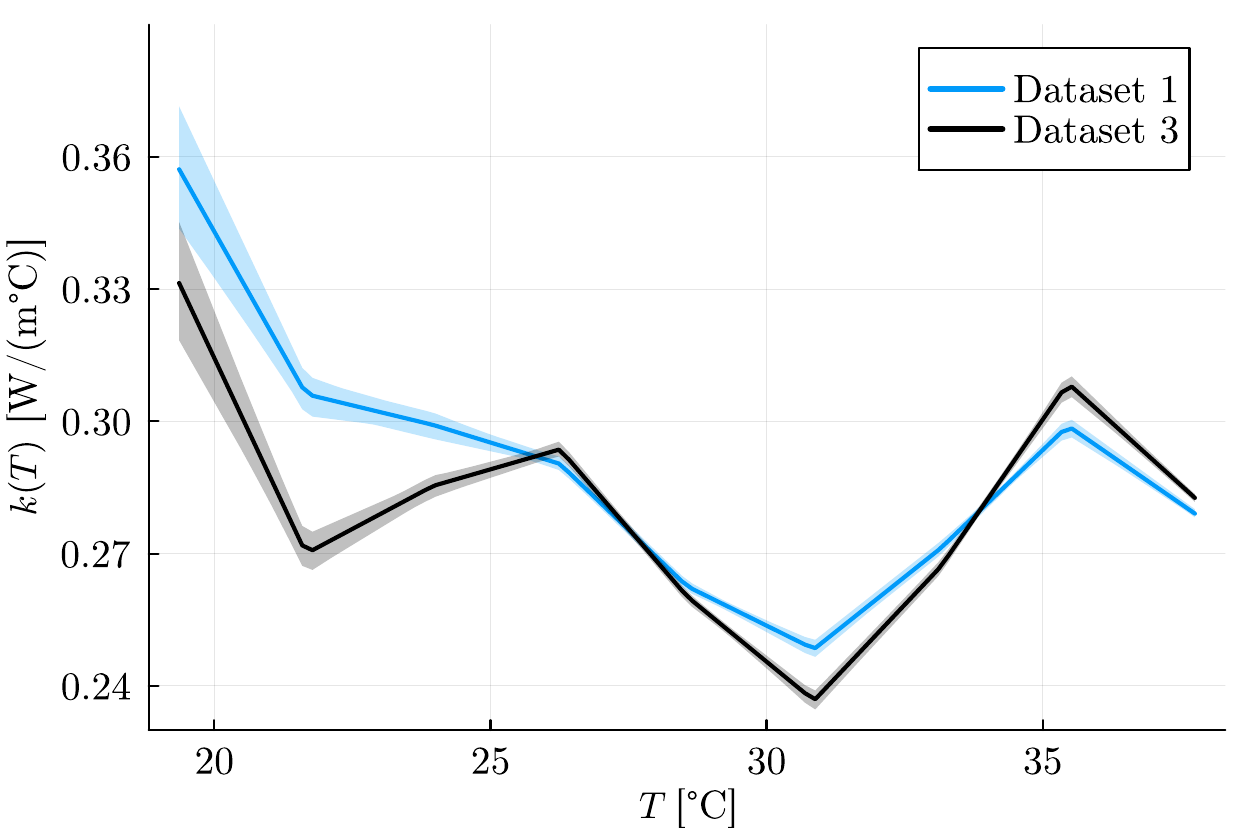}
            \caption{Datasets 1 and 3}
        \end{subfigure}
    \caption{Comparison between the inferred conductivity curves obtained with different datasets.}
    \label{fig-uq-multiple-runs}
    \end{figure}

    \textcolor{blue}{
    The conductivity curve inferred from Dataset~1 is used as reference and compared with the curves obtained when applying the same inference procedure to Datasets~2 and~3. 
    \autoref{fig-uq-multiple-runs} shows the posterior mean conductivity and associated uncertainty bands for the three datasets.
    The comparison between Dataset~1 and Dataset~2 shows good agreement across the considered temperature range. 
    The posterior means are close, and the corresponding credibility intervals exhibit substantial overlap.
    This indicates that the inferred conductivity curve is reproducible under comparable experimental conditions and is not dominated by a single realization of the data.
    For Dataset~3, a larger discrepancy is observed. 
    Differences in the posterior mean are more pronounced in certain temperature intervals, and the overlap between uncertainty bands is reduced.
    This behavior suggests that variations in ambient conditions can influence the inferred conductivity when the physical properties and boundary conditions are calibrated separately and subsequently kept fixed.
    }
    
    \textcolor{blue}{
    These results emphasize the sensitivity of the inverse problem to boundary and environmental variations, as presented in \ref{appendix-estimate-all-parameters}.
    Nevertheless, although this is a relevant topic of research, it is orthogonal to the focus of this work.
    }

\section{Joint estimation of physical properties and boundary conditions}
\label{appendix-estimate-all-parameters}

\begin{figure}
        \centering
        \includegraphics[width=0.5\linewidth]{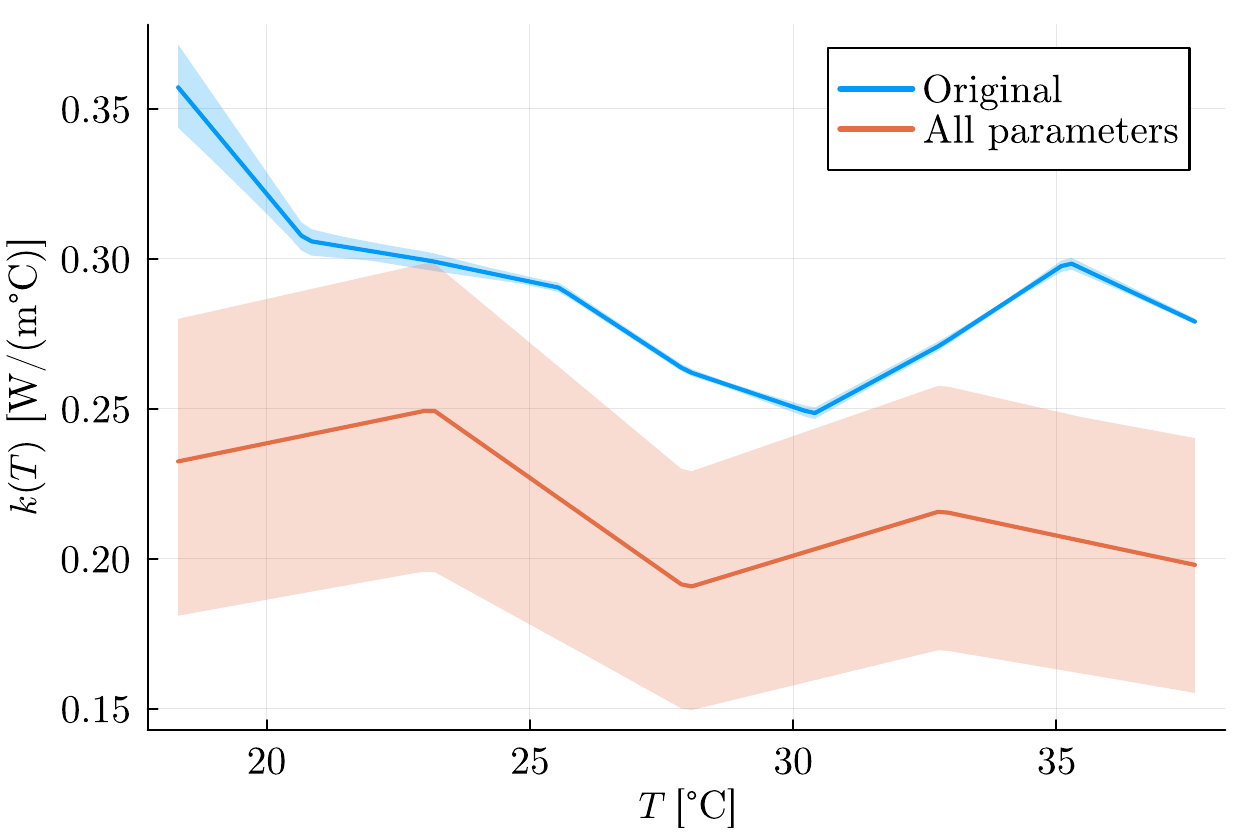}
        \caption{Comparison between the original inferred conductivity curve and the one obtained while estimating the physical properties and boundary conditions.}
        \label{fig-UQ-all-parameters}
    \end{figure}

    \textcolor{blue}{
    \autoref{fig-UQ-all-parameters} compares the conductivity curve obtained when the physical properties and boundary conditions are first calibrated and then kept fixed, with the result of a joint estimation using the complete dataset.
    In the joint case, the MCMC method requires \num{500000} samples, i.e., five times more than in the original setting, resulting in a substantial increase in simulation time.
    Next to that, the last \num{250000} samples are used to represent the posterior distribution.
    }

    \textcolor{blue}{
    We see the following key differences between the original estimation and the joint estimation case:
    \begin{itemize}
        \item The resulting credible intervals are visibly wider than the original one.
        This reflects the propagation of uncertainty from the physical properties and boundary conditions into the conductivity through the posterior distribution, as all parameters are inferred simultaneously rather than treated as fixed after calibration.
        \item Fewer segments for the piecewise linear functions are needed to model the temperature dependence.
        This is due to the fact that, in the joint estimation case, the information criteria indicate no further improvement already at a lower model complexity when compared to the original case.
        \item The estimated conductivity values are generally lower than those obtained in the original case.
        This shows that the choice to fix certain parameters can have a significant impact and care should be taken when calibrating them.
        Analyses such as the one in this appendix can be part of that, but they do not affect the format of the algorithm we propose.
    \end{itemize}
    }

